\documentclass[aps,10pt,notitlepage,nofootinbib]{revtex4-1}

\usepackage[utf8]{inputenc}
\usepackage{newtxtext,newtxmath}
\usepackage{bbold}
\usepackage{bm}
\usepackage{graphicx}
\usepackage[usenames,dvipsnames]{xcolor}
\usepackage{color}
\usepackage[colorlinks=true,linkcolor=blue,urlcolor=blue,citecolor=blue]{hyperref}
\usepackage{amsmath,amssymb,amsfonts}
\usepackage{slashed}
\usepackage[english]{babel}
\usepackage{dcolumn}
\usepackage{blindtext}
\usepackage{pifont}
\usepackage{dsfont,mathrsfs}
\usepackage{cancel}
\usepackage{bigints}
\usepackage{accents}
\usepackage{soul}
\usepackage{multirow}

\newcommand{\nn}{\nonumber}

\newcommand{\ensembleaverage}[1]{\left\langle#1\right\rangle}

\newcommand{\Ensembleaverage}[1]{\langle#1\rangle}

\newcommand{\FB}[1]{\left(#1\right)}

\newcommand{\SB}[1]{\left\{#1\right\}}
\newcommand{\TB}[1]{\left[#1\right]}

\newcommand{\scrL}{\mathscr{L}}

\newcommand{\scrD}{\mathscr{D}}
\newcommand{\sign}[1]{\text{sign}\left(#1\right)}
\newcommand{\munu}{{\mu\nu}}

\newcommand{\alphabeta}{{\alpha\beta}}
\newcommand{\IM}{\text{Im}}
\newcommand{\RE}{\text{Re}}
\newcommand{\Tr}{\text{Tr}}

\newcommand{\Sigmabar}{\overline{\Sigma}}
\newcommand{\Sigmabarbar}{\overline{\Sigmabar}}
\newcommand{\Pibar}{\overline{\Pi}}
\newcommand{\Pibarbar}{\overline{\Pibar}}
\newcommand{\Pbar}{\overline{P}}
\newcommand{\Pbarbar}{\overline{\Pbar}}

\newcommand{\Dbar}{\overline{D}}
\newcommand{\Dbarbar}{\overline{\Dbar}}

\newcommand{\kpll}{k_\parallel}
\newcommand{\qpll}{q_\parallel}
\newcommand{\ppll}{p_\parallel}

\newcommand{\kper}{k_\perp}
\newcommand{\qper}{q_\perp}
\newcommand{\gpll}{g_\parallel}
\newcommand{\gper}{g_\perp}

\newcommand{\utilde}{\tilde{u}}
\newcommand{\btilde}{\tilde{b}}

\newcommand{\qdotb}{q\cdot b}

\newcommand{\del}{\partial}
\newcommand{\identity}{\mathds{1}}

\begin{document}
\title{Thermo-magnetic spectral properties of neutral mesons in vector and axial-vector channels using NJL model}

\author{Snigdha Ghosh$^{a}$}
\email{snigdha.physics@gmail.com, snigdha.ghosh@saha.ac.in}
\author{Arghya Mukherjee$^{a,c}$}
\email{arghya.mukherjee@saha.ac.in}
\thanks{(Corresponding Author)}
\author{Nilanjan Chaudhuri$^{b,c}$}
\email{sovon.nilanjan@gmail.com}
\author{ Pradip Roy$^{a,c}$}
\email{pradipk.roy@saha.ac.in}
\author{Sourav Sarkar$^{b,c}$}
\email{sourav@vecc.gov.in}
\affiliation{$^a$Saha Institute of Nuclear Physics, 1/AF Bidhannagar, Kolkata - 700064, India}
\affiliation{$^b$Variable Energy Cyclotron Centre, 1/AF Bidhannagar, Kolkata 700 064, India}
\affiliation{$^c$Homi Bhabha National Institute, Training School Complex, Anushaktinagar, Mumbai - 400085, India}

\begin{abstract}
In this work the neutral meson properties have been investigated in the presence of thermo-magnetic background using two-flavor Nambu--Jona-Lasinio model. Mass, spectral function and dispersion relations are obtained in the scalar ($\sigma$) and pseudo-scalar ($\pi^0$) channels as well as in the vector ($\rho^0$) and axial vector ($a^0_1$) channels. The general Lorentz structures for the vector and axial-vector meson polarization functions have been considered in detail. The ultra-violet divergences appearing in this work have been regularized using a mixed regularization technique where the gamma functions arising in dimensional regularization are replaced with incomplete gamma functions as usually done in the proper time regularization procedure. The meson spectral functions obtained in the presence of magnetic field possess non-trivial oscillatory structure. Similar to the scalar and pseudo-scalar channel, the spectral functions for each of the modes of $\rho^0$ are observed to overlap with the corresponding modes of its chiral partner $a_1^0$ mesons in the chiral symmetry restored phase. We observe discontinuities in the masses of all the mesonic excitations for non-zero external magnetic field. 
\end{abstract}

\maketitle
%
\section{INTRODUCTION}\label{sec.intro}
Based on a considerable amount of research regarding the generation of  magnetic fields in non-central heavy ion collision (HIC), there exists a growing consensus  that an extremely strong transient magnetic  field of the order of $\sim10^{18}$ G or larger can be produced at RHIC and LHC~\cite{Lect_note,Kharzeev,Skokov2,Duncan:1992hi,PhysRevC.83.054911,refId0,PhysRevC.96.034902,PhysRevC.96.054909}. Being comparable to the energy scale of strong interaction,  though short lived, the produced field can impart significant modifications in the properties of strongly interacting matter ~\cite{PhysRevC.93.014905,PhysRevC.85.044907,PhysRevLett.110.192301,PhysRevC.87.024912,AdvHighEnergyPhys2013,PhysRevC.83.017901} resulting in a plethora of  novel phenomena like chiral magnetic effect~\cite{Fukushima,Kharzeev,Kharzeev2}, magnetic catalysis~\cite{Shovkovy,Gusynin1,Gusynin2,Gusynin3}, inverse magnetic catalysis~\cite{Bali,Preis,Preis2} electromagnetically induced superconductivity and superfluidity~\cite{Chernodub1,Chernodub2} and so on. The tools and techniques developed for studying such magnetic modifications in HIC experiments also bear significant importance for their applicability in  many different physical scenarios where strong magnetic field can be realized.  For example, in the early universe during the electroweak phase transition, magnetic fields as high as  $\sim10^{23}$G might have been produced.  Also, in case of \textit{magnetars} surface magnetic field is of the order of $\sim10^{15}$G.  In the  interior, the  field intensity is even higher reaching up to $\sim10^{18}$G. Such low temperature and high density extreme states are expected to be explored in the Compressed Baryonic Matter (CBM) experiment at Facility for Antiproton and Ion Research (FAIR).  On theoretical grounds, at lower temperatures, usual field theoretic approach of studying \textit{quantum chromodynamics} (QCD)  is not feasible due to  the confining nature of  strong interaction that severely restricts the applicability of perturbative analysis. In this scenario, an alternative to the non-perturbative Lattice QCD approach is provided by the QCD inspired effective models. The modification of such effective descriptions in presence of external magnetic field has gained significant research interests in recent times~\cite{,RevModPhys.88.025001}. One such model is the Nambu--Jona-Lasinio model ~\cite{Nambu1,Nambu2,Klevansky,Hatsuda1,Vogl,Buballa} which has been widely used in the studies of chiral symmetry breaking  as well as  meson properties  in presence of thermo-magnetic background~\cite{Klevansky2,Sadooghi1,Mao,Ruggieri}.

In the context of studying the mesonic properties in presence of magnetic field in NJL model, often the lightest mesons $\sigma$ and $\pi$ are considered \cite{PhysRevD.93.014010,PhysRevD.96.034004,AVANCINI2017247,PhysRevD.97.034026,PhysRevD.99.056005,PhysRevD.99.056009}. In some studies diquarks are also included \cite{PhysRevD.97.076008}. $\rho$ meson properties have been discussed in Ref.~\cite{zhang,Liu_PRD_2015,Liu_CPC_2016}. In Ref.~\cite{zhang}, it is observed that at vanishing magnetic field, there exists a temperature when $\rho$ mass coincides with  twice the mass of the  constituent quark and beyond that temperature no solution for the $\rho$ meson mass exists which is described as the $\rho$ melting. Even at  finite magnetic field the melting  persists and two different melting temperatures are  observed corresponding to the  charged and the  neutral $\rho$. Comparison with $\rho^0$ meson suggests  that  melting of  $\rho^{\pm}$ occurs at lower temperature in the  presence of magnetic field. For example, in case of  charged $\rho$, no solution exists beyond temperature 169 MeV for  $eB$ around 0.2 GeV$^2$. However, similar analysis on $\rho^{\pm}$ in Ritus formalism \cite{Liu_CPC_2016} does find non vanishing mass for charged $\rho$ even at much higher values of temperature for similar strength of the background magnetic field (see for example Fig.4 of Ref.~\cite{Liu_CPC_2016}).  The apparent ambiguity thus demands investigation of the properties of neutral $\rho$ meson in thermo-magnetic background which essentially will be an extension of the study presented in Ref.~\cite{Liu_CPC_2016}.

On a different note, one of the significant features of studying the meson properties is that at temperatures higher than a critical value, the masses of the chiral partners become degenerate. This degeneracy in the meson mass spectrum  serves as an important signature of the  chiral symmetry restoration. Therefore, the restoration of  chiral symmetry  in the vector channel  can  be shown explicitly only when one includes the  $a_1$ meson  along with $\rho$  which is missing in  the studies of $\rho$ mesons discussed earlier \cite{zhang,Liu_CPC_2016}. It should be mentioned here that in order to  investigate  the vector and pseudo-vector channel, proper  incorporation of the general structure of meson self-energy is required. The general Lorentz structure for the $\rho$ meson in presence of thermo-magnetic medium  has been recently reported in Ref.~\cite{Snigdha_2019}. One may note that  the Lorentz structure of $\rho$ meson polarization function has not been considered in Ref.~\cite{zhang}.

In this work the neutral meson properties in scalar ($\sigma$) and  pseudo-scalar ($\pi^0$) channels  as well as  vector ($\rho^0$) and axial vector ($a^0_1$) channels have been investigated in the framework of two-flavor NJL model in presence of constant background magnetic field. The detailed general structure for the vector and axial-vector meson polarization tensor  have been considered.  The Schwinger scheme has been implemented in the evaluation of the polarization tensors. However, as only the  neutral mesons are considered, the Schwinger phase vanishes and Schwinger and Ritus formalisms  are expected to provide identical results \cite{PhysRevD.99.056005}. It should be mentioned here that being an effective description of  QCD at low energy regime, NJL model is non-renormalizable and requires a regularization prescription. The most commonly used regularization technique is to use a three momentum cut-off which acts as a parameter of the theory and can be fixed  to reproduce some well-known phenomenological quantities, for example the pion-decay constant and the condensate value. However, to obtain the general structure of the self energy in  a consistent  way,  we take recourse to dimensional regularization (DR) technique. Now, the ultra-violet divergences   in dimensional regularization prescription occurs as pole of gamma function. In that procedure, one extra  parameter arises which is to be simultaneously fitted to reproduce the phenomenological quantities. Detailed description regarding the fitting procedure can be found in Refs.~\cite{KREWALD1992201,KOHYAMA2015682,PhysRevD.99.116002}.  However, in this work,  to obtain the finite contribution,   the gamma functions arising from DR  are replaced with incomplete gamma functions. We refer this replacement procedure   as incomplete gamma regularization (IGR). As a reward, though  the number of parameter set remains identical to that of usual regularization procedures, in this scheme, the general Lorentz structure for vector and axial vector polarization functions can be obtained systematically. The regularization scheme has been used to obtain the neutral meson properties like mass, spectral function and dispersion relations. Non-trivial mass jump  is observed in the spectrum for each of the modes in the vector and axial vector channel which bears similarity with earlier studies of pions in presence of magnetic field~\cite{PhysRevD.99.056005,PhysRevD.99.056009}.

The article is organized as follows. Sec.~\ref{sec.quark} describes the constituent quark mass and the dressed quark propagators in the real time formalism of thermal field theory whereas the gap equations and general structure are described in Sec.~\ref{sec.propagator}. 
In both the sections, vacuum, thermal and thermo-magnetic cases are considered in separate subsections. 
The main results for the real and imaginary parts of the meson polarization functions are listed in Sec.~\ref{sec.pola}. Sec.~\ref{sec.regular} describes the regularization procedure used in this work.
All the numerical results are presented in Sec.~\ref{sec.results} followed by a brief summary in Sec.~\ref{sec.summary}. 
Some of the relevant calculational details are provided in the appendices.

\section{THE CONSTITUENT QUARK MASS AND THE DRESSED QUARK PROPAGATOR} \label{sec.quark}
The standard expression of the two-flavor NJL Lagrangian is 
\begin{eqnarray}
\scrL_\text{NJL} = \overline{\psi}(i\gamma^\mu\del_\mu-m)\psi + g_s\SB{(\overline{\psi}\psi)(\overline{\psi}\psi) 
	- (\overline{\psi}\gamma^5\vec{\tau}\psi)\cdot(\overline{\psi}\gamma^5\vec{\tau}\psi)} \nn \\
- g_v\SB{(\overline{\psi}\gamma^\mu\vec{\tau}\psi)\cdot(\overline{\psi}\gamma_\mu\vec{\tau}\psi) 
	+ (\overline{\psi}\gamma^\mu\gamma^5\vec{\tau}\psi)\cdot(\overline{\psi}\gamma_\mu\gamma^5\vec{\tau}\psi)}
\label{eq.Lagrangian}
\end{eqnarray}
where, $\psi=\begin{pmatrix} u \\ d \end{pmatrix}$ is the quark isospin flavor doublet with $u$ and $d$ being the 
up and down quark fields respectively. Each of the up and down quark fields are $\TB{4\times1}\otimes\TB{3\times1}$ matrices 
corresponding to their orientation in Dirac and color spaces. 
In Eq.~\eqref{eq.Lagrangian}, $g_s$ and $g_v$ are respectively the coupling constants in the spin-0 and spin-1 channels 
for the four point contact interactions among the quark fields and $m$ is the current quark mass which is assumed to be equal for the 
up and down quarks to ensure isospin symmetry. 
In the NJL model the constituent quark mass is dynamically generated  as a consequence of the spontaneous 
breaking of chiral symmetry. In the following subsections, we briefly introduce the formalism required to obtain the constituent quark mass and the dressed quark propagator 
for three different cases separately: (i) $T=0$, $B=0$, (ii) $T\ne0$, $B=0$ and (iii) $T\ne0$, $B\ne0$.

\subsection{CASE-I: \boldmath{$T=0, B=0$}}
We first consider the \textit{pure vacuum} case  for which the temperature is zero and external magnetic field is 
switched off. 
The \textit{dressed} quark propagator $S'(q)$ is calculated from the Dyson-Schwinger equation
\begin{eqnarray}
S' = S - S\Sigma S'
\label{eq.DS}
\end{eqnarray}
where, $S(q,m)=\frac{-(\cancel{q}+m)}{q^2-m^2+i\epsilon}\otimes\identity_\text{Flavor}\otimes\identity_\text{Color}$ is the free quark 
propagator and $\Sigma$ is the one-loop self energy of quark. In the Mean Field Approximation(MFA), the quark self energy becomes diagonal in 
Dirac, color and flavor spaces as
\begin{eqnarray}
\Sigma=\Sigma_\text{MFA}\identity_\text{Dirac}\otimes\identity_\text{Flavor}\otimes\identity_\text{Color}.
\end{eqnarray} 
This enables one to solve Eq.~\eqref{eq.DS} trivially to get the complete propagator as
\begin{eqnarray}
S'(q,m) = S(q,M) = \frac{-(\cancel{q}+M)}{q^2-M^2+i\epsilon}\otimes\identity_\text{Flavor}\otimes\identity_\text{Color}
\label{eq.dressed.propagator}
\end{eqnarray}
where
\begin{eqnarray}
M = m + \RE\Sigma_\text{MFA}^\text{Pure-Vac}
\label{eq.gap}
\end{eqnarray}
is the \textit{`constituent quark mass'}. The above equation is the well known `\textit{gap equation}'. 
\begin{figure}[h]
	\begin{center}
		\includegraphics[angle=0,scale=0.50]{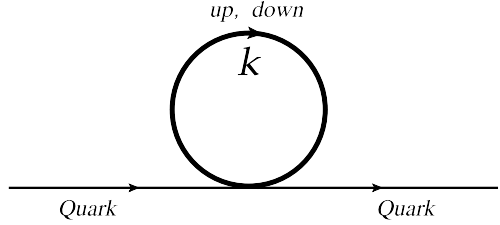}
	\end{center}
	\caption{Feynman diagram for one-loop quark self energy. The bold line corresponds to \textit{`complete/dressed'} quark 
		propagator obtained from the Dyson-Schwinger sum.}
	\label{Fig.DS}
\end{figure}

Our next task is to calculate the quantity $\Sigma_\text{MFA}$. 
Applying Feynman rules to Fig.~\ref{Fig.DS}, we get the one-loop self energy of quark in the MFA as
\begin{eqnarray}
\Sigma_\text{MFA}^\text{Pure-Vac}(M) = -2ig_s\int\frac{d^4k}{(2\pi)^4}\Tr_\text{c,f,d}\TB{S'(k,m)} 
= -2ig_s\int\frac{d^4k}{(2\pi)^4}\Tr_\text{c,f,d}\TB{S(k,M)}.
\label{eq.quark.self.energy}
\end{eqnarray}
It is to be noted that, the loop particle in the self energy  is dressed. 
In the above equation, the subscripts $c,d$ and $f$ in the $\Tr$ correspond to the traces taken over color, 
flavor and Dirac spaces respectively. Also note that, the quark self energy is a function of $M$ itself (since the loop particle 
is dressed) so that Eq.~\eqref{eq.gap} has to be solved self-consistently to calculate $M$.

Let us now explicitly evaluate the quantity $\Sigma^\text{Pure-Vac}_\text{MFA}$. 
Substituting Eq.~\eqref{eq.dressed.propagator} into Eq.~\eqref{eq.quark.self.energy}, 
we get 
\begin{eqnarray}
\Sigma^\text{Pure-Vac}_\text{MFA} = 8g_sN_cN_fM~ i\int \frac{d^4k}{(2\pi)^4}\frac{1}{k^2-M^2+i\epsilon}
\label{eq.quark.self.energy.1}
\end{eqnarray}
where, $N_c=3$ and $N_f=2$ are the number of colors and flavors respectively. 
The momentum integral in the above equation is ultra-violet (UV) divergent. The NJL model, being a non-renormalizable theory, requires 
a proper regularization scheme. There exists many such UV-regulators in the literature such as three-momentum cutoff, 
Euclidean four-momentum cutoff, Pauli-Villars, proper time and so on. The mostly used regulator is the momentum cutoff which breaks 
the Lorentz invariance and usually every symmetry of the theory. It will be demonstrated later in Sec.~\ref{sec.regular} that, 
the momentum cutoff regulator (or any other regulator which breaks Lorentz invariance) is not useful to study the vector 
meson $\rho$ in the NJL model. 
In this work, we will use `dimensional regularization' as our UV-regulator which respects all the symmetries of the theory. 
Going to $d$-dimension, Eq.~\eqref{eq.quark.self.energy.1} becomes
\begin{eqnarray}
\Sigma^\text{Pure-Vac}_\text{MFA} = 8g_sN_cN_fM\lambda^{2-d/2} ~i\int \frac{d^dk}{(2\pi)^d}\frac{1}{k^2-M^2+i\epsilon}
\Bigg|_{d\rightarrow4}.
\label{eq.quark.self.energy.2}
\end{eqnarray}
where $\lambda$ is a scale of dimension GeV$^2$ which has been introduced to keep the overall dimension of the equation consistent. 
Performing the momentum integral in the above equation, we get
\begin{eqnarray}
\RE\Sigma^\text{Pure-Vac}_\text{MFA} = 2g_s \frac{N_cN_fM^3}{4\pi^2}\FB{\frac{4\pi\lambda}{M^2}}^\varepsilon\Gamma\FB{\varepsilon-1}
\Bigg|_{\varepsilon\rightarrow0}
\label{eq.quark.self.energy.3}
\end{eqnarray}
where $\varepsilon=2-d/2$. It is to be noted that, the UV-divergence has been isolated as the pole of the Gamma function since 
$\Gamma(z)$ has simple poles at $z=0,-1,-2,....$. The regularization procedure of the above divergent quantity will be discussed in 
Sec.~\ref{sec.regular}. The above equation has  the following expansion about $\varepsilon=0$
\begin{eqnarray}
\RE\Sigma^\text{Pure-Vac}_\text{MFA} = 2g_s \frac{N_cN_fM^3}{4\pi^2}\TB{-\frac{1}{\varepsilon}+\gamma_\text{E}-1-\ln\FB{\frac{4\pi\lambda}{M^2}}}
\label{eq.sigma.purevac.expansion}
\end{eqnarray}
which will be used later.

\subsection{CASE-II: \boldmath{$T\ne0, B=0$}}
We now turn on the temperature and consider the case $T\ne0, B=0$. To include the effect of finite temperature, we will use 
the Real Time Formalism (RTF) of finite temperature field theory~\cite{Bellac_1996,Mallik_Sarkar_2016}. In the RTF, all the two point correlation functions including 
self energies and propagators become $2\times2$ matrices (will be denoted by boldface letters) in thermal space. 
As a result the Dyson-Schwinger equation generalizes to 
a matrix equation in thermal space
\begin{eqnarray}
\bm{S'} = \bm{S} - \bm{S\Sigma S'}
\label{eq.DS.T}
\end{eqnarray}
where each of the quantities is a $2\times2$ matrix. In the above equation, $\bm{S}$ is the free thermal quark propagator given by
\begin{eqnarray}
\bm{S}(q,m) = \bm{\mathcal{V}}
	\begin{pmatrix}
	S(q,m) & 0 \\ 
	0      & -\gamma^0 S^\dagger(q,m)\gamma^0
	\end{pmatrix}\bm{\mathcal{V}}.
\label{eq.free.thermal}
\end{eqnarray}
In the above equation, the diagonalizing matrix $\bm{\mathcal{V}}$ is given by,
\begin{eqnarray}
\bm{\mathcal{V}} = \begin{pmatrix}
N_2  & -N_1 \\
N_1  & N_2
\end{pmatrix}
\end{eqnarray}
with 
\begin{eqnarray}
N_2(q\cdot u) &=& \sqrt{1-f(q\cdot u)}\Theta(q\cdot u) + \sqrt{1-f(-q\cdot u)}\Theta(-q\cdot u), \\
N_1(q\cdot u) &=& \sqrt{f(q\cdot u)}\Theta(q\cdot u) + \sqrt{f(-q\cdot u)}\Theta(-q\cdot u)
\end{eqnarray}
where, $u^\mu$ is the four velocity of the thermal medium. 
In the Local Rest Frame (LRF), one has $u^\mu_\text{LRF}\equiv(1,\vec{0})$.
In the above equations, $\Theta(x)$ is the unit step function and $f(x)=\TB{e^{x/T}+1}^{-1}$ is the Fermi-Dirac thermal distribution 
function for the quarks. It is well known that, the complete thermal propagator matrix $\bm{S'}$ and the 
thermal self energy matrix $\bm{\Sigma}$ are diagonalized by $\bm{\mathcal{V}}$ and $\bm{\mathcal{V}^{-1}}$ respectively. 
Thus Eq.~\eqref{eq.DS.T} boils down to an algebraic equation in thermal space as 
\begin{eqnarray}
\overline{S'} = S - S~\Sigmabar~\overline{S'}
\label{eq.DS.T.Diagonal}
\end{eqnarray}
where $\overline{S'}$ and $\Sigmabar$ are respectively the $11$-component of the matrices $\bm{\mathcal{V}S'\mathcal{V}}$ 
and $\bm{\mathcal{V}}^{-1}\bm{\Sigma\mathcal{V}}^{-1}$. 
As before, in the MFA, the $\Sigmabar$ is diagonal in Dirac, color and flavor spaces 
$\Sigmabar=\Sigmabar_\text{MFA}\identity_\text{Dirac}\otimes\identity_\text{Flavor}\otimes\identity_\text{Color}$ 
so that Eq.~\eqref{eq.DS.T.Diagonal} can be trivially solved to obtain $\overline{S'}(q,m) = S(q,M(T))$ where the thermal 
constituent mass is given by
\begin{eqnarray}
M(T) = m + \Sigmabar_\text{MFA}.
\end{eqnarray}
It is easy to check that, $\RE\Sigmabar=\RE\bm{\Sigma}_{11}$, so that the knowledge of $\bm{S}'_{11}$ is sufficient to calculate 
the quantity $\RE\Sigmabar$. The explicit form of $\bm{S}'_{11}$ is given by
\begin{eqnarray}
\bm{S}'_{11}(q,m) &=& S(q,M) - \eta(q\cdot u) \TB{S(q,M)-\gamma^0S^\dagger(q,M)\gamma^0} \\
&=& (\cancel{q}+M)\TB{\frac{-1}{q^2-M^2+i\epsilon}-2\pi i\eta(q\cdot u)\delta(q^2-M^2)}\otimes\identity_\text{Flavor}\otimes\identity_\text{Colour}
\label{eq.s11.t}
\end{eqnarray}
where $\eta(x)=\Theta(x)f(x)+\Theta(-x)f(-x)$.

Let us now evaluate the thermal self energy function $\Sigmabar_\text{MFA}$ whose real part is obtained by replacing the 
vacuum complete propagator on the RHS of Eq.~\eqref{eq.quark.self.energy} by $S'\rightarrow \bm{S}'_{11}$ as
\begin{eqnarray}
\RE\Sigmabar_\text{MFA}(M,T) = -2g_s~\RE\TB{i\int\frac{d^4k}{(2\pi)^4}\Tr_\text{c,f,d}\TB{\bm{S'}_{11}(k,m)}}.
\label{eq.quark.self.energy.T}
\end{eqnarray}
Substituting Eq.~\eqref{eq.s11.t} into the above equation, we get after some simplification
\begin{eqnarray}
\RE\Sigmabar_\text{MFA}(M,T) = \RE\Sigma^\text{Pure-Vac}_\text{MFA}(M) 
-2g_s \frac{2N_cN_fM}{\pi^2}\int_{0}^{\infty}d|\vec{k}|\frac{\vec{k}^2}{\omega_k}f(\omega_k)
\label{eq.quark.self.energy.T.1}
\end{eqnarray}
where $\omega_k=\sqrt{\vec{k}^2+M^2}$.

\subsection{CASE-III: \boldmath{$T\ne0, B\ne0$}}
Finally, we consider the case of finite temperature and non-zero external magnetic field. In this case, 
the complete thermo-magnetic quark propagator $\bm{S'}_B$ satisfies the generalized Dyson-Schwinger equation 
\begin{eqnarray}
\bm{S'}_B = \bm{S}_B - \bm{S}_B\bm{\Sigma}_B\bm{ S'}_B
\label{eq.DS.TB}
\end{eqnarray}
where $\bm{\Sigma}_B$ is the thermo-magnetic quark one-loop self energy matrix and $\bm{S}_B$ is the free thermo-magnetic quark propagator. 
Analogous to Eq.~\eqref{eq.free.thermal}, $\bm{S}_B$ can be written explicitly as
\begin{eqnarray}
\bm{S}_B(q,m) = \bm{\mathcal{V}}
	\begin{pmatrix}
	S_B(q,m) & 0 \\ 
	0      & -\gamma^0S_B^\dagger(q,m)\gamma^0
	\end{pmatrix}\bm{\mathcal{V}}
\label{eq.free.thermo.magnetic}
\end{eqnarray}
where, 
\begin{eqnarray}
S_B(q,m) = \begin{pmatrix}
	S_\text{u} & 0 \\
	0 & S_\text{d}
	\end{pmatrix}
\end{eqnarray}
in which, $S_\text{u}$ and $S_\text{d}$ are respectively the Schwinger proper-time propagator for up and down quarks. 
They can be expressed as a sum over discrete Landau levels as
\begin{eqnarray}
S_f(q,m) = -\sum_{l=0}^{\infty}\frac{(-1)^le^{-\alpha_q^f}\mathscr{D}_{lf}(q)}{\qpll^2-M_{lf}^2+i\epsilon}\otimes\mathds{1}_\text{Color}
~~;~~f\in\{\text{u,d}\}
\end{eqnarray}
where, $\alpha_q^f = -\qper^2/|e_fB|$, 
\begin{eqnarray}
M_{lf} = \sqrt{M^2+2l|e_fB|}
\label{eq.Mlf}
\end{eqnarray}
and 
\begin{eqnarray}
\mathscr{D}_{lf}(q) = (\cancel{q}_\parallel+M)\TB{\SB{\mathds{1}_\text{Dirac}+\sign{e_f}i\gamma^1\gamma^2}L_l(2\alpha_q^f) 
- \SB{\mathds{1}_\text{Dirac}-\sign{e_f}i\gamma^1\gamma^2}L_{l-1}(2\alpha_q^f)} - 4\cancel{q}_\perp L^1_{l-1}(2\alpha_q^f)
\end{eqnarray}
with $e_f$ being the electric charge of flavor $f$ i.e. $e_\text{u}=\frac{2}{3}e$ and $e_\text{u}=-\frac{1}{3}e$; $e$ is the 
charge of a proton. In the above equation, $\sign{x}=\Theta(x)-\Theta(-x)$; $L_l^a(z)$ is the generalized 
Laguerre polynomial with the convention $L^a_{-1}=0$. The external magnetic field being in the positive z-direction, the metric 
tensor can be decomposed as $g^\munu=\gpll^\munu+\gper^\munu$ where $\gpll^\munu=\text{diag}(1,0,0,-1)$ and 
$\gper^\munu=\text{diag}(0,-1,-1,0)$ so that the parallel and perpendicular four vectors are defined as $\qpll^\mu = \gpll^\munu q_\nu$ 
and $\qper^\mu = \gper^\munu q_\nu$.

Similar to the thermal case, the Dyson-Schwinger equation in thermo-magnetic medium can be also represented in diagonal form as
\begin{eqnarray}
\overline{\overline{S'}} = S_B - S_B~\Sigmabarbar~\overline{\overline{S'}}
\label{eq.DS.TB.Diagonal}
\end{eqnarray}
Following the MFA, the $\Sigmabarbar$ is diagonal in Dirac, color and flavor spaces, 
$\Sigmabarbar=\Sigmabarbar_\text{MFA}\identity_\text{Dirac}\otimes\identity_\text{Flavor}
\otimes\identity_\text{Colour}$ 
so that Eq.~\eqref{eq.DS.TB.Diagonal} can be trivially solved to obtain $\overline{\overline{S'}}(q,m) = S_B(q,M(T,B))$ where the 
thermo-magnetic constituent quark mass is given by
\begin{eqnarray}
M(T,B) = m + \Sigmabarbar_\text{MFA}.
\end{eqnarray}

As before, because of the fact $\RE\Sigmabarbar=\RE\bm{\Sigma}_B^{11}$, the knowledge of $\bm{S'}_{B11}$ 
is sufficient to calculate the quantity $\RE\Sigmabarbar$. The explicit form of $\bm{S'}_{B11}$ is given by
\begin{eqnarray}
\bm{S'}_{B11}(q,m) &=& S_B(q,M) - \eta(q\cdot u) \TB{S_B(q,M)-\gamma^0 S_B^\dagger(q,M)\gamma^0} 
= \begin{pmatrix}
S_\text{u}^{11} & 0 \\
0 & S_\text{d}^{11}
\end{pmatrix}
\label{eq.s11.tb}
\end{eqnarray}
where 
\begin{eqnarray}
S_f^{11}(q,M) &=& S_f(q,M) - \eta(q\cdot u) \TB{S_f(q,M)-\gamma^0 S_f^\dagger(q,M)\gamma^0} \\
&=& \sum_{l=0}^{\infty}(-1)^le^{-\alpha_q^f}\mathscr{D}_{lf}(q)\TB{\frac{-1}{\qpll^2-M_{lf}^2+i\epsilon} 
- 2\pi i \eta(q\cdot u)\delta(\qpll^2-M_{lf}^2)}\otimes\mathds{1}_\text{Color}~~;~~f\in\{\text{u,d}\}.
\label{eq.sf11}
\end{eqnarray}

Let us now evaluate the thermo-magnetic self energy function $\Sigmabarbar_\text{MFA}$ whose 
real part is obtained by replacing the $11$-component of the 
complete thermal propagator on the RHS of Eq.~\eqref{eq.quark.self.energy.T} by $\bm{S'}_{11}\rightarrow \bm{S'}_{B11}$ as
\begin{eqnarray}
\RE\Sigmabarbar_\text{MFA}(M,T) = -2g_s~\RE\TB{i\int\frac{d^4k}{(2\pi)^4}\Tr_\text{c,f,d}\TB{\bm{S'}_{B11}(k,m)}}.
\label{eq.quark.self.energy.TB}
\end{eqnarray}
Substituting Eq.~\eqref{eq.s11.tb} into the above equation, we get after some simplification (see Appendix~\ref{app.condensate} for details)
\begin{eqnarray}
\RE\Sigmabarbar_\text{MFA}(M,B,T) = \RE\Sigma_\text{MFA}^\text{Pure-Vac}(M) + \RE\Sigma^\text{B-Vac}_\text{MFA}(M,B)
+ \RE\Sigma^\text{B-Med}_\text{MFA}(M,B,T)
\label{eq.quark.self.energy.TB.1}
\end{eqnarray}
where, $\RE\Sigma^\text{B-Vac}_\text{MFA}(M,B)$ is the real part of the magnetic field dependent vacuum contribution 
to the quark self energy and can be read off from Eq.~\eqref{eq.quark.self.energy.TB.final} as
\begin{eqnarray}
\Sigma^\text{B-Vac}_\text{MFA}(M,B) = -2g_s\frac{MN_c}{4\pi^2}\sum_{f\in\{\text{u,d}\}}^{}\TB{-M^2 + \FB{M^2-|e_fB|}\ln\FB{\frac{M^2}{2|e_fB|}}
-2|e_fB|\SB{\ln\Gamma\FB{\frac{M^2}{2|e_fB|}} - \ln\sqrt{2\pi}} }.
\label{eq.quark.self.energy.TB.2}
\end{eqnarray}
The temperature as well as magnetic field dependent contribution to the self energy, 
$\RE\Sigma^\text{B-Med}_\text{MFA}(M,B,T)$, can be obtained from Eq.~\eqref{eq.resigma.bmed.2} as
\begin{eqnarray}
\RE\Sigma^\text{B-Med}_\text{MFA}(M,B,T) = -2g_s \frac{N_cM}{\pi^2}\sum_{f\in\{\text{u,d}\}}^{}|e_fB|\sum_{l=0}^{\infty}(2-\delta_l^0)
\int_{0}^{\infty}dk_z \frac{1}{\omega_k^{lf}}f(\omega_k^{lf}).
\label{eq.quark.self.energy.TB.3}
\end{eqnarray}
It is interesting to note that in Eq.~\eqref{eq.quark.self.energy.TB.1},  the divergent pure vacuum contribution $\RE{\Sigma}_\text{MFA}(M)$ 
has completely been decoupled from the magnetic field and temperature dependent parts. 
One can notice from Eq.~\eqref{eq.quark.self.energy.TB.2} that, the quantity $\Sigma^\text{B-Vac}_\text{MFA}(M,B)$ is finite and thus 
the external magnetic field does not produce any additional divergences.

The formalism described in Appendix~\ref{app.condensate} to untangle the divergent pure vacuum contribution of the one-loop self energy graph is closely related  to the Magnetic Field Independent Regularization (MFIR) scheme as developed in Ref.~\cite{PhysRevD.99.116002}.  However, the methodology we have adopted is slightly different from that of MFIR. In our case, we have performed a dimensional regularization to the $d^d\kpll$ integral which leads to Hurwitz zeta function as a function of the dimension $d$.  An expansion of the Hurwitz zeta about its pole leads to the disentanglement of the pure vacuum part. On the other hand, in MFIR scheme, one does not change the space time dimension; rather one adds and subtracts the pure vacuum part.  Then,  using an integral representation of the Hurwitz zeta function,  the vacuum subtracted self energy is written as an integral over Schwinger proper-time   parameter . Finally the  proper-time  integral is performed to get the vacuum subtracted finite magnetic field dependent self-energy. Regardless of the methodology used, the two procedures lead to similar results. Specifically, the expressions in Eqs.~\eqref{eq.quark.self.energy.TB}-\eqref{eq.quark.self.energy.TB.3} are identical to the ones obtained in Refs.~\cite{PhysRevD.99.116002,PhysRevD.93.014010}.


\section{MESON PROPAGATORS IN RANDOM PHASE APPROXIMATION IN THE NJL MODEL} \label{sec.propagator}
Since mesons are the bound state of quarks and anti-quarks, their propagation can be studied from the scattering of quarks in 
different channels using the Bethe-Salpeter approach~\cite{Klevansky}. On the other hand, as discussed in Refs.~\cite{Liu_PRD_2015,Liu_CPC_2016}, 
the meson propagators can also be recast into the form of Dyson-Schwinger equations in the random phase approximation (RPA). 
Let us first consider the situation at vacuum (i.e. $T=0$ and $B=0$). In the scalar and pseudo-scalar channel, 
the $\pi$ and $\sigma$ meson propagators $D'_{h}$ satisfy the following Dyson-Schwinger equation
\begin{eqnarray}
D'_h = D - D\Pi_hD'_h~~;~~ h\in\{\pi,\sigma\}
\label{eq.ds.1}
\end{eqnarray}
where $D=(-2g_s)$ are the bare propagators and $\Pi_h$ are the one-loop polarization functions. 
The corresponding expression of the meson propagators $D'^\munu_{h}$ in the vector ($\rho$) and pseudo-vector ($a_1$) 
channels are given by
\begin{eqnarray}
D'^\munu_H = D^\munu - D^{\mu\alpha}\Pi_{H\alphabeta} D'^{\beta\nu}_H~~;~~ H\in\{\rho,a_1\}
\label{eq.ds.2}
\end{eqnarray}
where $D^\munu=(2g_vg^\munu)$ are the bare propagators and $\Pi_H^\munu$ are the one-loop polarization functions for the 
$\rho$ and $a_1$ mesons.

As already discussed in Sec.~\ref{sec.quark}, at finite temperature, all the real time two point correlation functions 
become $2\times2$ matrices in thermal space and will be denoted by boldface letters. 
Thus, at finite temperature, Eqs.~\eqref{eq.ds.1} and \eqref{eq.ds.2} 
generalize to
\begin{eqnarray}
\bm{D'}_h &=& \bm{D} - \bm{D}\bm{\Pi}_h\bm{D'}_h, \label{eq.ds.31} \\
\bm{D'}^\munu_H &=& \bm{D}^\munu - \bm{D}^{\mu\alpha}\bm{\Pi}_{H\alphabeta} \bm{D'}^{\beta\nu}_H.
\label{eq.ds.32}
\end{eqnarray}
However, each term of the above equations can be diagonalized to express them in terms of analytic functions~\cite{Mallik_Sarkar_2016} (will 
be denoted by bars) which in turn diagonalizes the Dyson-Schwinger equation making it an algebric equation in thermal space as
\begin{eqnarray}
\overline{D'}_h &=& D - D\Pibar_h\overline{D'}_h, \label{eq.ds.41}\\
\overline{D'}^\munu_H &=& D^\munu - D^{\mu\alpha}\Pibar_{H\alphabeta} \overline{D'}^{\beta\nu}_H.
\label{eq.ds.42}
\end{eqnarray}
In presence of both the finite temperature and external magnetic field, the generalization of Eqs.~\eqref{eq.ds.31} and \eqref{eq.ds.32} is
\begin{eqnarray}
\bm{D'}^B_{h} &=& \bm{D} - \bm{D}\bm{\Pi}^B_h\bm{D'}^B_h, \label{eq.ds.61}\\
\bm{D'}^{B\munu}_H &=& \bm{D}^\munu - \bm{D}^{\mu\alpha}\bm{\Pi}^B_{H\alphabeta} \bm{D'}^{B\beta\nu}_H.
\label{eq.ds.62}
\end{eqnarray}
so that, the the corresponding thermo-magnetic analytic functions denoted by a double-bar satisfy, 
\begin{eqnarray}
\overline{\overline{D'}}_h &=& D - D\Pibarbar_h\overline{\overline{D'}}_h \label{eq.ds.51} \\
\overline{\overline{D'}}^\munu_H &=& D^\munu - D^{\mu\alpha}\Pibarbar_{H\alphabeta} \overline{\overline{D'}}^{\beta\nu}_H.
\label{eq.ds.52}
\end{eqnarray}

Our next task is to solve the Dyson-Schwinger equations in order to express the complete meson propagators in terms of the 
polarization functions. It is trivial to solve Eqs.~\eqref{eq.ds.1}, \eqref{eq.ds.41} and \eqref{eq.ds.51} for the $\pi$ and $\sigma$ 
mesons as
\begin{eqnarray}
D'_h(q) = \FB{\frac{-2g_s}{1-2g_s\Pi_h}}~~,~~ \overline{D'}_h(q) = \FB{\frac{-2g_s}{1-2g_s\Pibar_h}}~~\text{and}~~ 
\overline{\overline{D'}}_h(q) = \FB{\frac{-2g_s}{1-2g_s\Pibarbar_h}}.
\label{eq.complete.prop.h}
\end{eqnarray}
However, for the $\rho$ and $a_1$ channels, additional complications arise because of the Lorentz indices in 
Eqs.~\eqref{eq.ds.2}, \eqref{eq.ds.42} and \eqref{eq.ds.52}. It is useful to decompose the polarization function and the 
complete propagator in terms of orthogonal tensor basis (constructed using the available vectors and tensors). This will 
enable one to solve the corresponding Dyson-Schwinger equation in a covariant way. 
We will discuss this in the following subsections.


\subsection{GENERAL LORENTZ STRUCTURE OF THE SPIN-1 POLARIZATION FUNCTION}
In order to decompose the polarization function into a suitable Lorentz basis, we use the fact that the 
polarization function is symmetric in its two Lorentz indices. 
We start with the simplest case of vacuum (i.e. $T=0$, $B=0$). The available quantities to construct a tensor basis are 
the momentum of the meson $q^\mu$ and the metric tensor $g^\munu$. Only two basis tensors can be constructed which are the following
\begin{eqnarray}
P_1^\munu &=& \FB{g^\munu - \frac{q^\mu q^\nu}{q^2}}, \label{eq.proj.vac.1}\\
P_2^\munu &=& \frac{q^\mu q^\nu}{q^2}. \label{eq.proj.vac.2}
\end{eqnarray}
It can easily be checked that, $P_i^\munu$ with $i=1,2$ satisfies all the properties of projection tensors i.e.
\begingroup
\renewcommand*{\arraystretch}{1.3}
\begin{eqnarray}
g_\alphabeta 
\begin{pmatrix}
	P_1^{\mu\alpha} \\ P_2^{\mu\alpha} 
	\end{pmatrix}
\begin{pmatrix}
	P_1^{\beta\nu} & P_2^{\beta\nu}
	\end{pmatrix}&=& 
\begin{pmatrix}
	P_1^\munu & 0 \\ 
	0  &  P_2^\munu
	\end{pmatrix}, \label{eq.proj.vac.3} \\
g_\munu g_\alphabeta 
\begin{pmatrix}
P_1^{\mu\alpha}  \\ P_2^{\mu\alpha} 
\end{pmatrix}
\begin{pmatrix}
P_1^{\beta\nu} &  P_2^{\beta\nu}
\end{pmatrix}&=& 
\begin{pmatrix}
3 & 0 \\ 
0 &  1
\end{pmatrix}. \label{eq.proj.vac.4}
\end{eqnarray} 
\endgroup

The vacuum polarization function $\Pi^\munu_H$ in this basis can be written as 
\begin{eqnarray}
\Pi^\munu_H = \sum_{i=1}^{2} \Pi_{Hi} P_i^\munu
\end{eqnarray}
where the form factors $\Pi_{Hi}$ are obtained using Eq.~\eqref{eq.proj.vac.4} as
\begin{eqnarray}
\Pi_{H1} = \frac{1}{3}P_1^\munu\Pi_{H\munu} ~~\text{and}~~ \Pi_{H2} = P_2^\munu\Pi_{H\munu}.
\label{eq.ff.1}
\end{eqnarray}
Note that, the form factors can be expressed in terms of the Lorentz invariants that can be formed by contracting $\Pi^\munu_H$ 
with the available tensors and vectors. In this case, we have $g^\munu$ and $q^\mu$ so that the form factors can be expressed 
in terms of $g_\munu\Pi^\munu_H$ and $q_\mu q_\nu \Pi^\munu_H$. See Appendix~\ref{app.form.factors} for details.

Let us now consider the case of finite temperature only (i.e. $T\ne0$ and $B=0$). Apart from $q^\mu$ and $g^\munu$, in this case we have 
an additional four-vector $u^\mu$. Thus one can choose the following four tensors as the basis
\begin{eqnarray}
\Pbar_1^\munu &=& \FB{g^\munu - \frac{q^\mu q^\nu}{q^2}- \frac{\utilde^\mu \utilde^\nu}{\utilde^2}}, \label{eq.proj.t.11} \\
\Pbar_2^\munu &=& \frac{q^\mu q^\nu}{q^2}, \label{eq.proj.t.12} \\
\Pbar_3^\munu &=& \frac{\utilde^\mu \utilde^\nu}{\utilde^2}, \label{eq.proj.t.13} \\
\Pbar_4^\munu &=& \frac{1}{\sqrt{q^2\utilde^2}}\FB{q^\mu\utilde^\nu+q^\nu\utilde^\mu} \label{eq.proj.t.14} 
\end{eqnarray}
where 
\begin{eqnarray}
\utilde^\mu = u^\mu - \frac{(q\cdot u)}{q^2}q^\mu
\label{eq.utilde}
\end{eqnarray}
is a vector orthogonal to $q^\mu$.
Similar to the vacuum case, one can verify that, the above tensors qualify to be the orthogonal projection tensors as they satisfy
\begingroup
\renewcommand*{\arraystretch}{1.4}
\begin{eqnarray}
g_\alphabeta 
\begin{pmatrix}
	\Pbar_1^{\mu\alpha} \\ \Pbar_2^{\mu\alpha} \\ \Pbar_3^{\mu\alpha} \\ \Pbar_4^{\mu\alpha}
	\end{pmatrix}
\begin{pmatrix}
	\Pbar_1^{\beta\nu} & \Pbar_2^{\beta\nu} & \Pbar_3^{\beta\nu} & \Pbar_4^{\beta\nu}
	\end{pmatrix} &=&
\begin{pmatrix}
	\Pbar_1^\munu & 0 & 0 & 0 \\
	0 & \Pbar_2^\munu & 0 & \Ensembleaverage{q^\mu\utilde^\nu} \\
	0 & 0 & \Pbar_3^\munu & \Ensembleaverage{q^\nu\utilde^\mu} \\
	0 & \Ensembleaverage{q^\nu\utilde^\mu} & \Ensembleaverage{q^\mu\utilde^\nu} & \Pbar_2^\munu + \Pbar_3^\munu
	\end{pmatrix}, \label{eq.proj.t.21} \\
%
g_\munu g_\alphabeta 
\begin{pmatrix}
	\Pbar_1^{\mu\alpha} \\ \Pbar_2^{\mu\alpha} \\ \Pbar_3^{\mu\alpha} \\ \Pbar_4^{\mu\alpha}
	\end{pmatrix}
\begin{pmatrix}
	\Pbar_1^{\beta\nu} & \Pbar_2^{\beta\nu} & \Pbar_3^{\beta\nu} & \Pbar_4^{\beta\nu}
	\end{pmatrix} &=&
\begin{pmatrix}
	2 & 0 & 0 & 0 \\
	0 & 1 & 0 & 0 \\
	0 & 0 & 1 & 0 \\
	0 & 0 & 0 & 2 
	\end{pmatrix}
\label{eq.proj.t.22}
\end{eqnarray} 
\endgroup
where the angular bracket is the shorthand notation for $\Ensembleaverage{A^\mu B^\nu} = A^\mu B^\nu/\sqrt{A^2B^2}$.

Now, the analytic thermal polarization function $\Pibar^\munu_H$ can be expanded in the above basis as 
\begin{eqnarray}
\Pibar^\munu_H = \sum_{i=1}^{4} \Pibar_{Hi} \Pbar_i^\munu
\end{eqnarray}
where the form factors $\Pibar_{Hi}$ are obtained using Eq.~\eqref{eq.proj.t.22} as
\begin{eqnarray}
\Pibar_{H1} = \frac{1}{2}\Pbar_1^\munu\Pibar_{H\munu} ~~,~~
\Pibar_{H2} = \Pbar_2^\munu\Pibar_{H\munu} ~~,~~
\Pibar_{H3} = \Pbar_3^\munu\Pibar_{H\munu} ~~\text{and}~~
\Pibar_{H4} = \frac{1}{2}\Pbar_4^\munu\Pibar_{H\munu}.
\label{eq.ff.t.1}
\end{eqnarray}
Note that, the form factors can be expressed in terms of the Lorentz invariants that can be formed by contracting $\Pibar^\munu_H$ 
with the available tensors and vectors. In this case, we have $g^\munu$, $q^\mu$ and $u^\mu$ so that the form factors can be expressed 
in terms of $g_\munu\Pibar^\munu_H$, $q_\mu q_\nu \Pibar^\munu_H$, $u_\mu u_\nu \Pibar^\munu_H$ and $q_\mu u_\nu \Pibar^\munu_H$. 
See Appendix~\ref{app.form.factors} for details.

Significant care has to be taken while considering the special case of $\vec{q}=\vec{0}$~\cite{Mallik_Sarkar_2016,Snigdha_2019}. 
To see this, let us write $\vec{q}=|\vec{q}|\hat{n}$ where $\hat{n}$ is the unit vector in the direction of $\vec{q}$. In the limit 
of $|\vec{q}|\rightarrow0$, we have
\begin{eqnarray}
\Pibar^{ij}_{H,\text{LRF}} (q^0,|\vec{q}|\rightarrow0) &=& g^{ij}\Pibar_{H1} + n^in^j\FB{\Pibar_{H1}-\Pibar_{H3}}, \\
\Pibar^{i0}_{H,\text{LRF}} (q^0,|\vec{q}|\rightarrow0) &=& \sqrt{-1}n^i\Pibar_{H4}
\end{eqnarray}
implying  that the above components of the polarization tensors depends of the direction of $\vec{q}$ even if $|\vec{q}|=0$. 
This ambiguity is rectified by imposing additional constraints on the form factors $\Pibar_i$ as
\begin{eqnarray}
\Pibar_{H1}(q^0,\vec{q}=\vec{0}) = \Pibar_{H3}(q^0,\vec{q}=\vec{0}) ~~\text{and}~~ \Pibar_{H4}(q^0,\vec{q}=\vec{0}) = 0.
\end{eqnarray}

Finally, we consider the general case of both finite temperature as well as finite external magnetic field. 
In this case, another four vector $b^\mu = \frac{1}{B}G^\munu u_\nu$ appears which specify the direction of the 
external magnetic field in the LRF where $G^\munu = \frac{1}{2}\epsilon^{\mu\nu\alpha\beta}F_\alphabeta$ is  
the dual of the field tensor $F_\alphabeta$ (we have used $\epsilon^{0123}=1$). In the LRF, we have $b^\mu_\text{LRF}\equiv(0,0,0,1)$. 
Thus using $q^\mu$, $u^\mu$, $b^\mu$ and $g^\munu$, we can construct the following seven orthogonal tensors:
\begin{eqnarray}
\Pbarbar_1^\munu &=& \FB{g^\munu - \frac{q^\mu q^\nu}{q^2}- \frac{\utilde^\mu \utilde^\nu}{\utilde^2}
- \frac{\btilde^\mu \btilde^\nu}{\btilde^2} }, \label{eq.proj.tb.11} \\
\Pbarbar_2^\munu &=& \frac{q^\mu q^\nu}{q^2}, \label{eq.proj.tb.12} \\
\Pbarbar_3^\munu &=& \frac{\utilde^\mu \utilde^\nu}{\utilde^2}, \label{eq.proj.tb.13} \\
\Pbarbar_4^\munu &=& \frac{\btilde^\mu \btilde^\nu}{\utilde^2}, \label{eq.proj.tb.14} \\
\Pbarbar_5^\munu &=& \frac{1}{\sqrt{q^2\utilde^2}}\FB{q^\mu\utilde^\nu+q^\nu\utilde^\mu}, \label{eq.proj.tb.15} \\
\Pbarbar_6^\munu &=& \frac{1}{\sqrt{q^2\btilde^2}}\FB{q^\mu\btilde^\nu+q^\nu\btilde^\mu}, \label{eq.proj.tb.16} \\
\Pbarbar_7^\munu &=& \frac{1}{\sqrt{\utilde^2\btilde^2}}\FB{\utilde^\mu\btilde^\nu+\utilde^\nu\btilde^\mu} \label{eq.proj.tb.17}  
\end{eqnarray}
where, 
\begin{eqnarray}
\btilde^\mu = b^\mu - \frac{(q\cdot b)}{q^2}q^\mu - \frac{(\utilde\cdot b)}{\utilde^2}\utilde^\mu.
\end{eqnarray}
It can be shown that, the tensors $\Pbarbar_i^\munu$ with $i=1,2,...,7$ satisfy all the properties of projection tensors as

\begingroup
\renewcommand*{\arraystretch}{1.4}
\begin{eqnarray}
g_\alphabeta 
\begin{pmatrix}
	\Pbarbar_1^{\mu\alpha} \\ \Pbarbar_2^{\mu\alpha} \\ \Pbarbar_3^{\mu\alpha} \\ \Pbarbar_4^{\mu\alpha} \\ 
	\Pbarbar_5^{\mu\alpha} \\ \Pbarbar_6^{\mu\alpha} \\	\Pbarbar_7^{\mu\alpha} 
	\end{pmatrix}
\begin{pmatrix}
	\Pbarbar_1^{\beta\nu} & \Pbarbar_2^{\beta\nu} & \Pbarbar_3^{\beta\nu} & 
	\Pbarbar_4^{\beta\nu} & \Pbarbar_5^{\beta\nu} & \Pbarbar_6^{\beta\nu} & 
	\Pbarbar_7^{\beta\nu} 
	\end{pmatrix}  &=& 
\begin{pmatrix}
	\Pbarbar_1^\munu & 0 & 0 & 0 & 0 & 0 & 0 \\
	0 & \Pbarbar_2^\munu & 0 & 0 & \Ensembleaverage{q^\mu\utilde^\nu} & \Ensembleaverage{q^\mu\btilde^\nu} & 0 \\
	0 & 0 & \Pbarbar_3^\munu & 0 & \Ensembleaverage{q^\nu\utilde^\mu} & 0 & \Ensembleaverage{\utilde^\mu\btilde^\nu} \\
	0 & 0 & 0 & \Pbarbar_4^\munu & 0 & \Ensembleaverage{q^\nu\btilde^\mu} & \Ensembleaverage{\utilde^\nu\btilde^\mu} \\
	0 & \Ensembleaverage{q^\nu\utilde^\mu} & \Ensembleaverage{q^\mu\utilde^\nu} & 0 & 
	\Pbarbar_2^\munu + \Pbarbar_3^\munu & \Ensembleaverage{\utilde^\mu\btilde^\nu} &	\Ensembleaverage{q^\mu\btilde^\nu} \\
	0 & \Ensembleaverage{q^\nu\btilde^\mu} & 0 & \Ensembleaverage{q^\mu\btilde^\nu} & \Ensembleaverage{\utilde^\nu\btilde^\mu} & 
	\Pbarbar_2^\munu + \Pbarbar_4^\munu & \Ensembleaverage{q^\mu\utilde^\nu} \\
	0 & 0 & \Ensembleaverage{\utilde^\nu\btilde^\mu} & \Ensembleaverage{\utilde^\mu\btilde^\nu} & \Ensembleaverage{q^\nu\btilde^\mu} & \Ensembleaverage{q^\nu\utilde^\mu} & 
	\Pbarbar_3^\munu + \Pbarbar_4^\munu 
	\end{pmatrix}, \nn \\ \label{eq.proj.tb.21} \\
	g_\munu g_\alphabeta 
	\begin{pmatrix}
			\Pbarbar_1^{\mu\alpha} \\ \Pbarbar_2^{\mu\alpha} \\ \Pbarbar_3^{\mu\alpha} \\	\Pbarbar_4^{\mu\alpha} \\ 
			\Pbarbar_5^{\mu\alpha} \\ \Pbarbar_6^{\mu\alpha} \\	\Pbarbar_7^{\mu\alpha} 
	\end{pmatrix}
	\begin{pmatrix}
			\Pbarbar_1^{\beta\nu} & \Pbarbar_2^{\beta\nu} & \Pbarbar_3^{\beta\nu} & 
			\Pbarbar_4^{\beta\nu} & \Pbarbar_5^{\beta\nu} & \Pbarbar_6^{\beta\nu} & 
			\Pbarbar_7^{\beta\nu} 
	\end{pmatrix} &=&
	\begin{pmatrix}
			1 & 0 & 0 & 0 & 0 & 0 & 0 \\
			0 & 1 & 0 & 0 & 0 & 0 & 0 \\
			0 & 0 & 1 & 0 & 0 & 0 & 0 \\
			0 & 0 & 0 & 1 & 0 & 0 & 0 \\
			0 & 0 & 0 & 0 & 2 & 0 & 0 \\
			0 & 0 & 0 & 0 & 0 & 2 & 0 \\
			0 & 0 & 0 & 0 & 0 & 0 & 2 
	\end{pmatrix}.
	\label{eq.proj.tb.22} 
\end{eqnarray} 
\endgroup

Now, the analytic thermo-magnetic polarization function $\Pibarbar^\munu_H$ can be expanded in the above basis as 
\begin{eqnarray}
\Pibarbar^\munu_H = \sum_{i=1}^{7} \Pibarbar_{Hi} \Pbarbar_i^\munu
\end{eqnarray}
where the form factors $\Pibarbar_{Hi}$ are obtained using Eq.~\eqref{eq.proj.tb.22} as
\begin{eqnarray}
\Pibarbar_{H1} = \Pbarbar_1^\munu\Pibarbar_{H\munu} ~~,~~
\Pibarbar_{H2} = \Pbarbar_2^\munu\Pibarbar_{H\munu} ~~,~~
\Pibarbar_{H3} = \Pbarbar_3^\munu\Pibarbar_{H\munu} ~~,~~
\Pibarbar_{H4} = \Pbarbar_4^\munu\Pibarbar_{H\munu} ~~,~~ \label{eq.ff.tb.1} \\
\Pibarbar_{H5} = \frac{1}{2}\Pbarbar_5^\munu\Pibarbar_{H\munu} ~~,~~
\Pibarbar_{H6} = \frac{1}{2}\Pbarbar_6^\munu\Pibarbar_{H\munu} ~~\text{and}~~
\Pibarbar_{H7} = \frac{1}{2}\Pbarbar_7^\munu\Pibarbar_{H\munu}. \label{eq.ff.tb.2}
\end{eqnarray}
As before, the form factors can be expressed in terms of the Lorentz invariants that can be formed by contracting $\Pibarbar^\munu_H$ 
with the available tensors and vectors. In this case, we have $g^\munu$, $q^\mu$, $u^\mu$ and $b^\mu$ so that the form factors 
can be expressed in terms of seven invariant quantities  
$g_\munu\Pibarbar^\munu_H$, $q_\mu q_\nu \Pibarbar^\munu_H$, $u_\mu u_\nu \Pibarbar^\munu_H$, 
$b_\mu b_\nu \Pibarbar^\munu_H$, $q_\mu b_\nu \Pibarbar^\munu_H$, $q_\mu u_\nu \Pibarbar^\munu_H$ and $u_\mu b_\nu \Pibarbar^\munu_H$.
See Appendix~\ref{app.form.factors} for details.

Similar to the thermal case, 
significant care has to be taken while considering the special case of $\vec{q}_\perp=\vec{0}$. To see this, let us 
write $\vec{q}_\perp=|\vec{q}_\perp|\hat{n}$ where $\hat{n}$ is the unit vector in the direction of $\vec{q}_\perp$. In the limit 
of $|\vec{q}_\perp|\rightarrow0$, we have
\begin{eqnarray}
\Pibarbar^{ij}_{H,\text{LRF}} (\qpll,|\vec{q}_\perp|\rightarrow0) &=& g^{ij}\Pibarbar_{H1} + n^in^j\FB{\Pibarbar_{H1}-\Pibarbar_{H4}}, \\
\Pibarbar^{i0}_{H,\text{LRF}} (\qpll,|\vec{q}_\perp|\rightarrow0) &=& n^i
\FB{\frac{q_z}{\sqrt{\qpll^2}}\Pibarbar_{H7} - \frac{q^0}{\sqrt{-\qpll^2}}\frac{|q_z|}{q_z}\Pibarbar_{H6} }, \\
\Pibarbar^{i3}_{H,\text{LRF}} (\qpll,|\vec{q}_\perp|\rightarrow0) &=& n^i
\FB{\frac{q^0}{\sqrt{\qpll^2}}\Pibarbar_{H7} - \frac{|q_z|}{\sqrt{-\qpll^2}}\Pibarbar_{H6} }
\end{eqnarray}
which implies that the above components of the thermo-magnetic polarization tensors depends of the 
direction of $\vec{q}_\perp$ even if $|\vec{q}_\perp|=0$. 
This ambiguity is rectified by imposing additional constraints on the form factors $\Pibarbar_i$ as
\begin{eqnarray}
\Pibarbar_{H1}(q^0,\vec{q}_\perp=\vec{0}) = \Pibarbar_{H4}(q^0,\vec{q}_\perp=\vec{0}) ~~\text{and}~~ 
\Pibarbar_{H6}(q^0,\vec{q}_\perp=\vec{0}) = \Pibarbar_{H7}(q^0,\vec{q}_\perp=\vec{0}) = 0.
\label{eq.special.case.2}
\end{eqnarray}


\subsection{SOLUTION OF THE DYSON-SCHWINGER EQUATION AND COMPLETE SPIN-1 PROPAGATORS}
Having  obtained the general Lorentz structure of the polarization functions in the previous subsection, we can now 
solve the Dyson-Schwinger Eqs.~\eqref{eq.ds.2}, \eqref{eq.ds.42} and \eqref{eq.ds.52} in order to calculate the complete 
propagators for $\rho$ and $a_1$ mesons.

Let us start with solving Eq.~\eqref{eq.ds.2}. We first write 
\begin{eqnarray}
D'^\munu_H = \sum_{i=1}^{2}D_{Hi}P_i^\munu
\end{eqnarray}
where the form factors $D_{Hi}$ are to be determined. Rewriting Eq.~\eqref{eq.ds.2} as 
\begin{eqnarray}
\FB{D'^\munu_H}^{-1} = \FB{D^\munu}^{-1} + \Pi^\munu_H = (2g_v)^{-1}g^\munu + \sum_{i=1}^{2} \Pi_{Hi} P_i^\munu 
=  \sum_{i=1}^{2} \TB{(2g_v)^{-1} + \Pi_{Hi}} P_i^\munu
\end{eqnarray}
and making use of $g_\alphabeta\FB{D'^{\mu\alpha}_H}^{-1} D'^{\beta\nu}_H = g^\munu = \sum_{i=1}^{2}P^\munu_i$ 
along with Eq.~\eqref{eq.proj.vac.3}, one obtains the form factors of the complete propagator as
\begin{eqnarray}
D_{Hi} = \FB{\frac{2g_v}{1+2g_v\Pi_{Hi}}}.
\end{eqnarray}

Let us now proceed to obtain the complete thermal propagator by solving Eq.~\eqref{eq.ds.42}. Expressing the 
complete propagator in the orthogonal tensor basis as
\begin{eqnarray}
\Dbar'^\munu_H = \sum_{i=1}^{4}\Dbar_{Hi}\Pbar_i^\munu
\end{eqnarray}
where the form factors $\Dbar_{Hi}$ are to be determined. Rewriting Eq.~\eqref{eq.ds.42} as 
\begin{eqnarray}
\FB{\Dbar'^\munu_H}^{-1} = \FB{\Dbar^\munu}^{-1} + \Pibar^\munu_H = (2g_v)^{-1}g^\munu + \sum_{i=1}^{4} \Pibar_{Hi} \Pbar_i^\munu 
=  \sum_{i=1}^{3} (2g_v)^{-1} \Pbar_i^\munu + \sum_{i=1}^{4} \Pibar_{Hi} \Pbar_i^\munu
\end{eqnarray}
and making use of $g_\alphabeta\FB{\Dbar'^{\mu\alpha}_H}^{-1} \Dbar'^{\beta\nu}_H = g^\munu = \sum_{i=1}^{3}\Pbar^\munu_i$ 
along with Eq.~\eqref{eq.proj.t.21}, one obtains the form factors of the complete thermal propagator as
\begin{eqnarray}
\Dbar_{H1} = \FB{\frac{2g_v}{1+2g_v\Pibar_{H1}}}~~,~~ 
\Dbar_{H2} = \frac{1}{\mathcal{A}_T}2g_v\FB{1+2g_v\Pibar_{H3}},\\
\Dbar_{H3} = \frac{1}{\mathcal{A}_T} 2g_v\FB{1+2g_v\Pibar_{H2}}~~\text{and}~~ 
\Dbar_{H4} = \frac{1}{\mathcal{A}_T} 2g_v\FB{-2g_v\Pibar_{H4}}
\end{eqnarray}
where $\mathcal{A}_T=\FB{1+2g_v\Pibar_{H2}}\FB{1+2g_v\Pibar_{H3}}- \FB{2g_v\Pibar_{H4}}^2$.

Finally we calculate complete thermo-magnetic propagator by solving Eq.~\eqref{eq.ds.52}. Expanding the 
complete propagator in the orthogonal tensor basis as
\begin{eqnarray}
\Dbarbar'^\munu_H = \sum_{i=1}^{7}\Dbarbar_{Hi}\Pbarbar_i^\munu
\label{eq.Dbarbar}
\end{eqnarray}
where the form factors $\Dbarbar_{Hi}$ are to be determined. Rewriting Eq.~\eqref{eq.ds.52} as 
\begin{eqnarray}
\FB{\Dbarbar'^\munu_H}^{-1} = \FB{\Dbarbar^\munu}^{-1} + \Pibarbar^\munu_H = (2g_v)^{-1}g^\munu + \sum_{i=1}^{7} \Pibarbar_{Hi} \Pbarbar_i^\munu 
=  \sum_{i=1}^{4} (2g_v)^{-1} \Pbarbar_i^\munu + \sum_{i=1}^{7} \Pibarbar_{Hi} \Pbarbar_i^\munu~,
\end{eqnarray}
and making use of $g_\alphabeta\FB{\Dbarbar'^{\mu\alpha}_H}^{-1} \Dbarbar'^{\beta\nu}_H = g^\munu = \sum_{i=1}^{4}\Pbarbar^\munu_i$ 
along with Eq.~\eqref{eq.proj.tb.21}, one obtains the form factors of the complete thermo-magnetic propagator as
\begin{eqnarray}
\Dbarbar_{H1} &=& \FB{\frac{2g_v}{1+2g_v\Pibarbar_{H1}}}, \label{eq.D1}\\
\Dbarbar_{H2} &=& \frac{1}{\mathcal{A}_{TB}}2g_v\TB{\FB{1+2g_v\Pibar_{H3}}\FB{1+2g_v\Pibar_{H4}}-\FB{2g_v\Pibar_{H7}}^2}, \label{eq.D2}\\
\Dbarbar_{H3} &=& \frac{1}{\mathcal{A}_{TB}}2g_v\TB{\FB{1+2g_v\Pibar_{H2}}\FB{1+2g_v\Pibar_{H4}}-\FB{2g_v\Pibar_{H6}}^2}, \label{eq.D3}\\
\Dbarbar_{H4} &=& \frac{1}{\mathcal{A}_{TB}}2g_v\TB{\FB{1+2g_v\Pibar_{H2}}\FB{1+2g_v\Pibar_{H3}}-\FB{2g_v\Pibar_{H5}}^2}, \label{eq.D4} \\
\Dbarbar_{H5} &=& \frac{1}{\mathcal{A}_{TB}}2g_v\TB{\FB{2g_v\Pibar_{H6}}\FB{2g_v\Pibar_{H7}}-\FB{1+2g_v\Pibar_{H4}}\FB{2g_v\Pibar_{H5}}},
\label{eq.D5}\\
\Dbarbar_{H6} &=& \frac{1}{\mathcal{A}_{TB}}2g_v\TB{\FB{2g_v\Pibar_{H5}}\FB{2g_v\Pibar_{H7}}-\FB{1+2g_v\Pibar_{H3}}\FB{2g_v\Pibar_{H6}}}, 
\label{eq.D6} \\
\Dbarbar_{H7} &=& \frac{1}{\mathcal{A}_{TB}}2g_v\TB{\FB{2g_v\Pibar_{H5}}\FB{2g_v\Pibar_{H6}}-\FB{1+2g_v\Pibar_{H2}}\FB{2g_v\Pibar_{H7}}}
\label{eq.D7}
\end{eqnarray}
where 
\begin{eqnarray}
\mathcal{A}_{TB} &=& \FB{1+2g_v\Pibarbar_{H2}}\FB{1+2g_v\Pibarbar_{H3}}\FB{1+2g_v\Pibarbar_{H4}} 
- \FB{1+2g_v\Pibarbar_{H2}} \FB{2g_v\Pibarbar_{H7}}^2 \nn \\ 
&& ~~ - \FB{1+2g_v\Pibarbar_{H3}} \FB{2g_v\Pibarbar_{H6}}^2 
- \FB{1+2g_v\Pibarbar_{H4}}\FB{2g_v\Pibarbar_{H5}}^2 + \FB{2g_v\Pibarbar_{H5}}\FB{2g_v\Pibarbar_{H6}}\FB{2g_v\Pibarbar_{H7}}.
\label{eq.ATB}
\end{eqnarray}


\section{POLARIZATION FUNCTIONS OF THE MESONS} \label{sec.pola}
In this section, we will explicitly calculate the polarization functions in various channels. 
In the current work, we only include the charge-neutral mesons i.e. $\pi^0$, $\sigma$, $\rho^0$ 
and $a_1^0$. Thus by $\pi, \rho$ and $a_1$ we will mean $\pi^0, \rho^0$ and $a_1^0$.  
We start with the well known expression for the vacuum polarization functions (at $T=0$ and $B=0$) of the 
charge-neutral mesons
\begin{eqnarray}
\Pi_\pi(q) &=& -i\int\frac{d^4k}{(2\pi)^4}\Tr_\text{d,f,c}\TB{\gamma^5\tau_3 S'(q+k,m)\gamma^5\tau_3S'(k,m)}, \label{eq.Pi.pi.1}\\
\Pi_\sigma(q) &=& i\int\frac{d^4k}{(2\pi)^4}\Tr_\text{d,f,c}\TB{ S'(q+k,m)S'(k,m)}, \label{eq.Pi.sigma.1} \\
\Pi^\munu_\rho(q) &=& -i\int\frac{d^4k}{(2\pi)^4}\Tr_\text{d,f,c}\TB{\gamma^\mu\tau_3 S'(q+k,m)\gamma^\nu\tau_3S'(k,m)}, \label{eq.Pi.rho.1} \\
\Pi^\munu_{a_1}(q) &=& -i\int\frac{d^4k}{(2\pi)^4}\Tr_\text{d,f,c}\TB{\gamma^\mu\gamma^5\tau_3 S'(q+k,m)\gamma^\nu\gamma^5\tau_3S'(k,m)} 
\label{eq.Pi.a1.1}
\end{eqnarray}
where $\tau_3 = \begin{pmatrix} 1 & 0 \\ 0 & -1 \end{pmatrix}$ is the third Pauli isospin matrix and $S'(q,m)$ is defined in 
Eq.~\eqref{eq.dressed.propagator}. Similar to the case of quark self energy calculation, we will use dimensional regularization 
for the evaluation of the above pure-vacuum polarization functions. 
The calculation has been briefly sketched in Appendix.~\ref{app.pola.vac} and the final result can be read off from 
Eqs.~\eqref{eq.Pi.pi.final}-\eqref{eq.Pi.a1.final} as
\begin{eqnarray}
\Pi_\pi(q) &=& \frac{N_c N_f}{4\pi^2}\TB{\frac{1}{2}q^2\Gamma(\varepsilon)\int_{0}^{1}dx\FB{\frac{4\pi\lambda}{\Delta}}^\varepsilon 
	+ M^2\Gamma(\varepsilon-1) \FB{\frac{4\pi\lambda}{M^2}}^\varepsilon}, \label{eq.Pi.pi} \\
\Pi_\sigma(q) &=& \frac{N_c N_f}{4\pi^2}\TB{\frac{1}{2}(q^2-4M^2)\Gamma(\varepsilon)\int_{0}^{1}dx\FB{\frac{4\pi\lambda}{\Delta}}^\varepsilon 
	+ M^2\Gamma(\varepsilon-1) \FB{\frac{4\pi\lambda}{M^2}}^\varepsilon}, \label{eq.Pi.sigma} \\
\Pi^\munu_\rho(q) &=& -\frac{N_c N_f}{2\pi^2}\Gamma(\varepsilon) \FB{g^\munu - \frac{q^\mu q^\nu}{q^2}} q^2
\int_{0}^{1}dxx(1-x)\FB{\frac{4\pi\lambda}{\Delta}}^\varepsilon,  \label{eq.Pi.rho} \\
\Pi^\munu_{a_1}(q) &=& \frac{N_c N_f}{2\pi^2} \Gamma(\varepsilon)\int_{0}^{1}dx
\TB{\FB{g^\munu - \frac{q^\mu q^\nu}{q^2}}\Delta + \frac{q^\mu q^\nu}{q^2} M^2 }\FB{\frac{4\pi\lambda}{\Delta}}^\varepsilon. \label{eq.Pi.a1}
\end{eqnarray}

As finite temperature, the analytic thermal polarization functions $\Pibar_h(q)$ and $\Pibar_H(q)$ are related to the 
$11$-components of respective thermal polarization matrices $\bm{\Pi}^{11}_h(q)$ and $\bm{\Pi}^{11}_H(q)$ via relations~\cite{Mallik_Sarkar_2016,Bellac_1996}
\begin{eqnarray}
\RE \Pibar_h(q) &=& \RE \bm{\Pi}^{11}_h(q) ~~,~~ \RE \Pibar^\munu_H(q) = \RE \bm{\Pi}^{\munu 11}_H(q), \label{eq.barto11.1} \\
\IM \Pibar_h(q) &=& \text{sign}(q^0)\tanh\FB{\frac{q^0}{2T}}\IM \bm{\Pi}^{11}_h(q) ~~\text{and}~~
\IM \Pibar^\munu_H(q) = \text{sign}(q^0)\tanh\FB{\frac{q^0}{2T}}\IM \bm{\Pi}^{\munu 11}_H(q).
\label{eq.barto11.2}
\end{eqnarray}
Now, the $11$-components of the thermal polarization functions are obtained by replacing the vacuum propagators on the RHS of 
Eqs.~\eqref{eq.Pi.pi.1}-\eqref{eq.Pi.a1.1} by $S'\rightarrow \bm{S'}_{11}$ where $\bm{S'}_{11}$ is defined in Eq.~\eqref{eq.s11.t}. 
Therefore,
\begin{eqnarray}
\bm{\Pi}^{11}_\pi(q) &=& -i\int\frac{d^4k}{(2\pi)^4}\Tr_\text{d,f,c}\TB{\gamma^5\tau_3 \bm{S'}_{11}(q+k,m)\gamma^5\tau_3\bm{S'}_{11}(k,m)}, \label{eq.Pi.pi.2}\\
\bm{\Pi}^{11}_\sigma(q) &=& i\int\frac{d^4k}{(2\pi)^4}\Tr_\text{d,f,c}\TB{ \bm{S'}_{11}(q+k,m)\bm{S'}_{11}(k,m)}, \label{eq.Pi.sigma.2} \\
\bm{\Pi}^{\munu 11}_\rho(q) &=& -i\int\frac{d^4k}{(2\pi)^4}\Tr_\text{d,f,c}\TB{\gamma^\mu\tau_3 \bm{S'}_{11}(q+k,m)\gamma^\nu\tau_3\bm{S'}_{11}(k,m)}, \label{eq.Pi.rho.2} \\
\bm{\Pi}^{\munu 11}_{a_1}(q) &=& -i\int\frac{d^4k}{(2\pi)^4}\Tr_\text{d,f,c}\TB{\gamma^\mu\gamma^5\tau_3 \bm{S'}_{11}(q+k,m)\gamma^\nu\gamma^5\tau_3\bm{S'}_{11}(k,m)}.
\label{eq.Pi.a1.2}
\end{eqnarray}
Substituting $\bm{S'}_{11}$ from Eq.~\eqref{eq.s11.t} into the above equation and making use of Eqs.~\eqref{eq.barto11.1} and 
\eqref{eq.barto11.2}, we get 
after some simplifications the real parts of the polarization functions as
\begin{eqnarray}
\RE \Pibar_h(q) = \RE\Pi_h(q) - \int\frac{d^3k}{(2\pi)^3}\mathcal{P}\TB{ 
\frac{N_h(k^0=-\omega_k)f(\omega_k)}{2\omega_k\SB{(q^0-\omega_k)^2-\omega_p^2}} 
+ \frac{N_h(k^0=\omega_k)f(\omega_k)}{2\omega_k\SB{(q^0+\omega_k)^2-\omega_p^2}} \right. \nn \\ \left.
+ \frac{N_h(k^0=-q^0-\omega_p)f(\omega_p)}{2\omega_p\SB{(q^0+\omega_p)^2-\omega_k^2}} 
+ \frac{N_h(k^0=-q^0+\omega_p)f(\omega_p)}{2\omega_p\SB{(q^0-\omega_p)^2-\omega_k^2}} 
}, \\
\RE \Pibar^\munu_H(q) = \RE\Pi^\munu_H(q) - \int\frac{d^3k}{(2\pi)^3}\mathcal{P}\TB{ 
	\frac{N^\munu_H(k^0=-\omega_k)f(\omega_k)}{2\omega_k\SB{(q^0-\omega_k)^2-\omega_p^2}} 
	+ \frac{N^\munu_H(k^0=\omega_k)f(\omega_k)}{2\omega_k\SB{(q^0+\omega_k)^2-\omega_p^2}} \right. \nn \\ \left.
	+ \frac{N^\munu_H(k^0=-q^0-\omega_p)f(\omega_p)}{2\omega_p\SB{(q^0+\omega_p)^2-\omega_k^2}} 
	+ \frac{N^\munu_H(k^0=-q^0+\omega_p)f(\omega_p)}{2\omega_p\SB{(q^0-\omega_p)^2-\omega_k^2}} 
}
\end{eqnarray}
and imaginary parts as
\begin{eqnarray}
\IM \Pibar_h(q) &=& -\text{sign}(q^0)\tanh\FB{\frac{q^0}{2T}}\pi \int\frac{d^3k}{(2\pi)^3}\frac{1}{4\omega_k\omega_p} \nn \\
&& \hspace{-1cm}\TB{ 
\SB{1-f(\omega_k)-f(\omega_p)+2f(\omega_k)f(\omega_p)}
\SB{N_h(k^0=-\omega_k)\delta(q^0-\omega_k-\omega_p)+ N_h(k^0=\omega_k)\delta(q^0+\omega_k+\omega_p)} \right. \nn \\ 
&& \hspace{-1cm} \left.
+ \SB{-f(\omega_k)-f(\omega_p)+2f(\omega_k)f(\omega_p)}
\SB{N_h(k^0=-\omega_k)\delta(q^0-\omega_k-\omega_p)+ N_h(k^0=\omega_k)\delta(q^0+\omega_k+\omega_p)}
}, \\
\IM \Pibar^\munu_H(q) &=& -\text{sign}(q^0)\tanh\FB{\frac{q^0}{2T}}\pi \int\frac{d^3k}{(2\pi)^3}\frac{1}{4\omega_k\omega_p} \nn \\
&& \hspace{-1cm}\TB{ 
	\SB{1-f(\omega_k)-f(\omega_p)+2f(\omega_k)f(\omega_p)}
	\SB{N^\munu_H(k^0=-\omega_k)\delta(q^0-\omega_k-\omega_p)+ N^\munu_H(k^0=\omega_k)\delta(q^0+\omega_k+\omega_p)} \right. \nn \\ 
	&& \hspace{-1cm} \left.
	+\SB{-f(\omega_k)-f(\omega_p)+2f(\omega_k)f(\omega_p)}
	\SB{N^\munu_H(k^0=-\omega_k)\delta(q^0-\omega_k-\omega_p)+ N^\munu_H(k^0=\omega_k)\delta(q^0+\omega_k+\omega_p)}
} 
\end{eqnarray}
where $N_h(q,k)$ and $N_H^\munu(q,k)$ are defined in Eqs.~\eqref{eq.Nh} and \eqref{eq.NH}.

Finally, we consider the case of both finite temperature as well as non-zero external magnetic field. 
The analytic thermo-magnetic polarization functions $\Pibarbar_h(q)$ and $\Pibarbar_H(q)$ are related to the 
$11$-components of respective thermo-magnetic polarization matrices $\bm{\Pi}^{B 11}_h(q)$ and $\bm{\Pi}^{B 11}_H(q)$ via 
similar relations as in Eqs.~\eqref{eq.barto11.1} and \eqref{eq.barto11.2}. 
Thus, the $11$-components of the thermo-magnetic polarization functions are obtained by replacing the vacuum propagators on the RHS of 
Eqs.~\eqref{eq.Pi.pi.1}-\eqref{eq.Pi.a1.1} by $S'\rightarrow \bm{S'}_{B11}$ where $\bm{S'}_{B11}$ is defined in Eq.~\eqref{eq.s11.tb}. 
Therefore,
\begin{eqnarray}
\bm{\Pi}^{B 11}_\pi(q) &=& -i\int\frac{d^4k}{(2\pi)^4}\Tr_\text{d,f,c}\TB{\gamma^5\tau_3 \bm{S'}_{B11}(q+k,m)\gamma^5\tau_3\bm{S'}_{B11}(k,m)}. \label{eq.Pi.pi.3}\\
\bm{\Pi}^{B 11}_\sigma(q) &=& i\int\frac{d^4k}{(2\pi)^4}\Tr_\text{d,f,c}\TB{ \bm{S'}_{B11}(q+k,m)\bm{S'}_{B11}(k,m)}, \label{eq.Pi.sigma.3} \\
\bm{\Pi}^{B \munu 11}_\rho(q) &=& -i\int\frac{d^4k}{(2\pi)^4}\Tr_\text{d,f,c}\TB{\gamma^\mu\tau_3 \bm{S'}_{B 11}(q+k,m)\gamma^\nu\tau_3\bm{S'}_{B11}(k,m)}, \label{eq.Pi.rho.3} \\
\bm{\Pi}^{B \munu 11}_{a_1}(q) &=& -i\int\frac{d^4k}{(2\pi)^4}\Tr_\text{d,f,c}\TB{\gamma^\mu\gamma^5\tau_3 \bm{S'}_{B 11}(q+k,m)\gamma^\nu\gamma^5\tau_3\bm{S'}_{B11}(k,m)}. 
\label{eq.Pi.a1.3}
\end{eqnarray}
Substituting $\bm{S''}_{11}$ from Eq.~\eqref{eq.s11.tb} into the above equation and making use of 
analogous relations to Eqs.~\eqref{eq.barto11.1} and \eqref{eq.barto11.2}, we will obtain the 
real and imaginary parts of the analytic thermo-magnetic polarization functions. 
For the simplicity in analytic calculations, we take $\qper=0$ for which the corresponding calculations are provided in 
Appendix~\ref{app.pola.vbt} and below we only give the final expressions. 
From Eqs.~\eqref{eq.repi.tb.31}, \eqref{eq.repi.tb.32} and \eqref{eq.pih.B}-\eqref{eq.pia1.bvac.final}, 
we get the real parts of the analytic thermo-magnetic polarization functions as
\begin{eqnarray}
\RE \Pibarbar_h(\qpll) &=& \RE\Pi_h(\qpll) + \RE\Pi_{h\text{B}}(\qpll,B) - \sum_{l=0}^{\infty}\sum_{n=(l-1)}^{(l+1)}\sum_{f\in\{\text{u,d}\}} 
\int_{-\infty}^{\infty}\frac{dk_z}{2\pi}\mathcal{P}\TB{ 
	\frac{N_h^{lnf}(k^0=-\omega_k^{lf})f(\omega_k^{lf})}{2\omega_k^{lf}\SB{(q^0-\omega_k^{lf})^2-(\omega_p^{nf})^2}} \right. \nn \\ 
	&& \left. \hspace{-0.5cm} + \frac{N_h^{lnf}(k^0=\omega_k^{lf})f(\omega_k^{lf})}{2\omega_k^{lf}\SB{(q^0+\omega_k^{lf})^2-(\omega_p^{nf})^2}} 
	+ \frac{N_h^{lnf}(k^0=-q^0-\omega_p^{nf})f(\omega_p^{nf})}{2\omega_p^{nf}\SB{(q^0+\omega_p^{nf})^2-(\omega_k^{lf})^2}} 
	+ \frac{N_h^{lnf}(k^0=-q^0+\omega_p^{nf})f(\omega_p^{nf})}{2\omega_p^{nf}\SB{(q^0-\omega_p^{nf})^2-(\omega_k^{lf})^2}} 
}, \label{eq.repi.tb.1}\\ 
\RE \Pibarbar^\munu_H(\qpll) &=& \RE\Pi^\munu_H(\qpll) + \RE\Pi^\munu_{H\text{B}}(\qpll,B) - \sum_{l=0}^{\infty}\sum_{n=(l-1)}^{(l+1)}\sum_{f\in\{\text{u,d}\}} 
\int_{-\infty}^{\infty}\frac{dk_z}{2\pi}\mathcal{P}\TB{ 
	\frac{N_H^{lnf\munu}(k^0=-\omega_k^{lf})f(\omega_k^{lf})}{2\omega_k^{lf}\SB{(q^0-\omega_k^{lf})^2-(\omega_p^{nf})^2}} 
	\right. \nn \\ &&
	\left. 	\hspace{-0.5cm} + \frac{N_H^{lnf\munu}(k^0=\omega_k^{lf})f(\omega_k^{lf})}{2\omega_k^{lf}\SB{(q^0+\omega_k^{lf})^2-(\omega_p^{nf})^2}} 
	+ \frac{N_H^{lnf\munu}(k^0=-q^0-\omega_p^{nf})f(\omega_p^{nf})}{2\omega_p^{nf}\SB{(q^0+\omega_p^{nf})^2-(\omega_k^{lf})^2}} 
	+ \frac{N_H^{lnf\munu}(k^0=-q^0+\omega_p^{nf})f(\omega_p^{nf})}{2\omega_p^{nf}\SB{(q^0-\omega_p^{nf})^2-(\omega_k^{lf})^2}} 
} \label{eq.repi.tb.2}
\end{eqnarray}
where the novel magnetic field dependent vacuum contributions are
\begin{eqnarray}
\Pi_{\pi\text{B}}(\qpll,B) &=&  \frac{N_c}{4\pi^2}\sum_{f\in\{\text{u,d}\}} \Bigg[ M^2 + (|e_fB|-M^2)\ln\FB{\frac{M^2}{2|e_fB|}}
+ 2|e_fB|\SB{\ln\Gamma\FB{\frac{M^2}{2|e_fB|}}-\ln\sqrt{2\pi}}  \nn \\
&& + \frac{1}{2}\qpll^2\int_{0}^{1}dx\SB{\ln z-\psi(z)-\frac{1}{2z}}
\Bigg], \label{eq.pipi.bvac}\\
\Pi_{\sigma\text{B}}(\qpll,B) &=& \frac{N_c}{4\pi^2}\sum_{f\in\{\text{u,d}\}} \Bigg[ M^2 + (|e_fB|-M^2)\ln\FB{\frac{M^2}{2|e_fB|}}
+ 2|e_fB|\SB{\ln\Gamma\FB{\frac{M^2}{2|e_fB|}}-\ln\sqrt{2\pi}}  \nn \\
&& + \frac{1}{2}\FB{\qpll^2-4M^2}\int_{0}^{1}dx\SB{\ln z-\psi(z)-\frac{1}{2z}}
\Bigg], \label{eq.pisigma.bvac} \\
\Pi^\munu_{\rho\text{B}}(\qpll,B) &=& -\frac{N_c}{4\pi^2}\sum_{f\in\{\text{u,d}\}}\int_{0}^{1}dx \Bigg[
\FB{\qpll^2g^\munu-\qpll^\mu\qpll^\nu}2x(1-x)\ln z 
- \FB{\qpll^2\gpll^\munu-\qpll^\mu\qpll^\nu}x(1-x)\FB{2\psi(z)+1/z} \nn \\
&& + \TB{\FB{\Delta-2M^2}\psi\FB{z+x} + \Delta 
	+ 2|e_fB|\SB{\ln\Gamma\FB{z+x}-\ln\sqrt{2\pi}}}\gper^\munu
\Bigg] \label{eq.pirho.bvac}, \\
\Pi^\munu_{a_1\text{B}}(\qpll,B) &=& -\frac{N_c}{4\pi^2}\sum_{f\in\{\text{u,d}\}}\int_{0}^{1}dx \Bigg[
\FB{g^\munu\Delta+x(1-x)\qpll^\mu\qpll^\nu}(-2\ln z)
+ \FB{\gpll^\munu\Delta+x(1-x)\qpll^\mu\qpll^\nu}\FB{2\psi(z)+1/z} \nn \\
&& + \TB{\Delta\psi\FB{z+x} + \Delta 
	+ 2|e_fB|\SB{\ln\Gamma\FB{z+x}-\ln\sqrt{2\pi}}}\gper^\munu
\Bigg] \label{eq.pia1.bvac}.
\end{eqnarray}
The imaginary parts are to be read off from Eqs.~\eqref{eq.impi.tb.1.final} and \eqref{eq.impi.tb.2.final} as
\begin{eqnarray}
\IM \Pibarbar_h(\qpll) &=& -\text{sign}(q^0)\tanh\FB{\frac{q^0}{2T}}\pi\sum_{l=0}^{\infty}\sum_{n=(l-1)}^{(l+1)}\sum_{f\in\{\text{u,d}\}} 
\int_{-\infty}^{\infty}\frac{dk_z}{2\pi} \frac{1}{4\omega_k^{lf}\omega_p^{nf}} \nn \\
&& \hspace{-1.5cm}\TB{ 
	\SB{1-f(\omega_k^{lf})-f(\omega_p^{nf})+2f(\omega_k^{lf})f(\omega_p^{nf})}
	\SB{N^{lnf}_h(k^0=-\omega_k^{lf})\delta(q^0-\omega_k^{lf}-\omega_p^{nf})+ N_h^{lnf}(k^0=\omega_k^{lf})\delta(q^0+\omega_k^{lf}+\omega_p^{nf})} \right. \nn \\ 
	&& \hspace{-1.9cm} \left.
	+ \SB{-f(\omega_k^{lf})-f(\omega_p^{nf})+2f(\omega_k^{lf})f(\omega_p^{nf})}
	\SB{N^{lnf}_h(k^0=-\omega_k^{lf})\delta(q^0-\omega_k^{lf}-\omega_p^{nf})+ N^{lnf}_h(k^0=\omega_k^{lf})\delta(q^0+\omega_k^{lf}+\omega_p^{nf})}
}, \label{eq.impi.tb.1} \\
\IM \Pibarbar_H^\munu(\qpll) &=& -\text{sign}(q^0)\tanh\FB{\frac{q^0}{2T}}\pi\sum_{l=0}^{\infty}\sum_{n=(l-1)}^{(l+1)}\sum_{f\in\{\text{u,d}\}} 
\int_{-\infty}^{\infty}\frac{dk_z}{2\pi} \frac{1}{4\omega_k^{lf}\omega_p^{nf}} \nn \\
&& \hspace{-1.8cm}\TB{ 
	\SB{1-f(\omega_k^{lf})-f(\omega_p^{nf})+2f(\omega_k^{lf})f(\omega_p^{nf})}
	\SB{N^{lnf\munu}_H(k^0=-\omega_k^{lf})\delta(q^0-\omega_k^{lf}-\omega_p^{nf})+ N_H^{lnf\munu}(k^0=\omega_k^{lf})\delta(q^0+\omega_k^{lf}+\omega_p^{nf})} \right. \nn \\ 
	&& \hspace{-2.3cm} \left.
	+ \SB{-f(\omega_k^{lf})-f(\omega_p^{nf})+2f(\omega_k^{lf})f(\omega_p^{nf})}
	\SB{N^{lnf\munu}_H(k^0=-\omega_k^{lf})\delta(q^0-\omega_k^{lf}-\omega_p^{nf})+ N^{lnf\munu}_H(k^0=\omega_k^{lf})\delta(q^0+\omega_k^{lf}+\omega_p^{nf})}
}  \label{eq.impi.tb.2}
\end{eqnarray}
where $N_h^{lnf}(q,k)$ and $N_H^{lnf\munu}(q,k)$ are defined in Eqs.~\eqref{eq.Nh.tb} and \eqref{eq.NH.tb}.

  It may be emphasized that though the present work uses Real Time version of thermal field theory, use of the more popular imaginary time formalism (ITF) leads to the same expressions. For example, the expression of the thermo-magnetic quark self energies or the polarization functions of $\pi^0$ and $\sigma$ obtained here are identical to the ones obtained in Refs.~\cite{Klevansky,zhang,PhysRevD.93.014010} earlier using the ITF.


\section{REGULARIZATION PROCEDURE FOR THE NJL MODEL} \label{sec.regular}
As already mentioned in the previous sections, that the NJL model requires a proper 
regularization procedure. Using the dimensional regularization technique, we have been able to isolate the UV-divergences as the 
pole of Gamma functions in Eqs.~\eqref{eq.sigma.purevac.expansion} and \eqref{eq.Pi.pi}-\eqref{eq.Pi.a1}. Now, in order to obtain 
a finite contributions from these equations, we first note that the integral representation of the Gamma functions can be written as
\begin{eqnarray}
\Gamma(z) &=& \int_{0}^{\infty} dt e^{-t} t^{z-1} =  \int_{0}^{r} dt e^{-t} t^{z-1} + \int_{r}^{\infty} dt e^{-t} t^{z-1} \nn \\
&=&  \gamma(z,r) + \Gamma(z,r)
\end{eqnarray}
where, $\gamma(z,r)$ is the lower incomplete gamma function and $\Gamma(z,r)$ is the (upper) incomplete Gamma function.  In the evaluation of  loop diagrams in the NJL model using Schwinger proper-time method,  one often encounters integrals which can be written in terms of  $\Gamma$ functions with  negative integer argument. Clearly, those are divergent quantities and need to be regulated. One possible way is to introduce proper-time regulator where the lower incomplete gamma function  containing  the divergence is discarded  and only the  $\Gamma(z,r)$ part is retained (see for example  Eq.~(3.15) in \cite{Klevansky} which is the proper-time regularized version of Eq.~(3.13) there in).   Following the similar procedure, in our regularization scheme,  
  the divergent Gamma functions  obtained from dimensional regularization are replaced with the  incomplete Gamma function i.e.
\begin{eqnarray}
\Gamma(0) \to \Gamma\FB{0,\frac{M^2}{\Lambda^2}} ~~~ \text{and}~~~ 
\Gamma(-1) \to \Gamma\FB{-1,\frac{M^2}{\Lambda^2}}
\end{eqnarray}
where $\Lambda$ is a scale parameter to be determined. Thus our regularization scheme is a mixed procedure where, though the  dimensional regularization is used at first to obtain the consistent Lorentz structure, the divergences appeared are regulated following the proper-time regularization. 
After these replacements, Eqs.~\eqref{eq.sigma.purevac.expansion} and \eqref{eq.Pi.pi}-\eqref{eq.Pi.a1} can be simplified to 
\begin{eqnarray}
\RE\Sigma^\text{Pure-Vac}_\text{MFA} = 2g_s \frac{N_cN_fM^3}{4\pi^2}\Gamma\FB{-1,\frac{M^2}{\Lambda^2}}
\label{eq.quark.self.energy.4}
\end{eqnarray}
and 
\begin{eqnarray}
\Pi_\pi(q) &=& \frac{N_c N_f}{4\pi^2}\TB{\frac{1}{2}q^2\Gamma\FB{0,\frac{M^2}{\Lambda^2}}
	+ M^2\Gamma\FB{-1,\frac{M^2}{\Lambda^2}}}, \label{eq.Pi.pi.vac} \\
\Pi_\sigma(q) &=& \frac{N_c N_f}{4\pi^2}\TB{\frac{1}{2}(q^2-4M^2)\Gamma\FB{0,\frac{M^2}{\Lambda^2}} 
	+ M^2\Gamma\FB{-1,\frac{M^2}{\Lambda^2}}}, \label{eq.Pi.sigma.vac} \\
\Pi^\munu_\rho(q) &=& -\frac{N_c N_f}{12\pi^2}\Gamma\FB{0,\frac{M^2}{\Lambda^2}}q^2 \FB{g^\munu - \frac{q^\mu q^\nu}{q^2}},
 \label{eq.Pi.rho.vac} \\
\Pi^\munu_{a_1}(q) &=& -\frac{N_c N_f}{12\pi^2} \Gamma\FB{0,\frac{M^2}{\Lambda^2}}
\TB{\FB{q^2-6M^2}\FB{g^\munu - \frac{q^\mu q^\nu}{q^2}} - 6M^2 \frac{q^\mu q^\nu}{q^2}  }. \label{eq.Pi.a1.vac}
\end{eqnarray}

It can be notice in Eqs.~\eqref{eq.Pi.pi.vac}-\eqref{eq.Pi.a1.vac}, that if the chiral symmetry is completely restored (i.e. $M=0$), 
then the polarization functions of $\sigma$ and $a_1$ become identical to that of $\pi$ and $\rho$ respectively.
Moreover, observing the Lorentz structure in Eq.~\eqref{eq.Pi.rho.vac}, it immediately follows that, the polarization 
function of $\rho$ is transverse i.e.
\begin{eqnarray}
q_\mu \Pi^\munu_\rho(q) = 0.
\end{eqnarray}
The reason behind this transversality is the conservation of the vector current $J^\mu(x) = \overline{\Psi}(x)\gamma^\mu\Psi$ 
which is the Noether's current corresponding to the $U(1)$ symmetry of the NJL Lagrangian in Eq.~\eqref{eq.Lagrangian}. 
Similar arguments also hold for the Lorentz structure of the polarization function of $a_1$ in which the non-transverse 
piece is proportional to the constituent quark mass $M$. This is because of the non-conservation of the axial-vector current 
$J^{5\mu} = \overline{\Psi}(x)\gamma^\mu\gamma^5\Psi$ whose four-divergence is 
\begin{eqnarray}
\del_\mu J^{5\mu} \propto M.
\end{eqnarray}
In the chiral limit ($M=0$), the axial-vector current is conserved leading to a transverse polarization function of $a_1$.

It is worth mentioning that, the consistent Lorentz structure of the polarization functions of $\rho$ and $a_1$ could be obtained only because we have used dimensional regularization technique which respects the Lorentz symmetry. 
Any other regulator such as three-momentum cutoff, Euclidean four-momentum cutoff and Schwinger proper-time regulator 
will spoil the Lorentz structures and $\Pi^\munu_\rho(q)$ will no longer be transverse.

We now fix the parameters for the  NJL model. For this we need the expression of pion decay constant ($f_\pi^2$) which comes out to be
\begin{eqnarray}
f_\pi^2 = \frac{N_cM^2}{4\pi^2}\Gamma\FB{0,\frac{M^2}{\Lambda^2}}
\label{eq.fpi2}
\end{eqnarray}
using the dimensional regularization. By simultaneously fitting the vacuum quark condensate and pion decay constant values as
\begin{eqnarray}
\frac{1}{N_f}\ensembleaverage{\overline{\psi}\psi} = -\frac{1}{2g_s N_f}\RE\Sigma^\text{Pure-Vac}_\text{MFA}  = -(230)^3~\text{MeV}^3 
~~~ \text{and}~~~ f_\pi = 95~\text{MeV}
\end{eqnarray}
we find $\Lambda = 936$ MeV and $M = 226$ MeV. Next, considering the current quark mass $m=6.6$ MeV and vacuum pion mass $m_\pi = 135$ MeV, 
the scalar coupling comes out to be $g_s=4.5126$ GeV$^{-2}$. Finally, $g_v=4.289$ GeV$^{-2}$ is chosen to reproduce the 
vacuum mass of $\rho$ meson as $m_\rho=770$ MeV.

It should  be mentioned here  that, the expressions of $\RE\Sigma^\text{Pure-Vac}_\text{MFA}$ and $f_\pi^2$ in Eqs.~\eqref{eq.quark.self.energy.4} and \eqref{eq.fpi2} are  same as those  obtained using the proper-time regularization technique~\cite{Klevansky,PhysRevD.99.116002}. However,  the expressions of the polarization functions will be different if one uses the proper-time regulator. For example, in that case,  the consistent Lorentz structures of the polarization functions of $\rho^0$ and $a_1^0$ as in Eqs.~\eqref{eq.Pi.rho.vac} and \eqref{eq.Pi.a1.vac} will not appear automatically as appears in dimensional regularization.  Moreover, the transversality condition $q_\mu \Pi^\munu_\rho(q) = 0$ is not satisfied if the proper-time regularization is used.


\section{NUMERICAL RESULTS}\label{sec.results}
\begin{figure}[h]
	\begin{center}
		\includegraphics[angle=-90,scale=0.35]{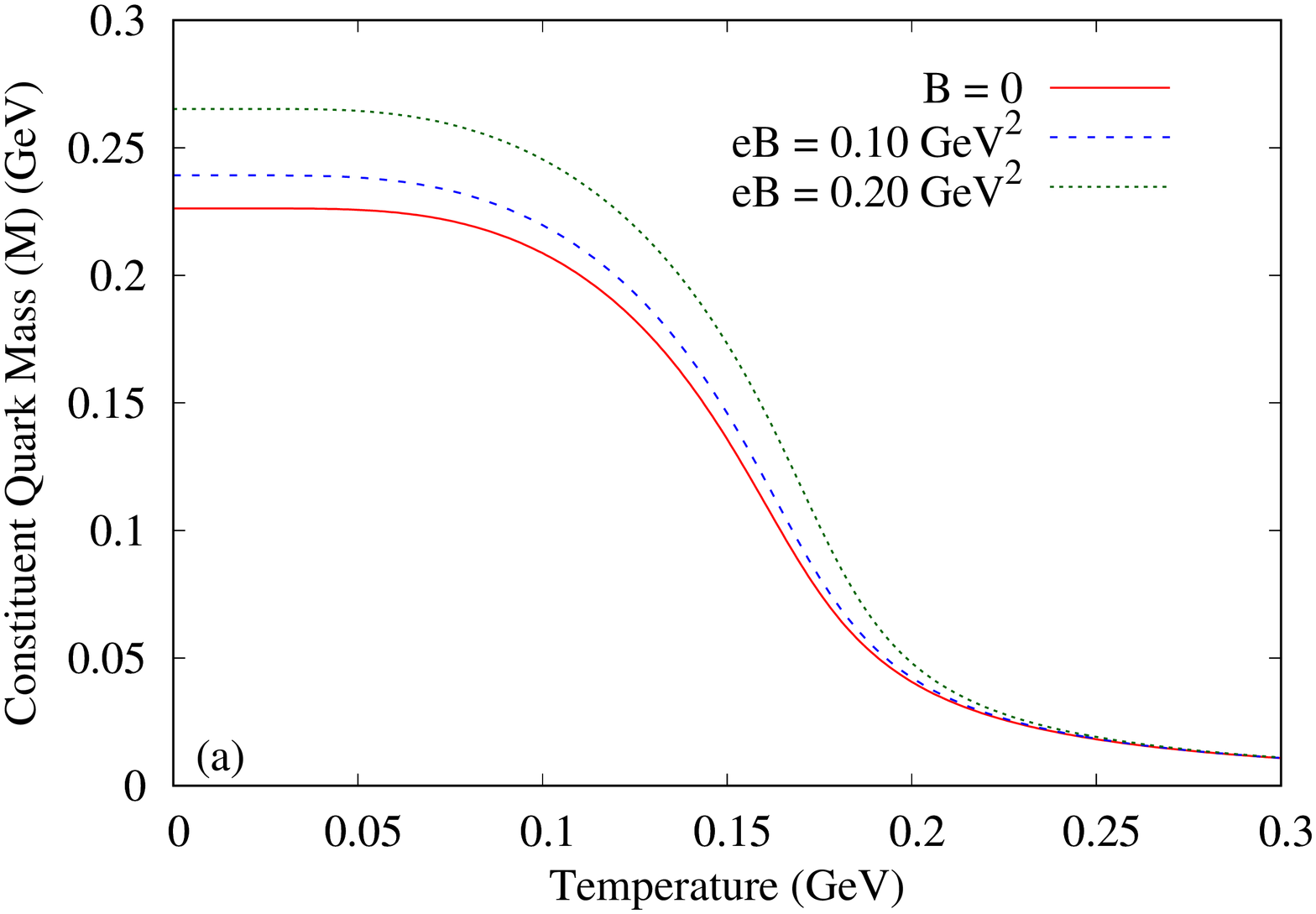}  \includegraphics[angle=-90,scale=0.35]{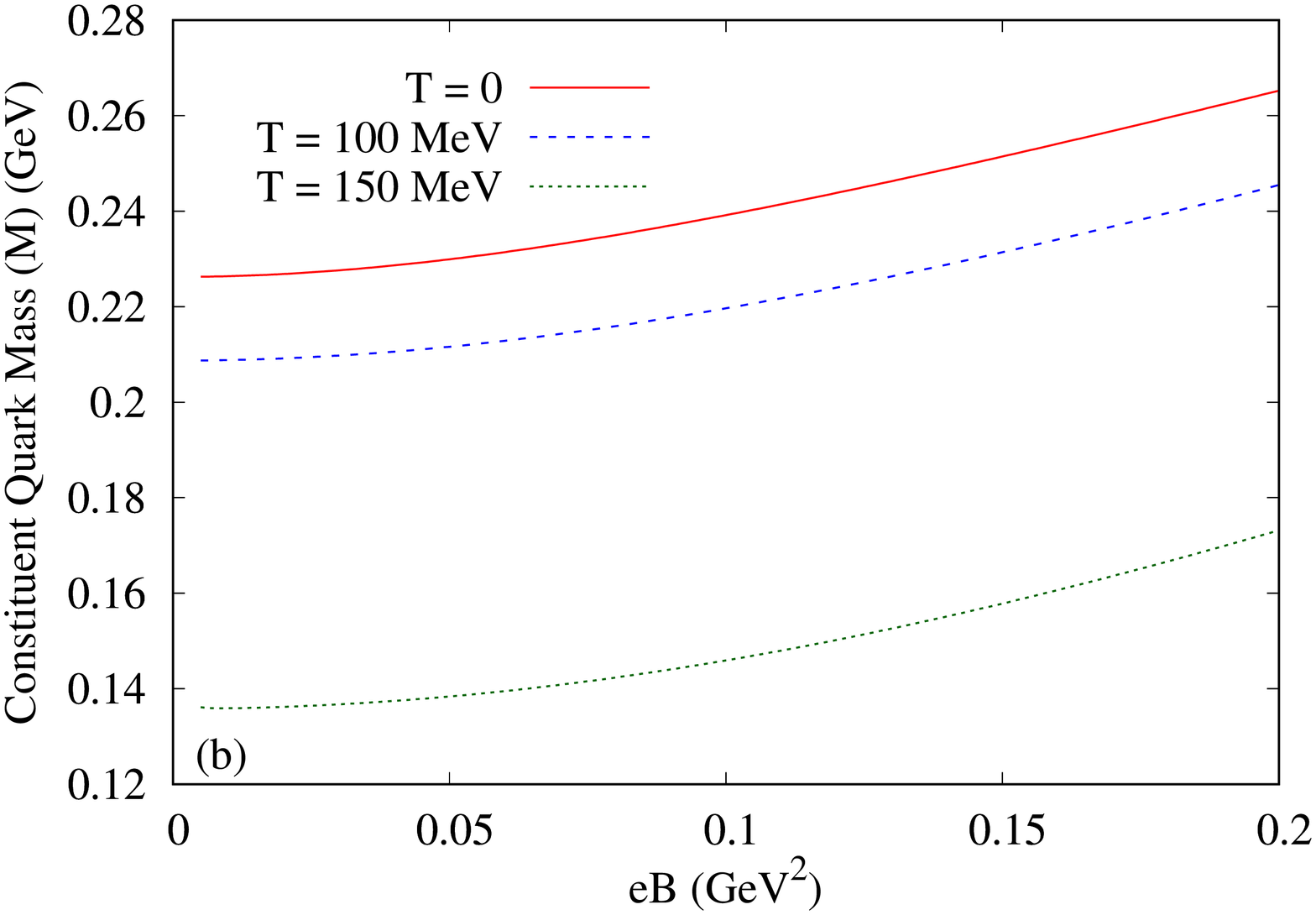}
	\end{center}
	\caption{Variation of the constituent quark mass ($M$) as a function of 
		(a) temperature for different values of external magnetic field and 
		(b) external magnetic field for different values of temperature.}
	\label{fig.M}
\end{figure}

We start this section by showing the variation of the constituent quark mass as a function of temperature for different values 
of external magnetic field in Fig.~\ref{fig.M}(a). 
As can be seen in the figure, $M$  remains almost constant in the low temperature region. However, with further increase in temperature, 
 the constituent quark mass decreases substantially signifying a phase transition. 
Throughout the whole temperature range $M$ remains single-valued depicting the smooth crossover nature of the phase transition. 
Since we are working with finite current quark mass $m\ne0$, the chiral symmetry is only partially restored. 
To obtain the transition temperature, one can use various susceptibilities which will be discussed in the next paragraph. 
For a particular value of temperature, the constituent quark mass increases with the  external magnetic field as  shown in Fig.~\ref{fig.M}(b).

\begin{figure}[h]
	\begin{center}
		\includegraphics[angle=-90,scale=0.35]{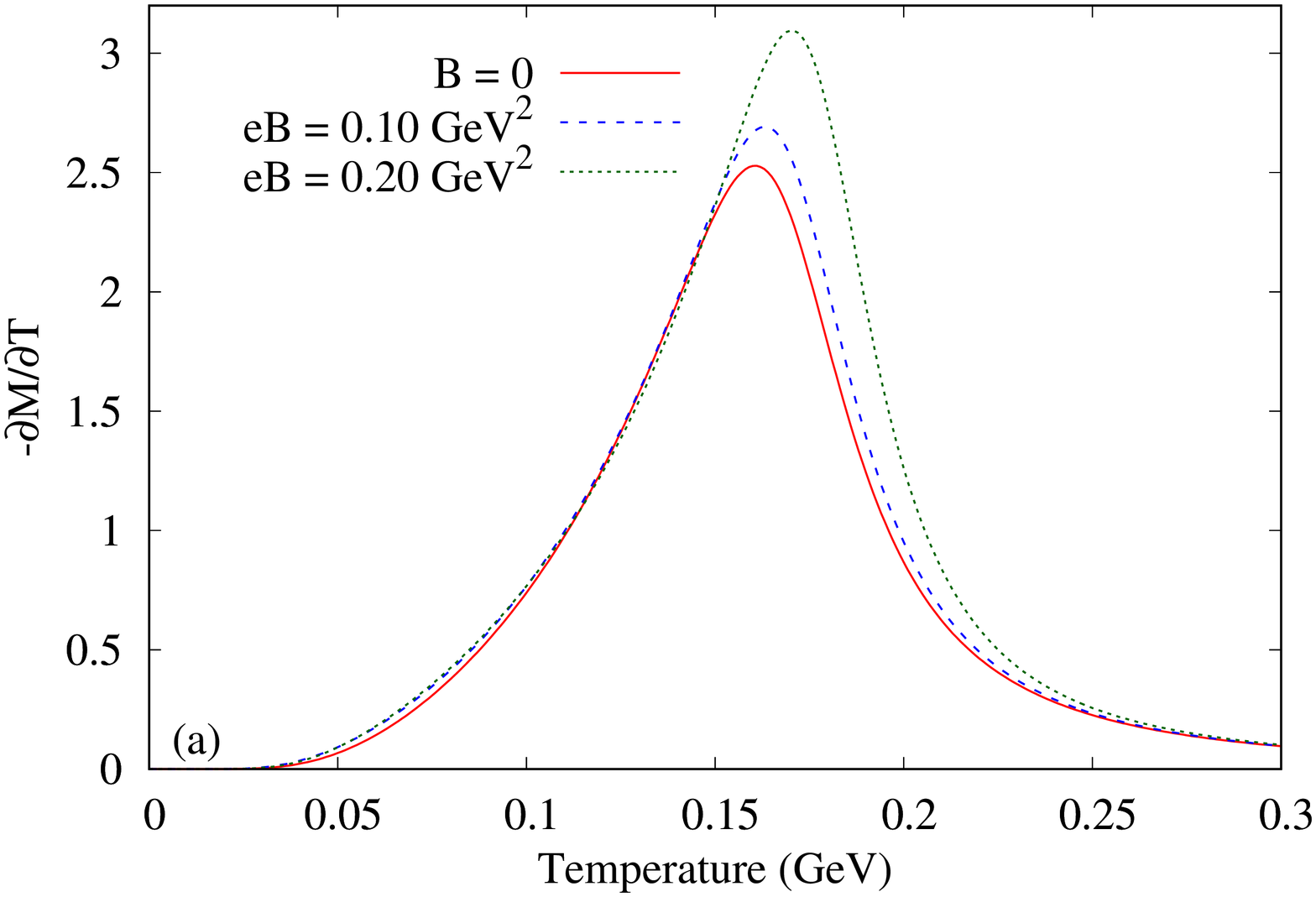}  \includegraphics[angle=-90,scale=0.35]{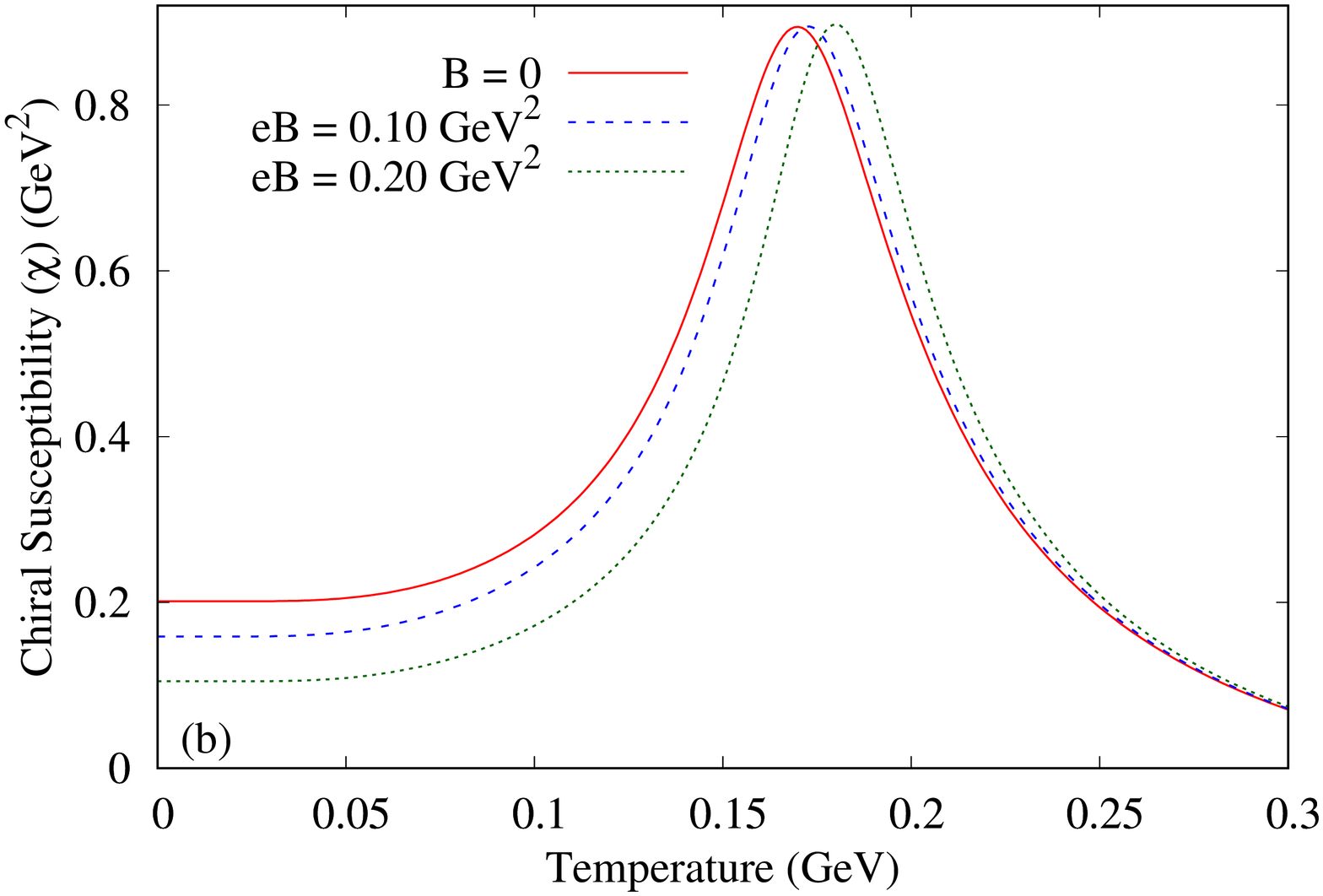}
	\end{center}
	\caption{(a) Variation of the $-\partial M/\partial T$ and (b) the chiral susceptibility ($\chi$) as a function 
		of temperature for different values of external magnetic field. 
	 }
	\label{fig.Sus}
\end{figure}
The transition temperature corresponding to the  partial restoration of chiral symmetry can be obtained from various susceptibilities. 
The calculation of the susceptibility $-\del M/\del T$ and chiral susceptibility $\chi = \frac{1}{2g_s}\FB{\frac{\del M}{\del m}-1}$ 
have been provided in Appendix~\ref{app.sus}.  
In Figs.~\ref{fig.Sus}(a) and (b), $-\del M/\del T$ and $\chi$ are respectively plotted as a function of temperature 
for different values of the external magnetic field. The position of the peak of $-\del M/\del T$ or $\chi$ represents the transition 
temperature. As can be noticed from the plots, with the increase in external magnetic field the peak of the susceptibilities moves 
towards higher values of temperature. Thus, in this framework, the transition temperature increases with  $B$. 
This may be identified as \textit{magnetic catalysis} (MC) in the NJL model where the external magnetic field catalyzes  the 
spontaneous breaking of chiral symmetry~\cite{Shovkovy,Gusynin1,Gusynin2,Gusynin3}. Moreover, as the susceptibilities remain continuous and finite with the change in temperature, 
the nature of the phase transition can be inferred as smooth crossover.

\begin{figure}[h]
	\begin{center}
		\includegraphics[angle=-90,scale=0.35]{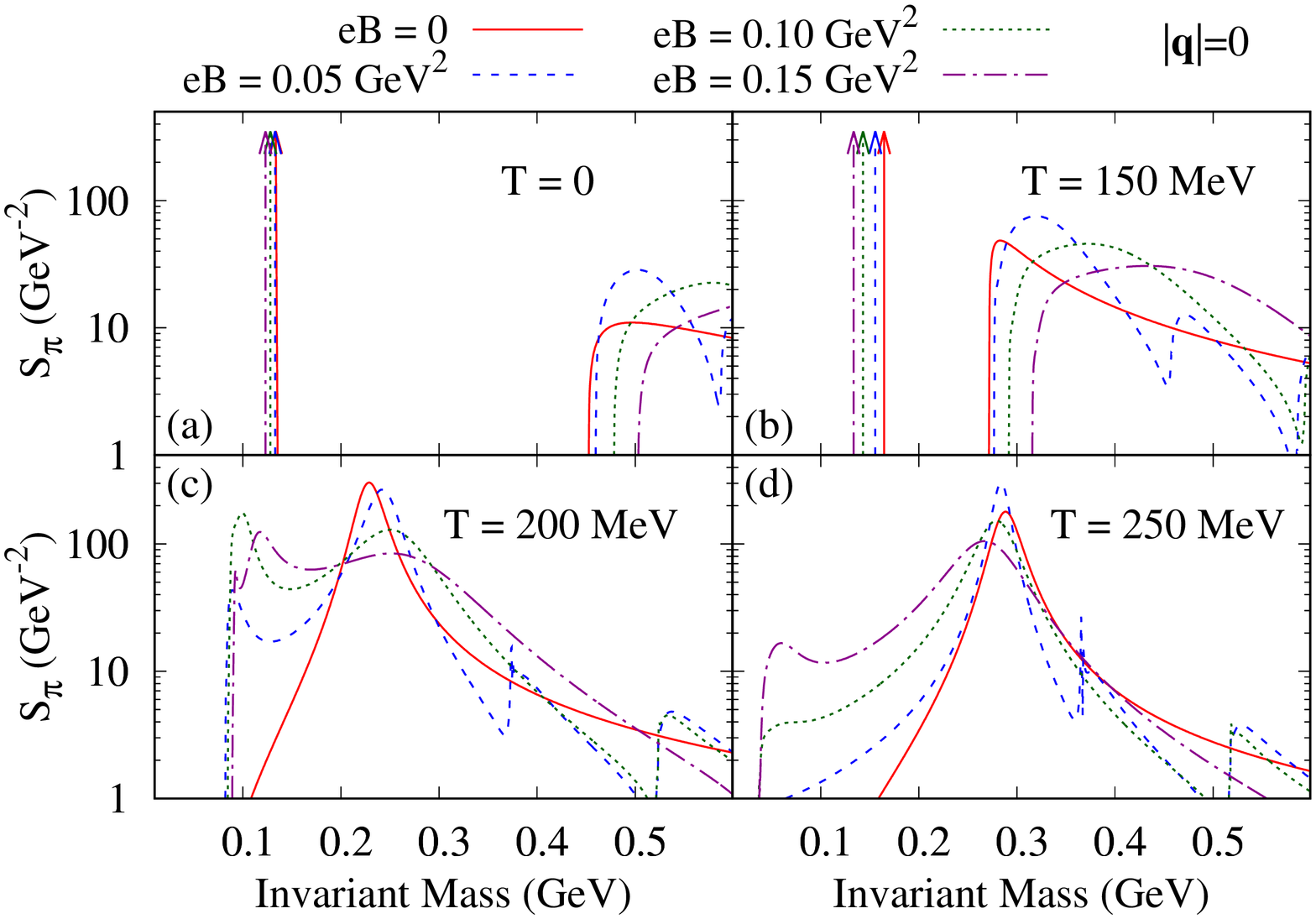} \includegraphics[angle=-90,scale=0.35]{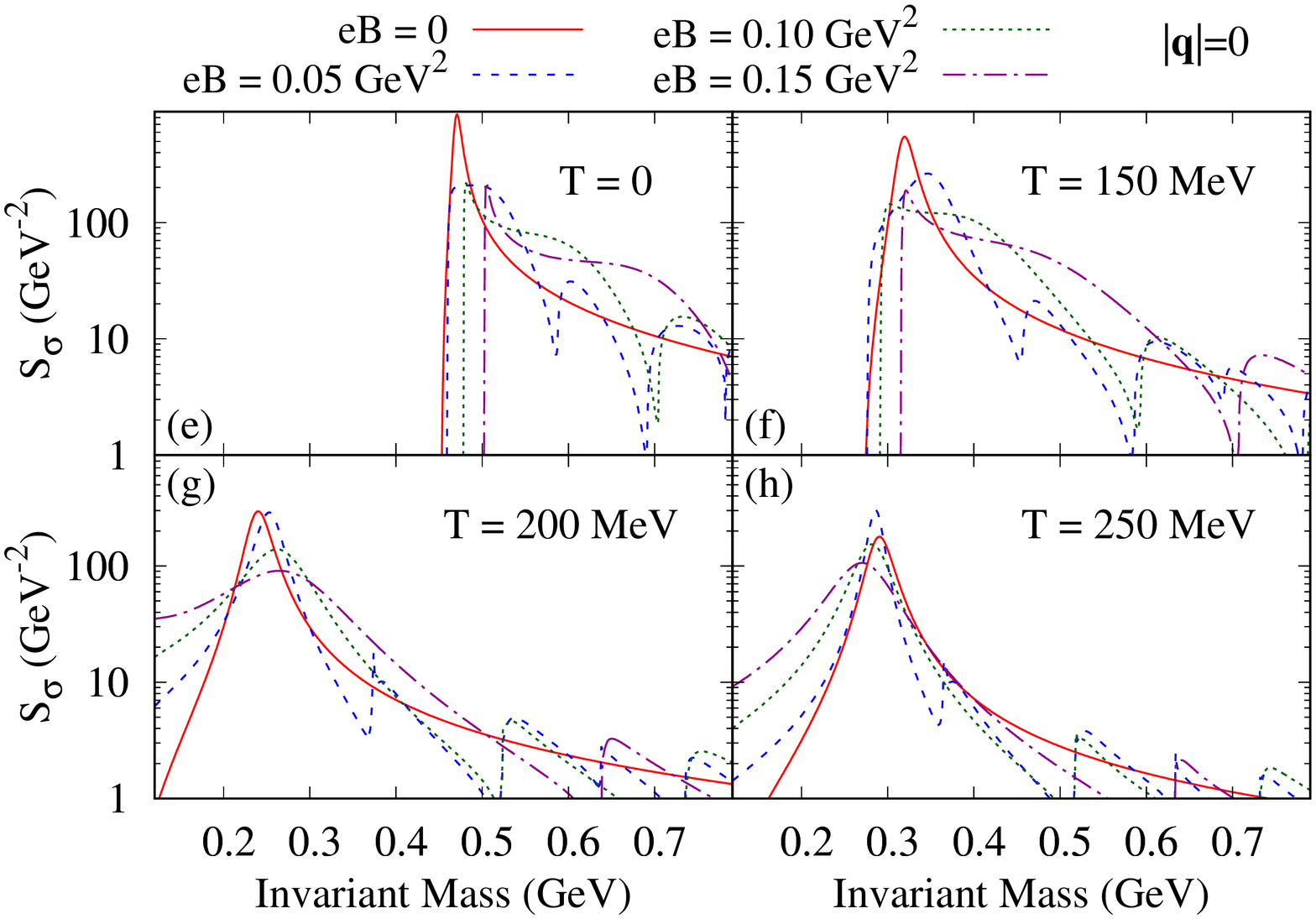}
	\end{center}
	\caption{Spectral function of $\pi^0$ and $\sigma$ mesons as a function of invariant mass for $\vec{q}=\vec{0}$ at different 
	values of temperature and external magnetic field. The arrows represent Dirac delta functions.}
	\label{fig.s1}
\end{figure}
We now turn our attention to the mesonic properties. We define the spectral functions of mesons as the imaginary part of the 
respective complete propagators. From Eq.~\eqref{eq.complete.prop.h}, the spectral function for the $\pi$ and $\sigma$ mesons can be written as 
\begin{eqnarray}
S_h(q) = \IM\overline{\overline{D'}}_h(q) = \IM\TB{\frac{-2g_s}{1-2g_s\Pibarbar_h(q)}} 
= (-2g_s)\TB{\frac{2g_s\IM\Pibarbar_h(q)}{\FB{1-2g_s\RE\Pibarbar_h(q)}^2+\FB{2g_s\IM\Pibarbar_h(q)}^2}}.
\end{eqnarray}
In Figs.~\ref{fig.s1}(a)-(d), the spectral functions of $\pi$ have been shown as a function of its 
invariant mass $\sqrt{q^2}$ for different values of temperature and external magnetic field in the rest 
frame of $\pi$ (i.e. $\vec{q}=\vec{0}$). 
Let us first consider the $B=0$ cases which are shown as solid-red curves in Figs.~\ref{fig.s1}(a)-(d). At zero temperature, $S_\pi$ is a Dirac delta function at its pole mass ($\sqrt{q^2}=135$ MeV) along with a two-quark continuum starting at 
$\sqrt{q^2}>2M$. It can be  noticed from Fig.~\ref{fig.s1}(b), that at $T=150$ MeV, the Dirac delta function moves
 towards the higher invariant mass and the two-quark continuum threshold has significantly decreased which 
is due to the decrease in $M$ with temperature. Yet, the  delta function is well separated from the continuum 
revealing the fact that  $\pi$ is still a bound state. With further increase in temperature, as shown in Figs.~\ref{fig.s1}(c)-(d),  
 the Dirac delta function disappears and the shape of spectral function becomes a Breit-Wigner. These imply that, the 
pion has now become a resonant state with finite decay width. 
Let us now discuss the effect of external magnetic field on $S_\pi$. For the lower temperature ($T=0$ and $150$ MeV), 
the Dirac delta functions move towards higher values of the invariant mass with the increase in external magnetic field. For higher values of 
temperature ($T=200$ and $250$ MeV),  the spectral functions at non-zero $B$ are observed to oscillate about the $B=0$ curve and the peak of the Breit-Wigner  shifts significantly  towards higher invariant mass. The oscillation 
frequency (amplitude) is observed to be  large (small)  at lower values of $B$ as compared to its higher values.

The situation is quite different in case of $\sigma$ meson. 
In Figs.~\ref{fig.s1}(e)-(h), the spectral functions of $\sigma$ have been shown as a function of its 
invariant mass for different values of temperature and external magnetic field for $\vec{q}=\vec{0}$. 
In this case, the spectral function is always Breit-Wigner shaped implying that the $\sigma$ remains always a resonant excitation. 
As shown in Figs.~\ref{fig.s1}(e)-(g), with the increase in temperature (up to $T=200$ MeV), 
the peak of $S_\sigma$ moves towards lower invariant mass. However in Fig.~\ref{fig.s1}(h), (at $T=250$ MeV), the peak again start moving towards higher $\sqrt{q^2}$ values. The effect of external 
magnetic field on $S_\sigma$ is similar to that of $\pi$ showing oscillations in  $S_\sigma$ at non-zero  $B$  about the $B=0$ curve. The oscillation 
frequency (amplitude) follows the similar trend as described for pion.

Let us now consider the propagation of $\rho$ and $a_1$ meson. Since we will be considering the special case $\qper=0$, 
we have significant simplifications of the complete propagators of $\rho$ and $a_1$. As given in Eq.~\eqref{eq.special.case.2}, 
we have for $\qper=0$,
\begin{eqnarray}
\Pibarbar_{H1}(q^0,\vec{q}_\perp=\vec{0}) = \Pibarbar_{H4}(q^0,\vec{q}_\perp=\vec{0}) ~~\text{and}~~ 
\Pibarbar_{H6}(q^0,\vec{q}_\perp=\vec{0}) = \Pibarbar_{H7}(q^0,\vec{q}_\perp=\vec{0}) = 0.
\end{eqnarray}
Moreover, we find in our numerical calculations that $\Pibarbar_{H5}(q^0,\vec{q}_\perp=\vec{0})=0$. Thus, the form factors for 
the complete thermo-magnetic propagators in Eqs.~\eqref{eq.D1}-\eqref{eq.D7} simplify to
\begin{eqnarray}
\Dbarbar_{H1} &=& \Dbarbar_{H4} = \FB{\frac{2g_v}{1+2g_v\Pibarbar_{H1}}} \label{eq.D14.1},\\
\Dbarbar_{H2} &=& \FB{\frac{2g_v}{1+2g_v\Pibarbar_{H2}}} \label{eq.D2.1},\\
\Dbarbar_{H3} &=& \FB{\frac{2g_v}{1+2g_v\Pibarbar_{H3}}} \label{eq.D3.1},\\
\Dbarbar_{H5} &=& \Dbarbar_{H6} = \Dbarbar_{H7} = 0 \label{eq.D567.1}.
\end{eqnarray}
\begin{figure}[h]
	\begin{center}
		\includegraphics[angle=-90,scale=0.35]{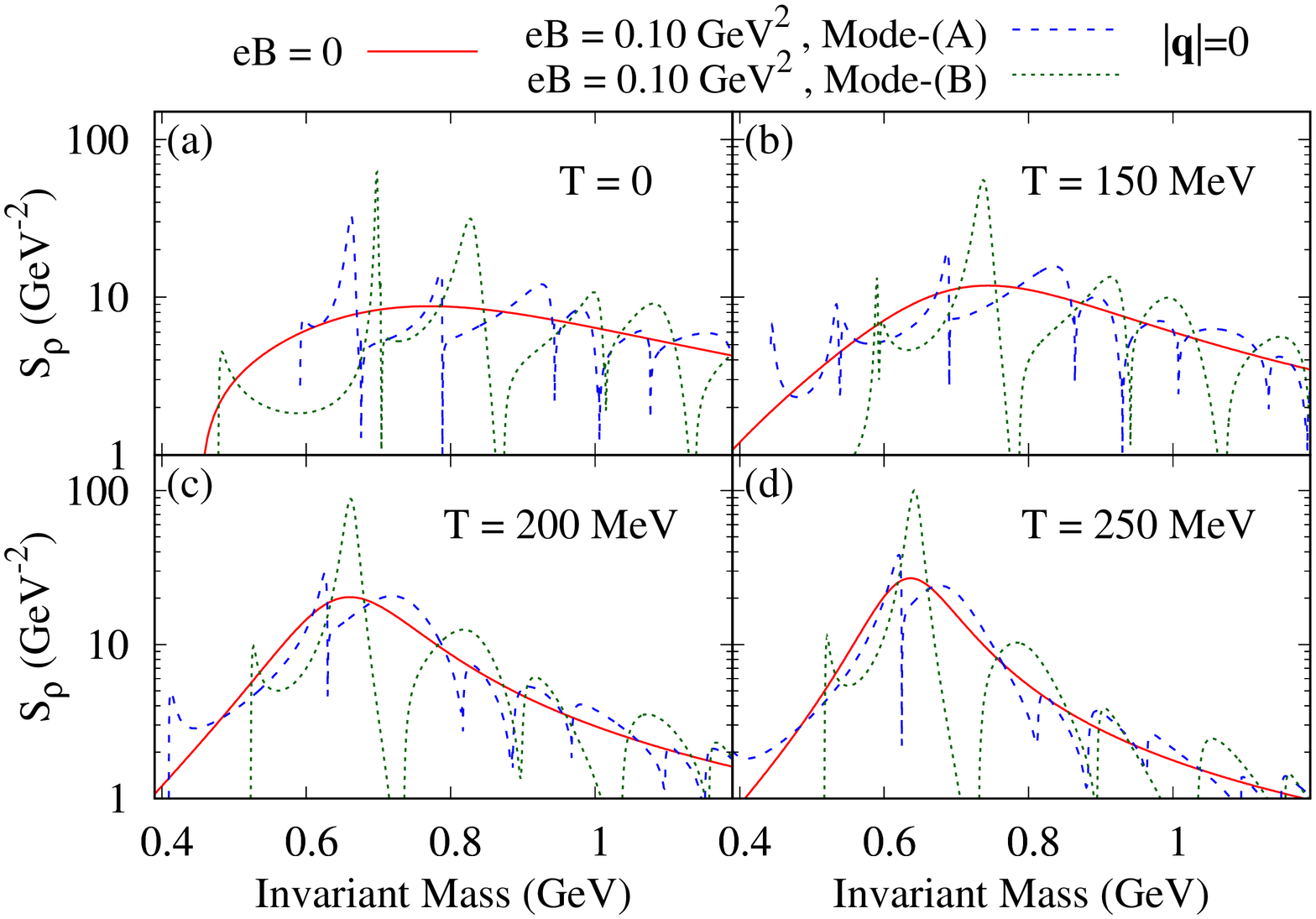} \includegraphics[angle=-90,scale=0.35]{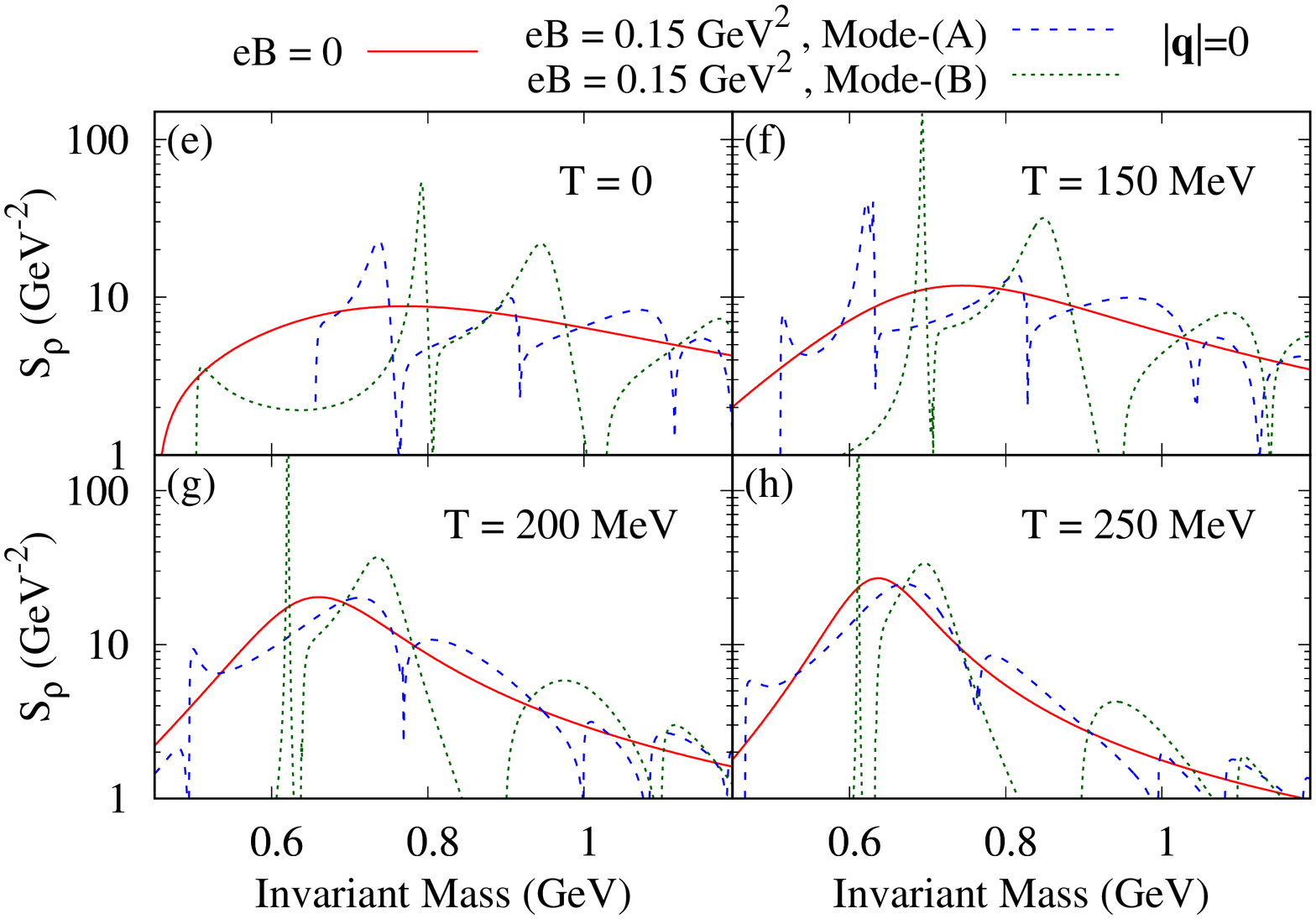}
		\includegraphics[angle=-90,scale=0.35]{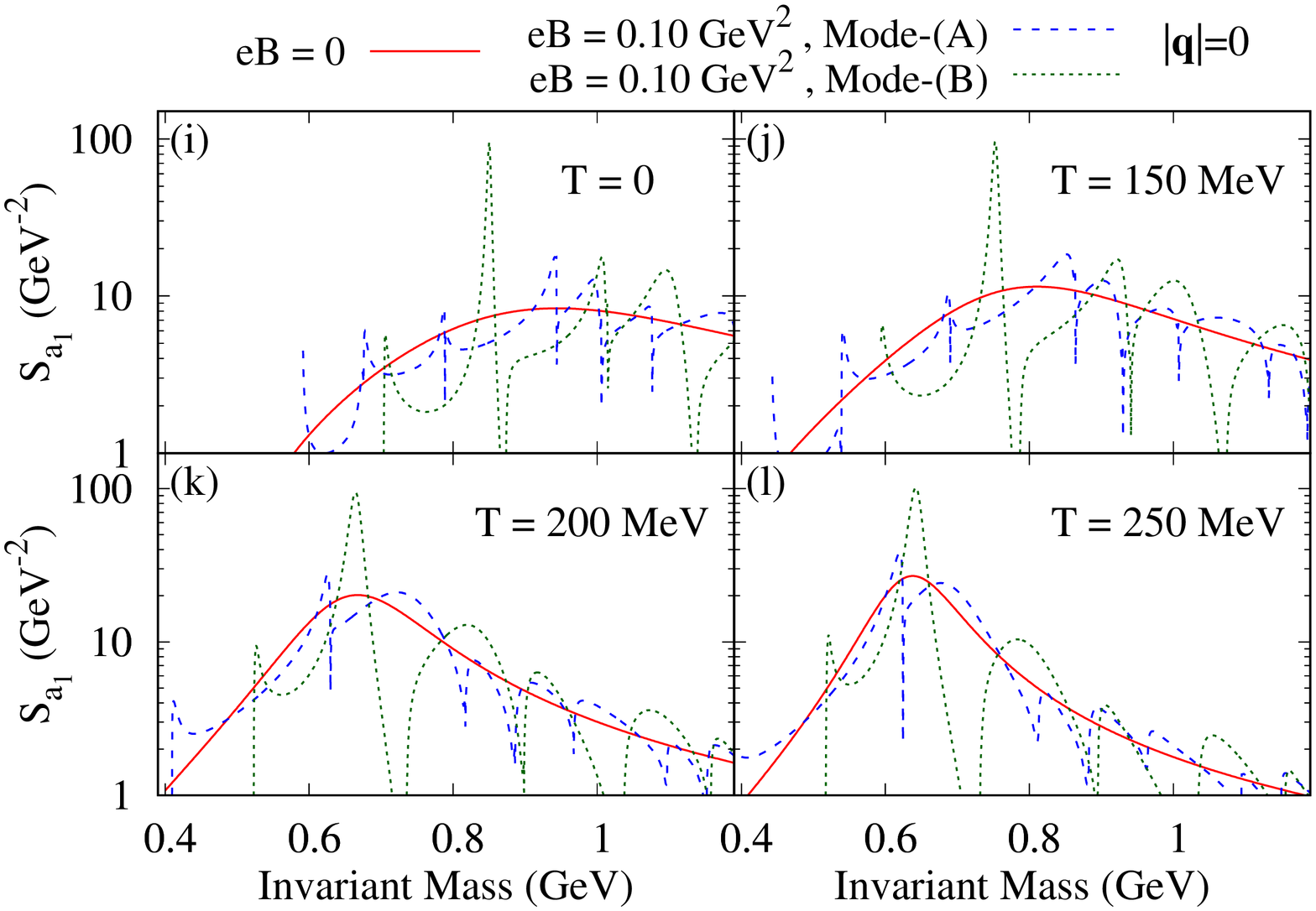} \includegraphics[angle=-90,scale=0.35]{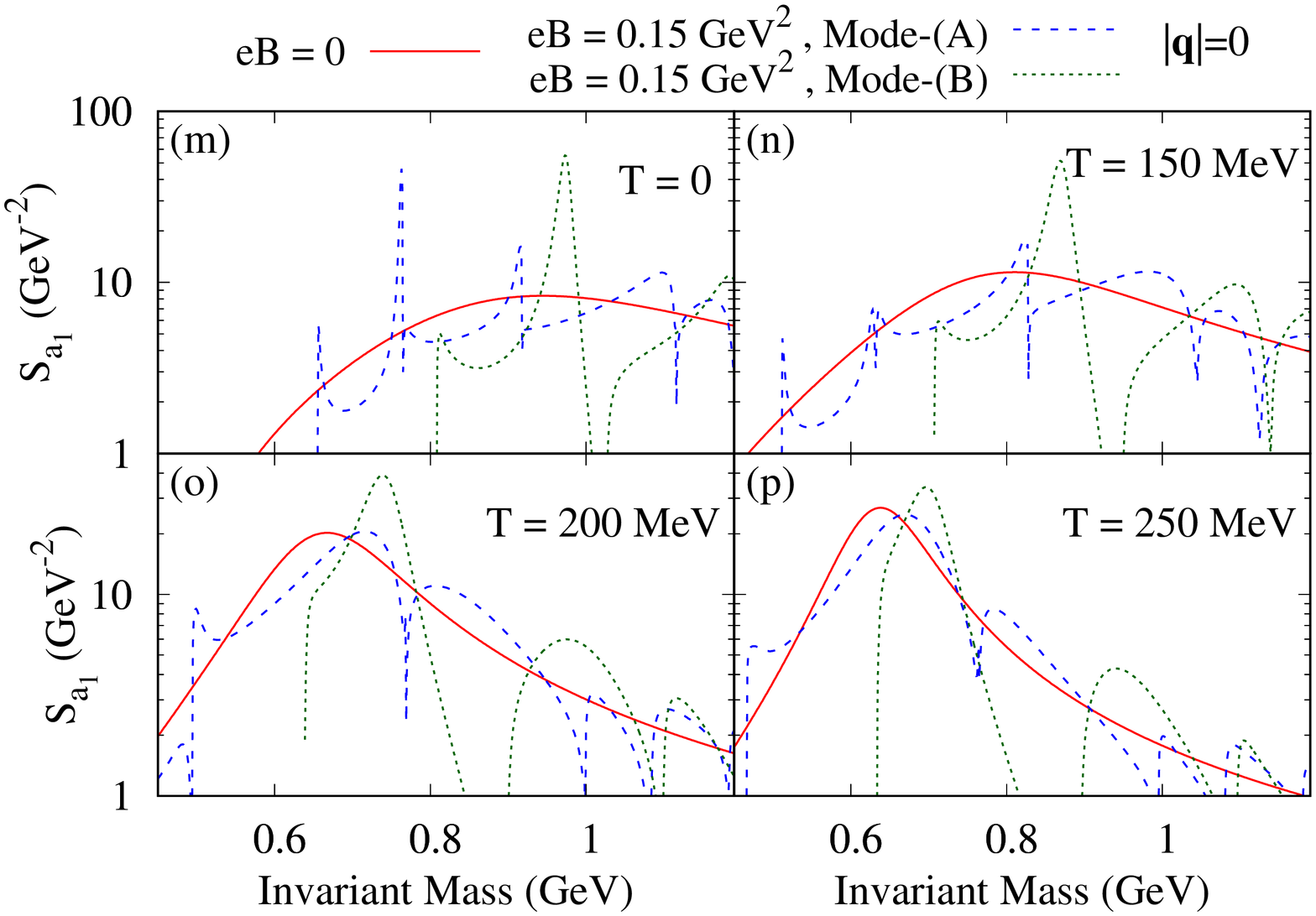}
	\end{center}
	\caption{Spectral functions of $\rho^0$ and $a_1^0$ mesons as a function of invariant mass for $\vec{q}=\vec{0}$ at different 
		values of temperature and external magnetic field.}
	\label{fig.s2}
\end{figure}
Therefore, the complete thermo-magnetic propagator from Eq.~\eqref{eq.Dbarbar} becomes 
\begin{eqnarray}
\Dbarbar'^\munu_H(\qpll,\qper=0) &=& \sum_{i=1}^{7}\Dbarbar_{Hi}(\qpll,\qper=0)\Pbarbar_i^\munu \nn \\
 &=& \FB{\frac{2g_v}{1+2g_v\Pibarbar_{H1}}}\FB{\Pbarbar_1^\munu+\Pbarbar_4^\munu} 
 + \FB{\frac{2g_v}{1+2g_v\Pibarbar_{H2}}}\Pbarbar_2^\munu 
 + \FB{\frac{2g_v}{1+2g_v\Pibarbar_{H3}}}\Pbarbar_3^\munu.
\label{eq.Dbarbar.1}
\end{eqnarray}
The second term on the RHS of the above equation containing the non-transverse tensor $\Pbarbar_2^\munu$ corresponds 
to a non-propagating mode as the corresponding form factor does not have any pole. Thus, we find three modes of 
propagation of $\rho$ and $a_1$ mesons in the thermo-magnetic medium; two of them are found to be degenerate (corresponding 
to $\Pbarbar_1^\munu$ and $\Pbarbar_4^\munu$). This degeneracy is solely due to our special choice of $\qper=0$. 
Thus, we are left with two distinct modes for the $\rho$ and $a_1$ propagations. We call them as Mode-(A) and Mode-(B) respectively. 
The spectral functions for these two modes are therefore defined as,
\begin{eqnarray}
S_H^\text{A} &=& \IM\TB{\frac{2g_v}{1+2g_v\Pibarbar_{H1}}} 
= (-2g_v)\TB{\frac{2g_v\IM\Pibarbar_{H1}(q)}{\FB{1+2g_s\RE\Pibarbar_{H1}(q)}^2+\FB{2g_s\IM\Pibarbar_{H1}(q)}^2}}, \\
S_H^\text{B} &=& \IM\TB{\frac{2g_v}{1+2g_v\Pibarbar_{H3}}} 
= (-2g_v)\TB{\frac{2g_v\IM\Pibarbar_{H3}(q)}{\FB{1+2g_s\RE\Pibarbar_{H3}(q)}^2+\FB{2g_s\IM\Pibarbar_{H3}(q)}^2}}.
\end{eqnarray}

In Figs.~\ref{fig.s2}(a)-(p), we have presented the spectral functions of $\rho$ and $a_1$ mesons as a function of their invariant mass 
at $\vec{q}=\vec{0}$ for different temperature and external magnetic field. Similar to the  case of $\sigma$, the $\rho$ and $a_1$ are 
 always in resonant state so that the shape of their spectral functions remains  Breit-Wigner. 
Since we have taken $\vec{q}=\vec{0}$ in these plots, the two modes are degenerate for $B=0$ (the solid-red curves). 
The external magnetic field breaks this degeneracy and we find two distinct modes of $\rho$ and $a_1$ propagations even in 
their rest frames for non-zero values of $B$. 
With the increase in temperature, the peaks of the spectral functions move towards lower values of invariant mass. 
Moreover, the spectral functions at non-zero external magnetic field show highly oscillatory behaviour about the $B=0$ curves. 
Similar to the case of $\pi$ and $\sigma$, we observe higher(lower)  oscillation frequency (amplitude)  at lower values 
of $B$.

\begin{figure}[h]
	\begin{center}
		\includegraphics[angle=-90,scale=0.35]{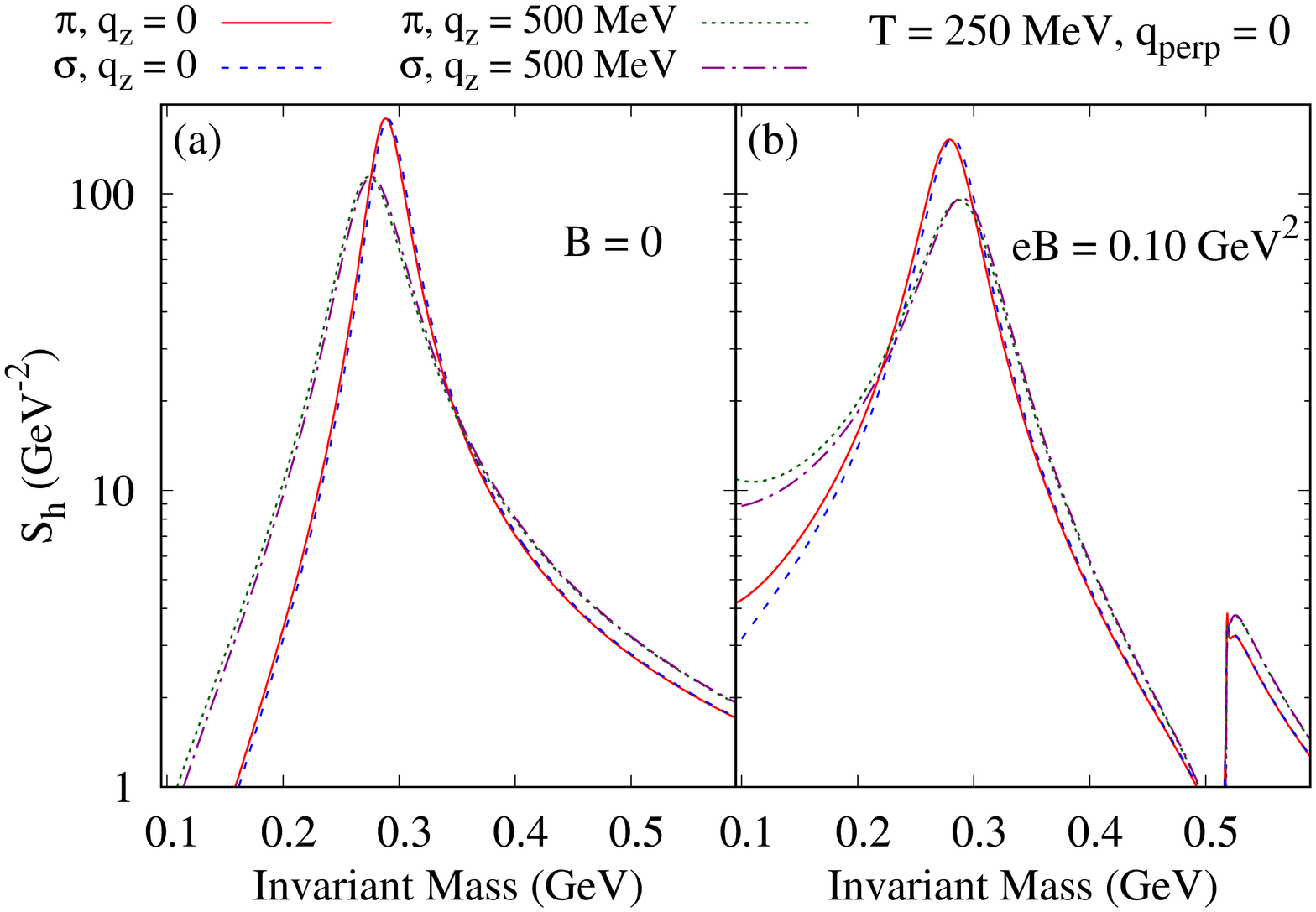} \includegraphics[angle=-90,scale=0.35]{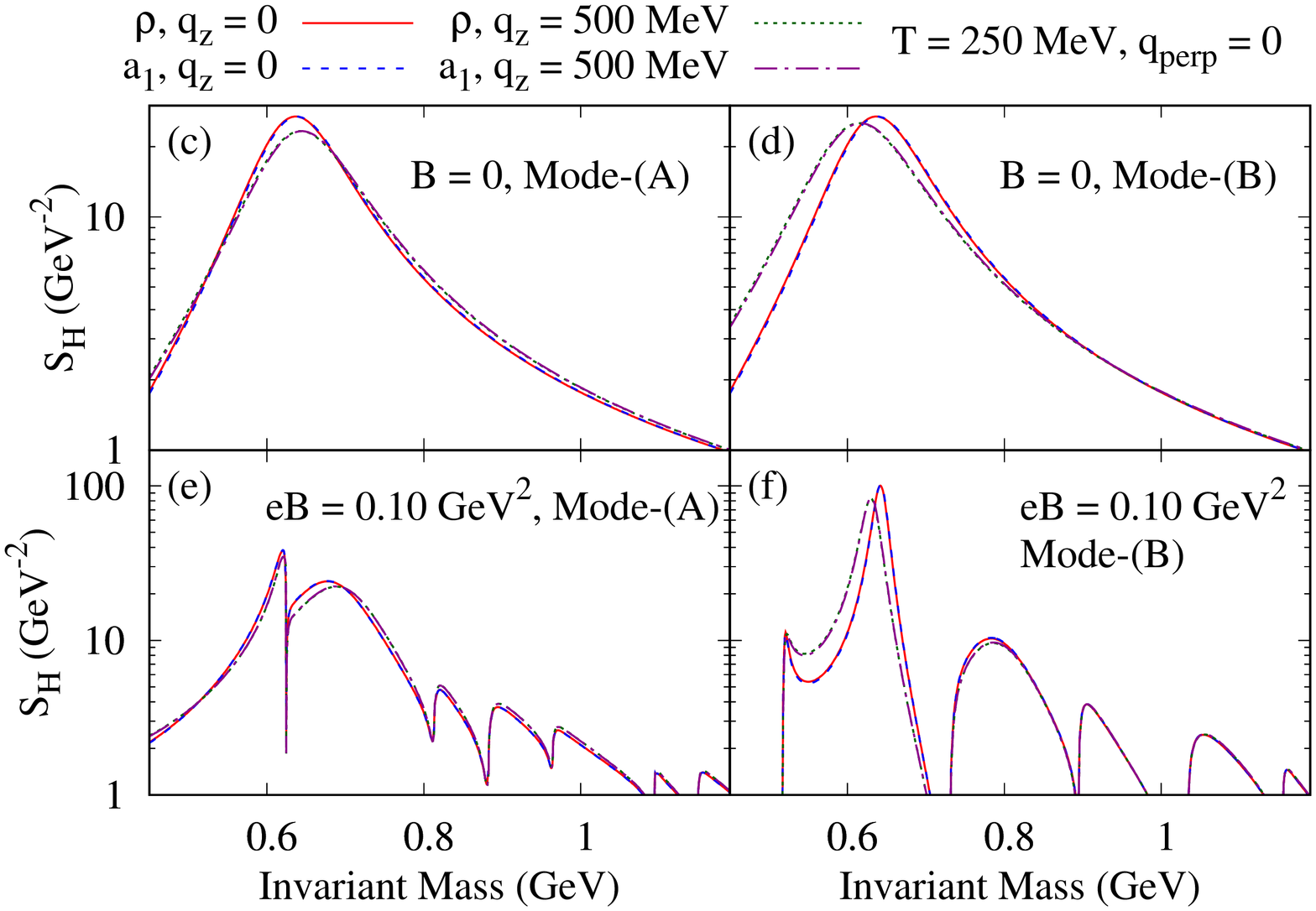}
	\end{center}
	\caption{Comparison of the spectral functions of $\pi^0$ with $\sigma$ and $\rho^0$ with $a_1^0$ at $T=250$ MeV, $q_\perp=0$ 
		for different values of their longitudinal momentum ($q_z=0$ and 500 MeV). }
	\label{fig.scompare}
\end{figure}
Till now, we have taken $\vec{q}=\vec{0}$. To see the effect of longitudinal momentum on the spectral function, we have plotted 
the spectral functions of the mesons as a function of invariant mass for $T=250$ MeV and $\qper=0$  with different values of $q_z$ and external magnetic field in Figs.~\ref{fig.scompare}(a)-(f). First of all, it can be observed that, the spectral functions of $\sigma$ and 
$a_1$  become identical to that of $\pi$ and $\rho$ respectively in all the cases as a consequence of the chiral symmetry restoration 
. In all the cases, the effect of increase in the $q_z$ decreases the height of spectral 
functions  with a marginal change of their peak positions. 
Moreover, comparing the green-dot and violet-dash-dot curves in Figs.~\ref{fig.scompare}(c) and (d), it can be noticed 
that, a non-zero value of $q_z$ lifts the degeneracy of the two modes of $\rho$ and $a_1$ at $B=0$.

\begin{figure}[h]
	\begin{center}
		\includegraphics[angle=-90,scale=0.35]{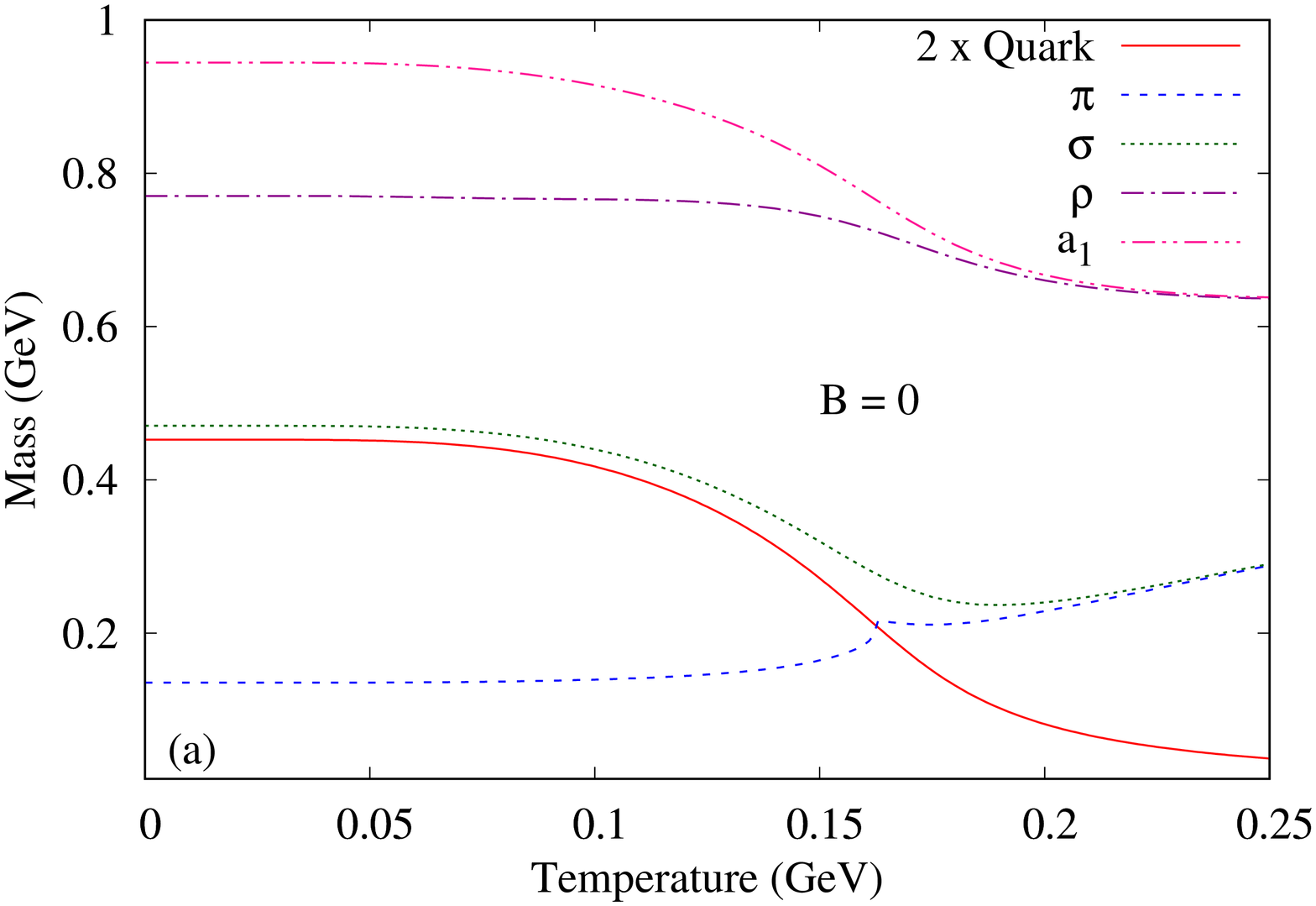}  \includegraphics[angle=-90,scale=0.35]{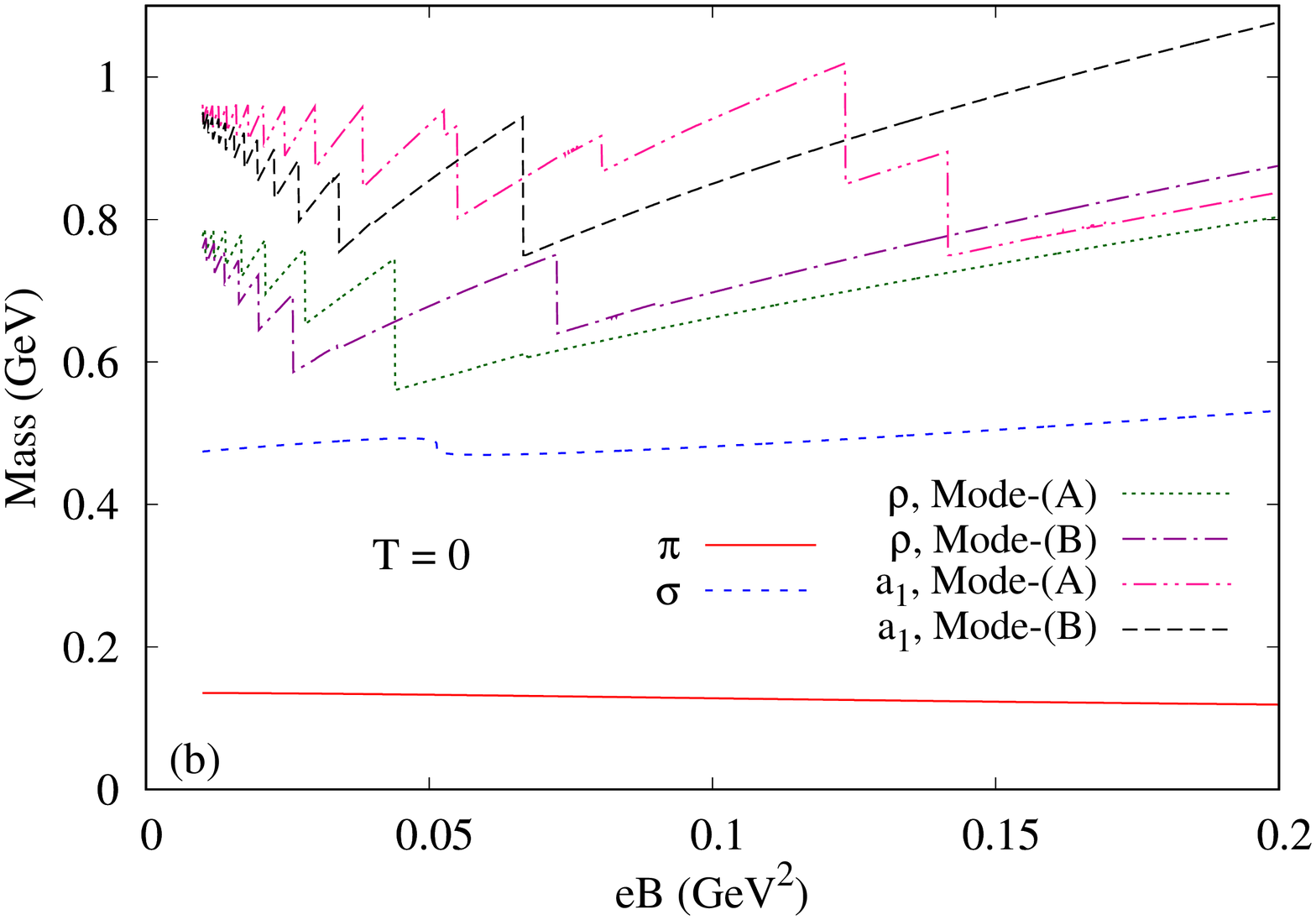}
	\end{center}
	\caption{Variation of masses of $\pi^0$, $\sigma$, $\rho^0$ and $a_1^0$ as a function of (a) temperature at $B=0$ and 
		(b) external magnetic field at $T=0$.  Two times the value of the constituent quark mass is also shown in (a).}
	\label{fig.mass.1}
\end{figure}
We now turn our attention to the study of the effect of temperature and external magnetic field on the meson masses and dispersion relations. 
We define the dispersion relations of the mesons as the value of $\omega(\vec{q})$ at which the spectral function $S(q^0=\omega,\vec{q})$ 
has a peak (global maxima) or in other words the locus ($q^0=\omega,\vec{q}$) of the peak of the spectral function gives the dispersion relations. Thus, the (effective) masses of the mesons are obtained by putting $\vec{q}=\vec{0}$ in the dispersion relation i.e. $m_{h,H} = \omega_{h,H}(\vec{q}=\vec{0}).$ 
\begin{figure}[h]
	\begin{center}
		\includegraphics[angle=-90,scale=0.35]{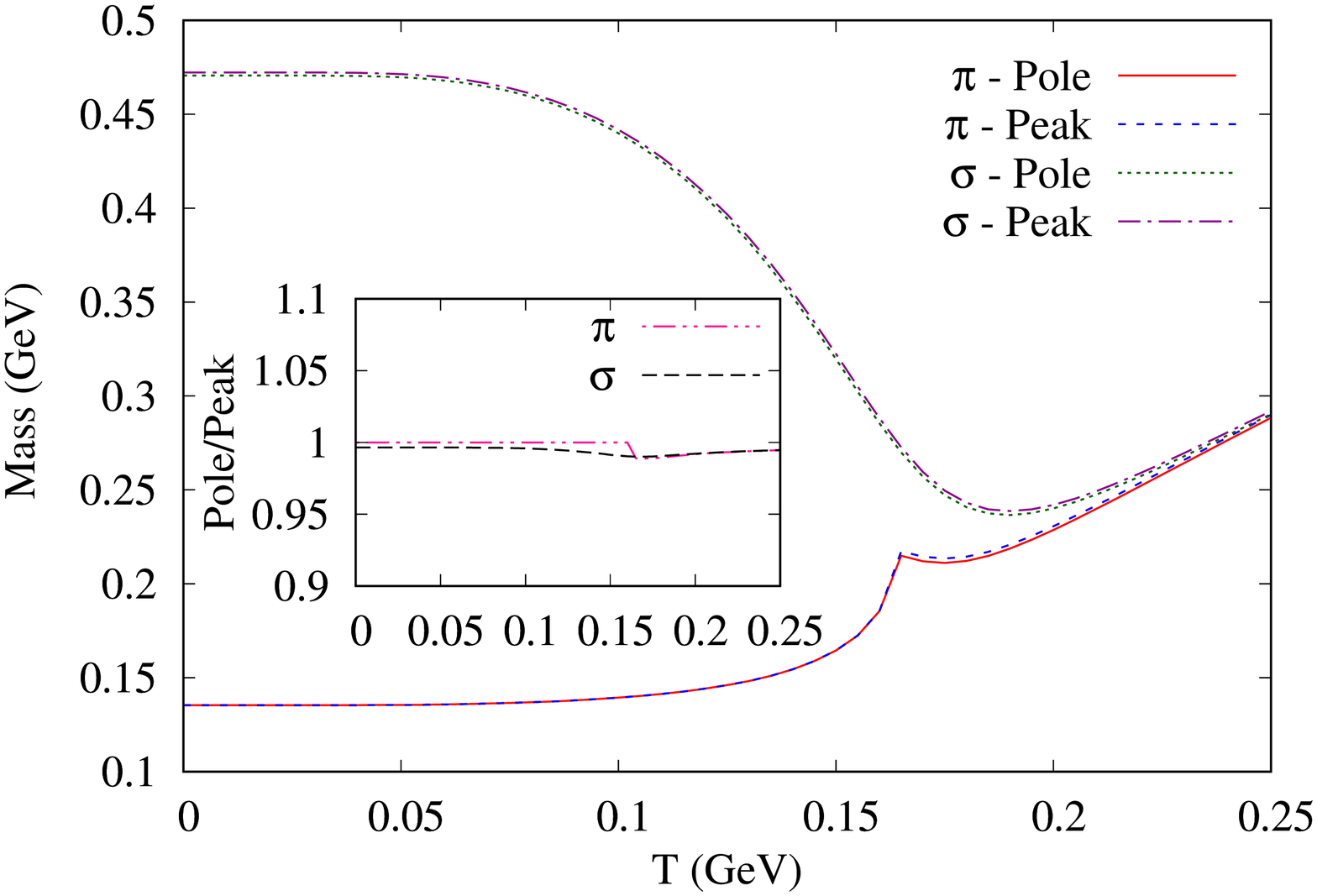} 
	\end{center}
	\caption{The masses of $\sigma$ and $\pi^0$ calculated from the pole of the propagator and peak of the spectral function as a function of temperature at zero magnetic field. The inset plot shows the ratio of masses obtained from pole and peak.}
	\label{fig.polepeak}
\end{figure} 
\begin{figure}[h]
	\begin{center}
		\includegraphics[angle=-90,scale=0.35]{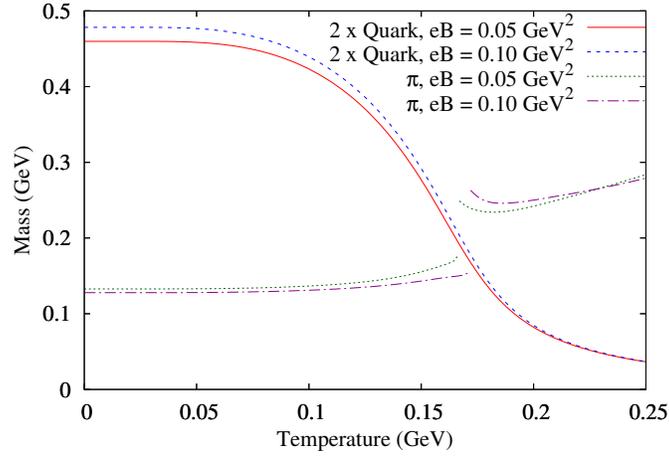}		
	\end{center}
	\caption{ Neutral pion mass is plotted   as a function of temperature for 
		different values of external magnetic field. Twice of the constituent quark mass is also shown for comparison. }
	\label{fig.mass.2_new}
\end{figure}

In Fig.~\ref{fig.mass.1}(a), the masses of the mesons are plotted as a function of temperature at vanishing external magnetic field. 
Twice of  the constituent quark mass  is also shown for comparison. 
In the lower temperature region, the meson masses remains almost constant. However,  $m_\pi$ starts increasing monotonically with temperature beyond $T\simeq150$ MeV and eventually it becomes larger than $2M$. On the other hand, $m_\sigma$ first decreases 
to attain a minimum  after which it increases. In the whole temperature range, $m_\sigma$ remains always greater than $2M$ 
maintaining its resonant signature. At high temperature, the mass of $\pi$ and $\sigma$ merge with each other as a consequence 
of the chiral symmetry restoration. Similar behaviour can also be noticed for $m_\rho$ and $m_{a_1}$ where both decrease 
with  temperature followed by a merging of their masses in the chiral symmetry restored phase.

 It is to be noted that, the mass/dispersion relation of the meson (or of any unstable resonance particle) can  have  different definition. The mass/dispersion relation can either be obtained from the locus ($q^0,\vec{q}$) of the pole of the propagator or of the peak of the spectral function. In the current work, we have used the peak of the spectral function for the definition of mass/dispersion relation. However, to check how these two  differ from each other, we have plotted the masses of $\sigma$ and $\pi^0$ as a function of temperature at $B=0$ in Fig.~\ref{fig.polepeak}. As can be seen from the Fig.~\ref{fig.polepeak}, the two different 
definitions of mass lead to no noticeable difference. Moreover, the ratio of the masses calculated from the pole to that from the peak is exactly unity when the particle  has zero decay width (for example the $\pi^0$ mass at the low temperature).

Now, keeping the temperature fixed at $T=0$,  the variation of meson masses as a function of external magnetic field are plotted in   Fig.~\ref{fig.mass.1}(b). 
Frequent mass jumps are observed for the  distinct modes of  $\rho$ and $a_1$. In between the two successive  discontinuities, the effective mass increases with $eB$. 
 It can be noticed that the frequency of oscillation decreases with the external field. In other words, separation between the two successive  discontinuities increases with $eB$. Also in case of $\sigma$ mesons, the effective mass shows increasing trend between the successive discontinuities.  However, only one  mass jump can be seen within the plotted range of the magnetic field.  Pion mass on the other hand remain continuous and is observed to decrease slowly with the external field which is consistent with  
 Refs.~\cite{PhysRevD.99.056005,PhysRevD.96.034004}.   
\begin{figure}[h]
	\begin{center}
		\includegraphics[angle=-90,scale=0.23]{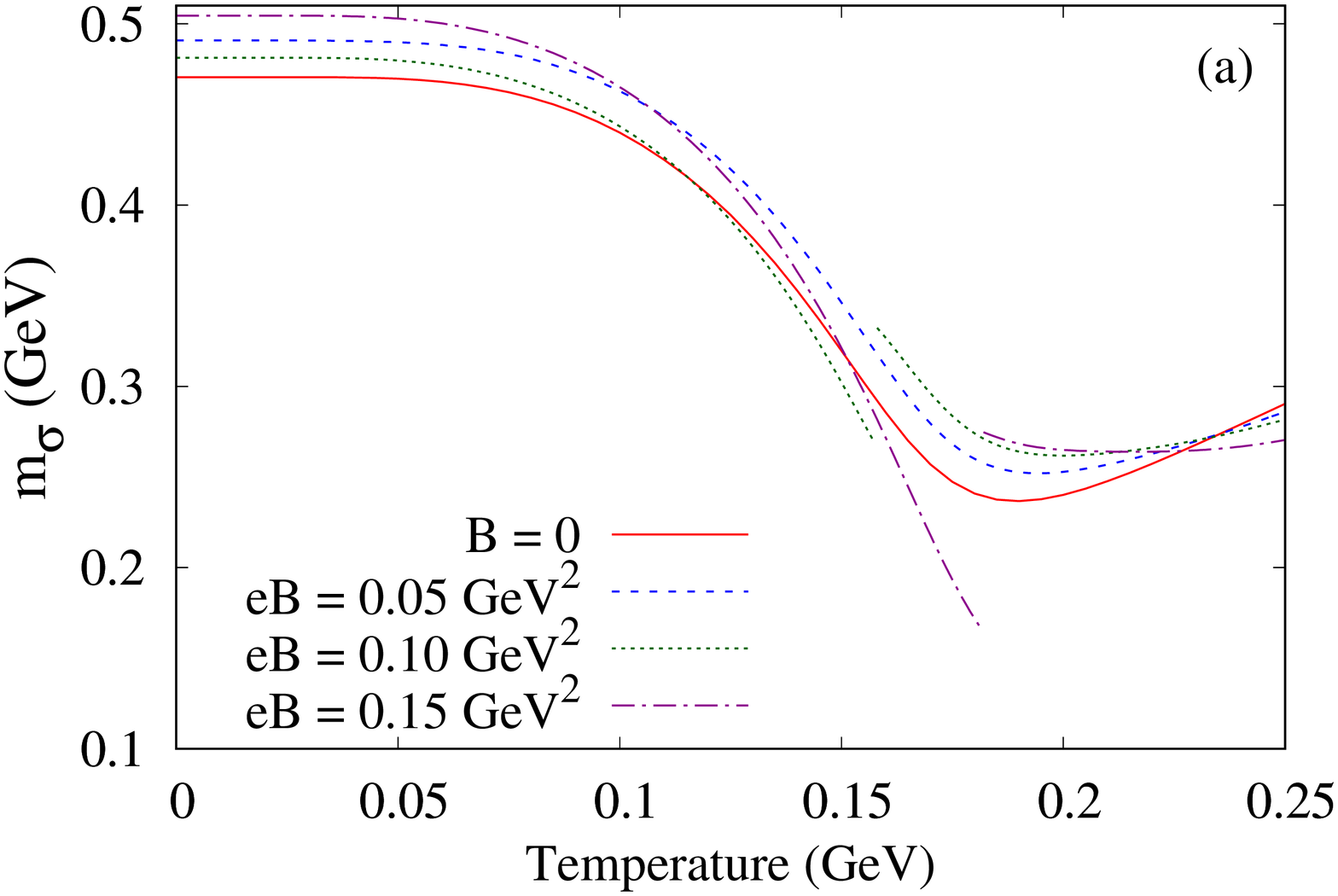} 
		\includegraphics[angle=-90,scale=0.23]{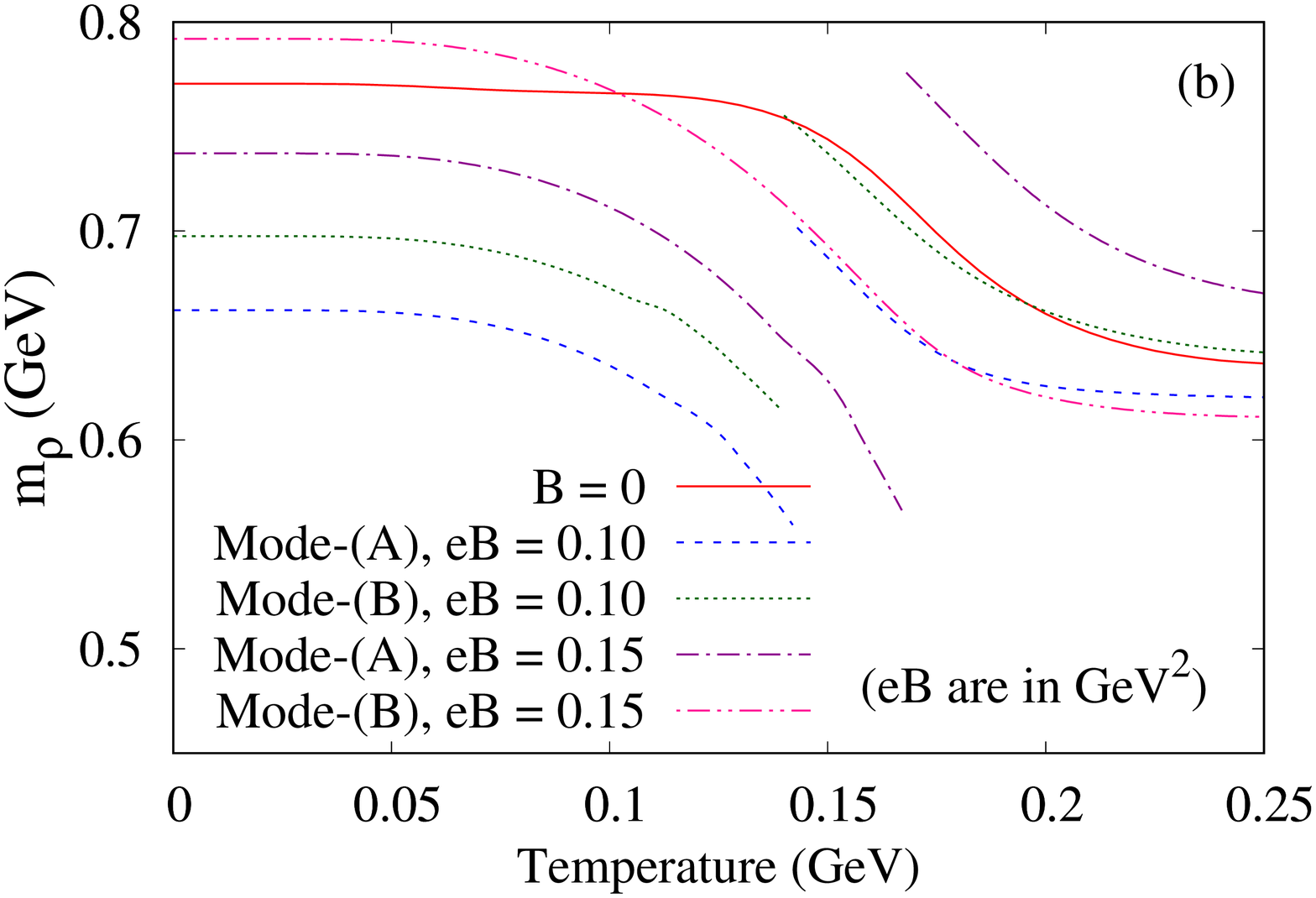} 
		\includegraphics[angle=-90,scale=0.23]{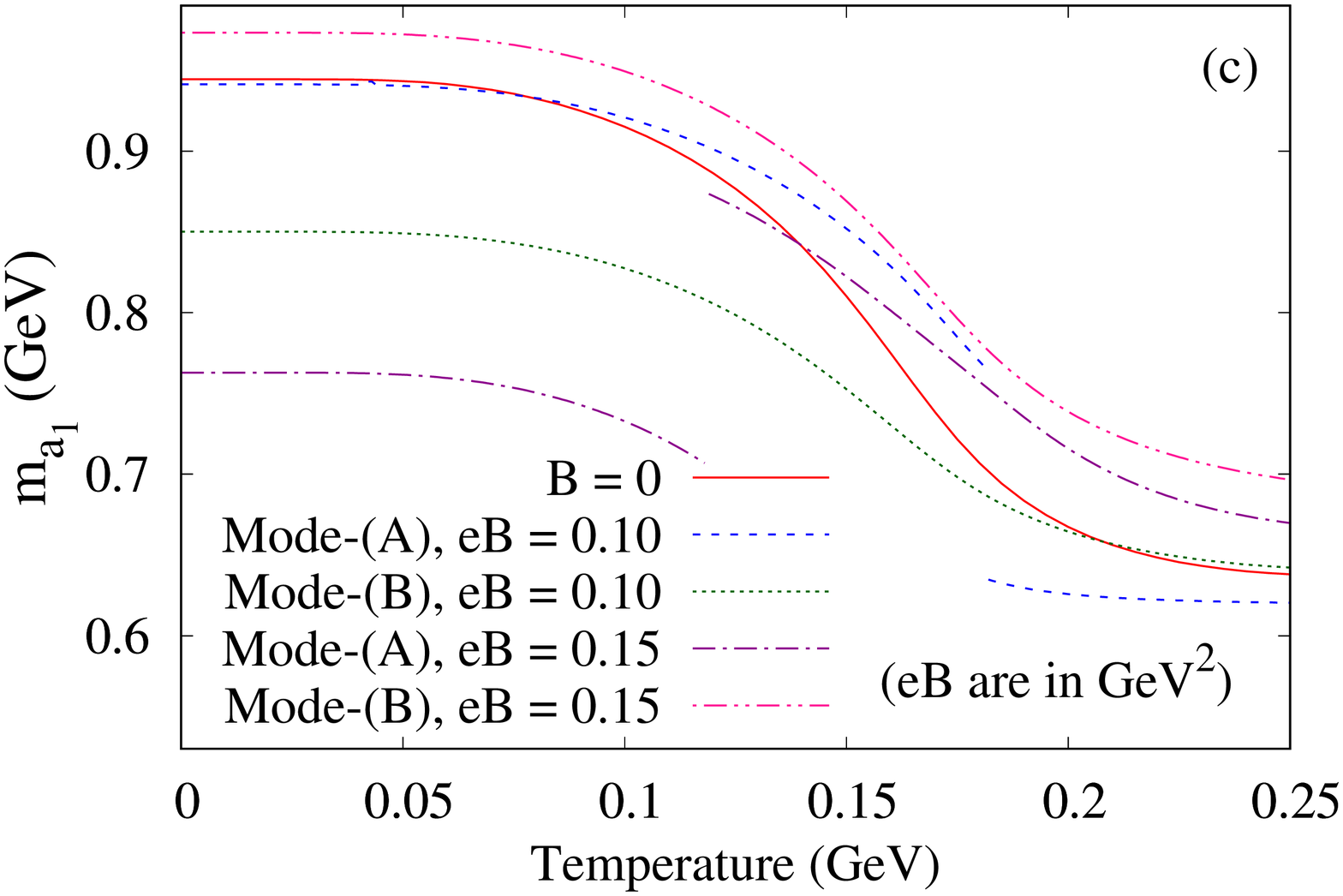}
	\end{center}
	\caption{Variation of masses of (a) $\sigma$, (b) $\rho^0$ and (c) $a_1^0$ as a function of temperature for 
		different values of external magnetic field.}
	\label{fig.mass.2}
\end{figure} 

If Fig.~\ref{fig.mass.2_new}, we have shown the variation of $m_\pi$ as a function of temperature at two different values of external 
magnetic field. At lower values of temperature, mass of pions are almost independent of $T$. 
At some particular temperature,  $m_\pi$ suffers a sudden jump (discontinuity) corresponding to Mott transition~\cite{PhysRevD.99.056005,PhysRevD.99.056009,PhysRevD.99.116025,Dubinin:2016wvt,MOTT:1968zz}. The jump structure is in qualitative 
agreement with most of the studies. However, there exists differences in  quantitative nature of the jump structure. For  example, the amount of discontinuity obtained here is  smaller in comparison to~\cite{PhysRevD.99.056009} which itself is different from~\cite{PhysRevD.99.116025}
as well as~\cite{PhysRevD.99.056005}. One should observe that  different parameter sets have been chosen in all these cases along with different regularization procedures. 
%

 Temperature dependence of  $m_\sigma$  is shown in Fig.~\ref{fig.mass.2}(a) at different values of the external magnetic field.  
 At lower  values of temperature, nature of $m_\sigma$  is dominated by its $eB$ dependence. 
 Because of the mass jump present at $T=0$,  $m_\sigma$ shows non-monotonic behaviour with respect to $eB$ variation. For example, effective mass at  $eB=0.10$GeV$^2$ is smaller than the effective mass  at $eB=0.05$ GeV$^2$ whereas the corresponding value of $m_\sigma$ at  $eB=0.15$GeV$^2$ remains well above the former two cases.  As a result, with the increase of  temperatures, when $m_\sigma$ decreases,
 crossing between fixed  $eB$ curves develops. With
 further increase  of temperature,  effective mass shows discontinuous jump structure for $eB=0.10$ and $0.15$ GeV$^2$. This 
 mass jump signifies the fact that even in case of sigma meson, there exist certain set of $T$ and $eB$ values for which no  solution exists for the pole of $\sigma$  propagator.  The pole reappears  at a higher value giving rise to a discontinuous jump. In general, this behaviour can be attributed to the oscillatory nature of the polarization function. One important feature to be noted is that at $eB=0.05$ GeV$^2$, the effective mass of $\sigma$ does not possess any discontinuous jump within the plotted temperature range.    
  We have also checked in our numerical calculations that, at finite temperature as well as at non-zero magnetic field, the relation $m_\sigma^2(T,B)=m_\pi^2(T,B)+4M^2(T,B)$ is in agreement with Refs.~\cite{Klevansky,zhang,PhysRevD.93.014010}.

In Fig.~\ref{fig.mass.2}(b), $m_\rho$ is plotted as a function of temperature for different values of external magnetic field. 
The $B=0$ curve is degenerate for the two modes. The degeneracy is lifted once the external magnetic field is turned on. 
For a given  value of $eB$, $m_\rho$ shows a decreasing trend with  temperature except at  particular 
values where discontinuous  jump occurs. The nature of the discontinuities is similar to that of $m_\pi$ and $m_\sigma$ i.e at the point of discontinuity, the solution for the pole position  always jumps to higher  values. Also in this case, one can observe that there exists  certain magnetic fields for which  no discontinuity appears within the plotted temperature range (see for example,  Mode-(B) at $eB=0.15$ GeV$^2$). On the other hand,   
for a particular temperature,  $m_\rho$ is found to be oscillatory with the change in $eB$. In other words, the effective mass can go to higher as well lower values depending upon the external magnetic field. 
This is again expected from the  highly oscillatory nature of the effective mass at $T=0$ (shown in Fig.~\ref{fig.mass.1}(b) ).  
Analogous feature is observed for the case of $a_1$ meson as shown in Fig.~\ref{fig.mass.2}(d). However, in this case, 
the effective mass of $a^0_1$ can jump to lower values  as well (see for example, Mode-(A) at 0.10 GeV$^2$).
\begin{figure}[h]
	\begin{center}
		\includegraphics[angle=-90,scale=0.35]{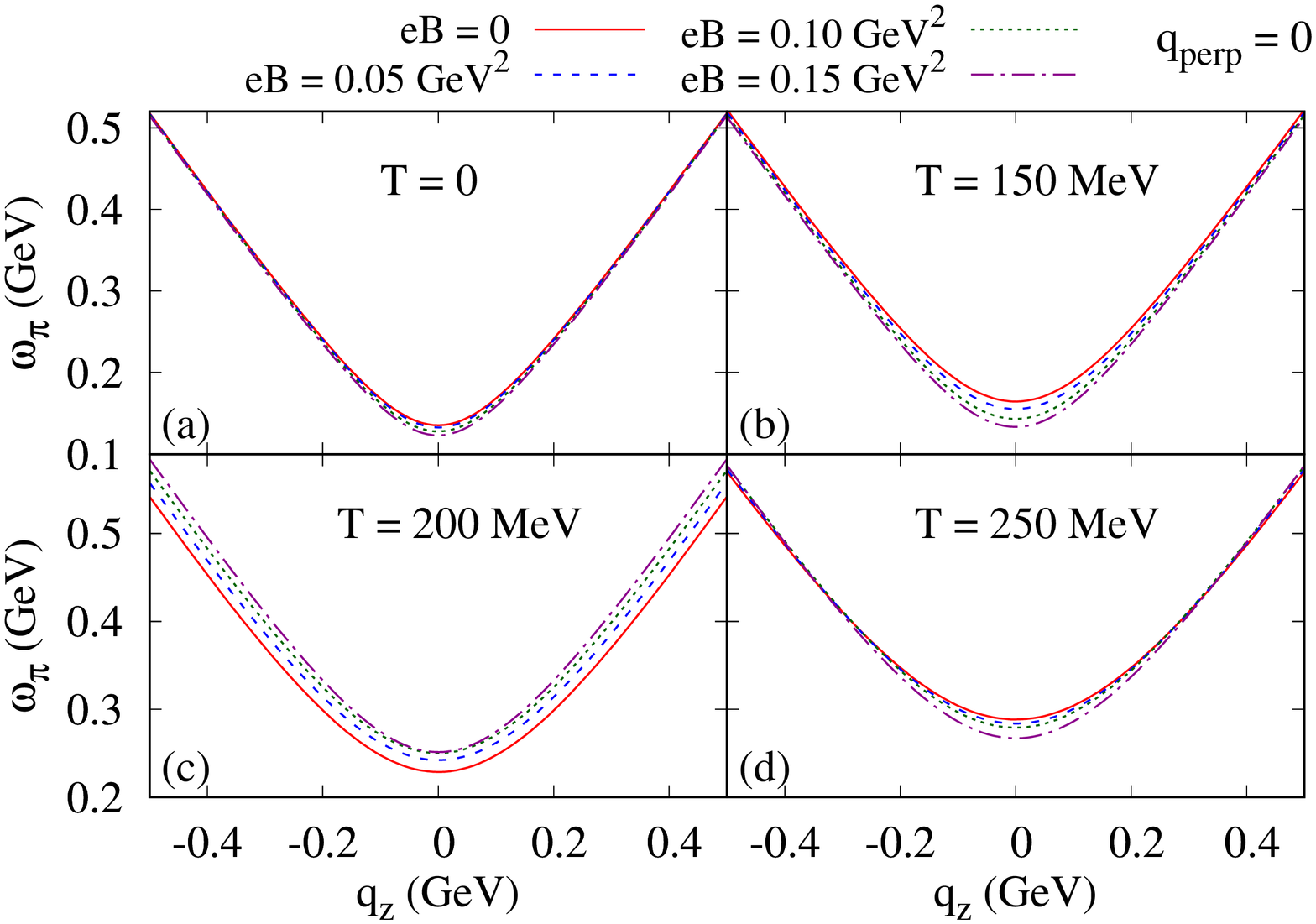} 
		\includegraphics[angle=-90,scale=0.35]{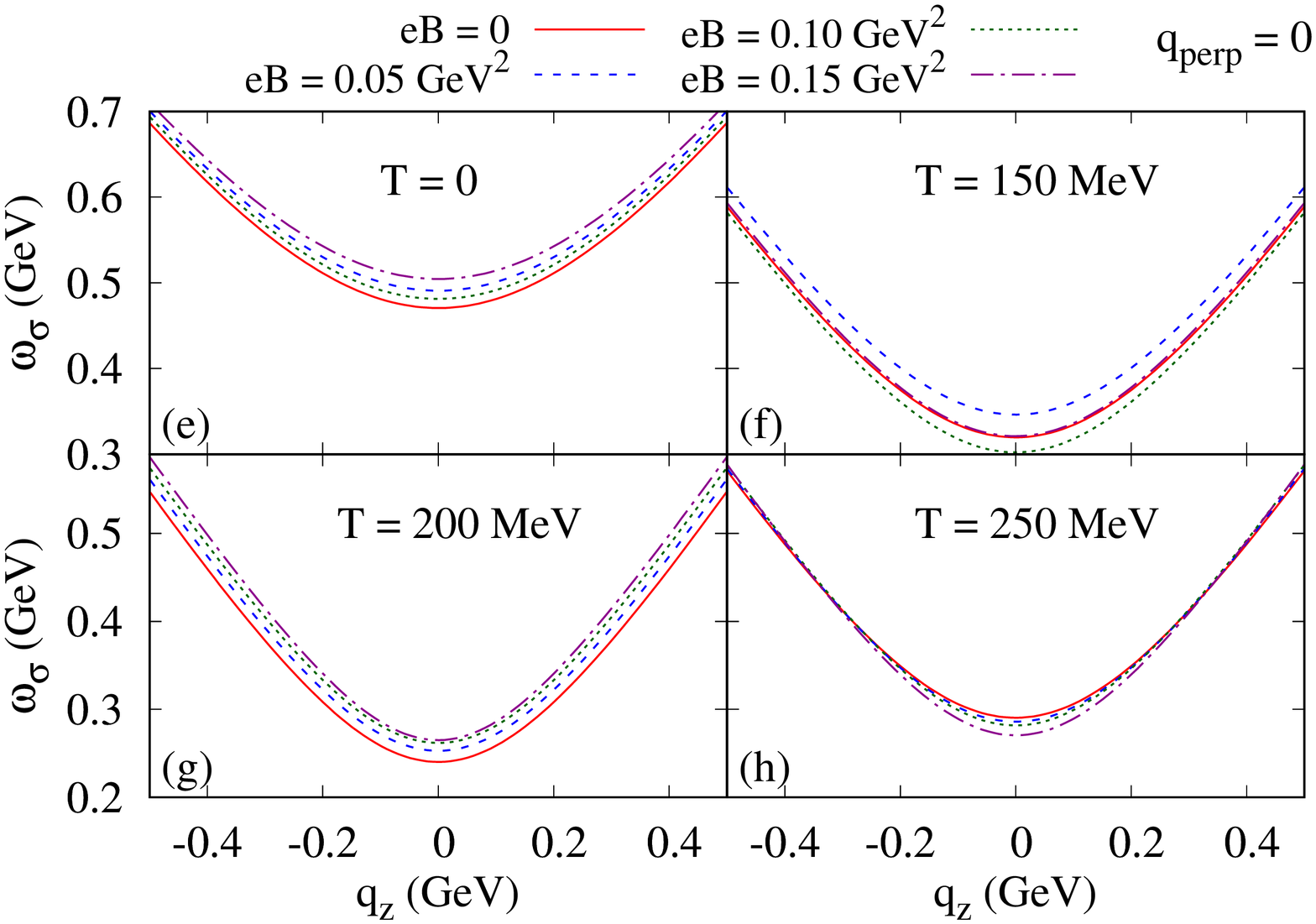}
	\end{center}
	\caption{The dispersion curves of $\pi^0$ and $\sigma$ mesons with vanishing transverse momentum ($q_\perp=0$) for different 
	values of temperature and external magnetic field.}
	\label{fig.disp.1}
\end{figure}
Finally, we concentrate on the dispersion relations of the mesons in the thermo-magnetic medium. 
In Figs.~\ref{fig.disp.1}(a)-(d), we have plotted $\omega_\pi$ as a function of longitudinal momentum ($q_z$) at different values of 
temperature and external magnetic field. For a particular temperature, the dispersion curves are mostly separated around $q_z\simeq0$. 
With the increase in $q_z$, the quantum corrections become sub-leading as compared to the kinetic energy which in turn leads to 
a light like dispersion and the dispersion curves of different $eB$ tend to merge with each other at high values of $q_z$. 
Moreover, the separation among the curves at different values of $eB$ is highest at the lower temperature as compared to higher temperature. 
An asymmetry of the dispersion curves for non-zero $eB$ about $q_z=0$ can be noticed as a consequence of breaking of rotational 
symmetry by the external magnetic field. 
The corresponding dispersion curves for the $\sigma$ meson is depicted in Figs.~\ref{fig.disp.1}(e)-(h). The nature of $\omega_\sigma$ 
is similar to that of $\omega_\pi$. 
\begin{figure}[h]
	\begin{center}
		\includegraphics[angle=-90,scale=0.35]{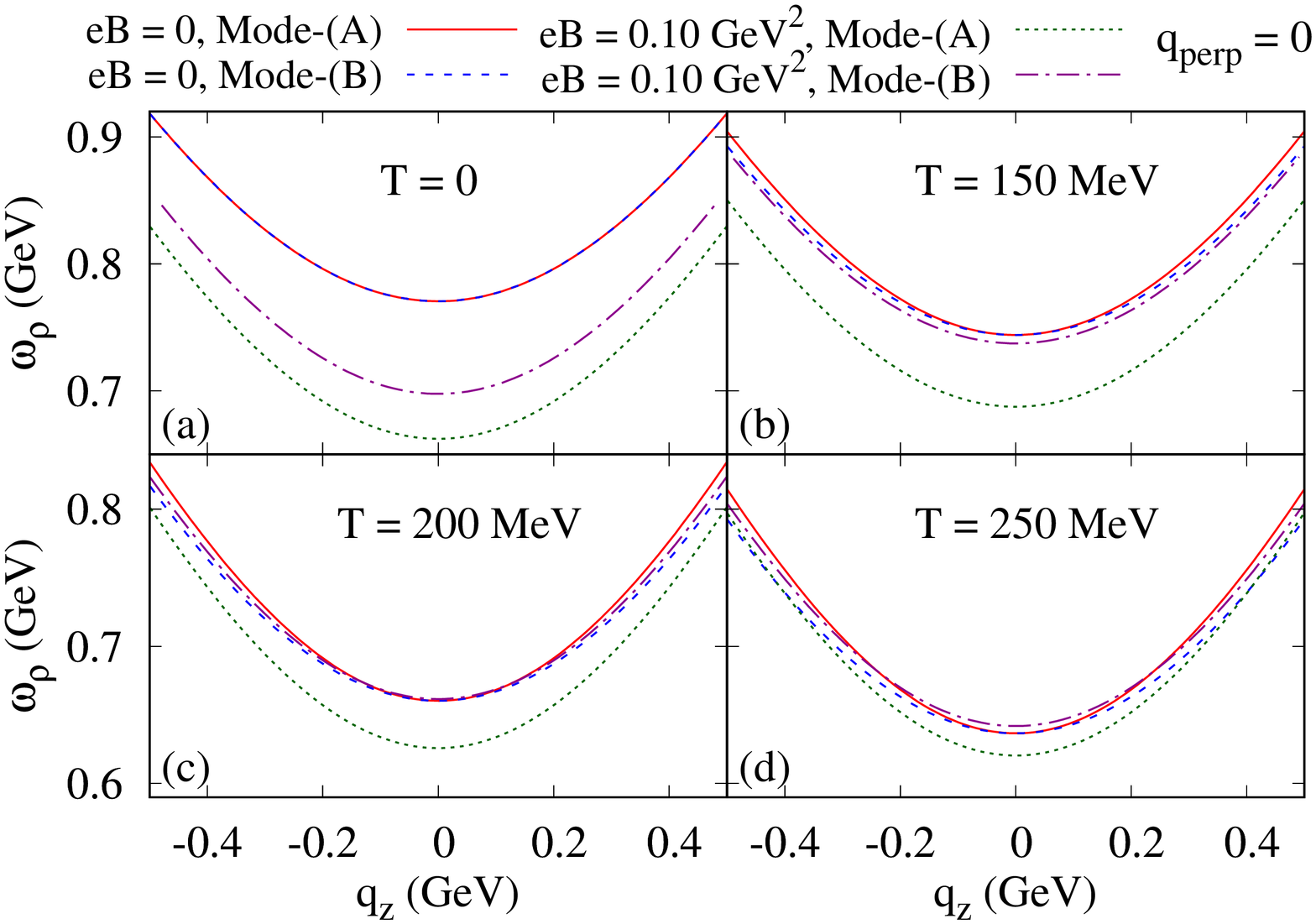} \includegraphics[angle=-90,scale=0.35]{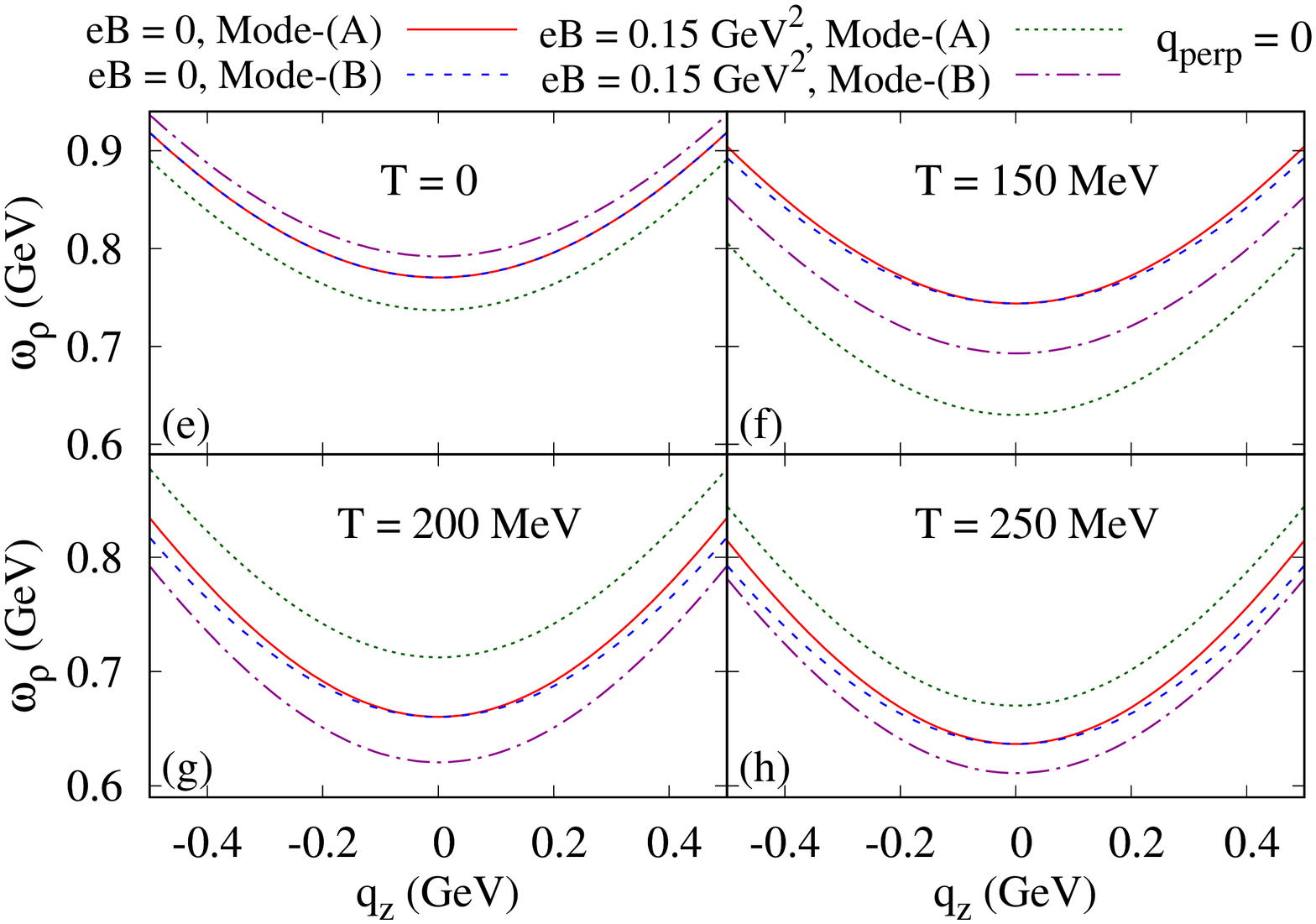}
		\includegraphics[angle=-90,scale=0.35]{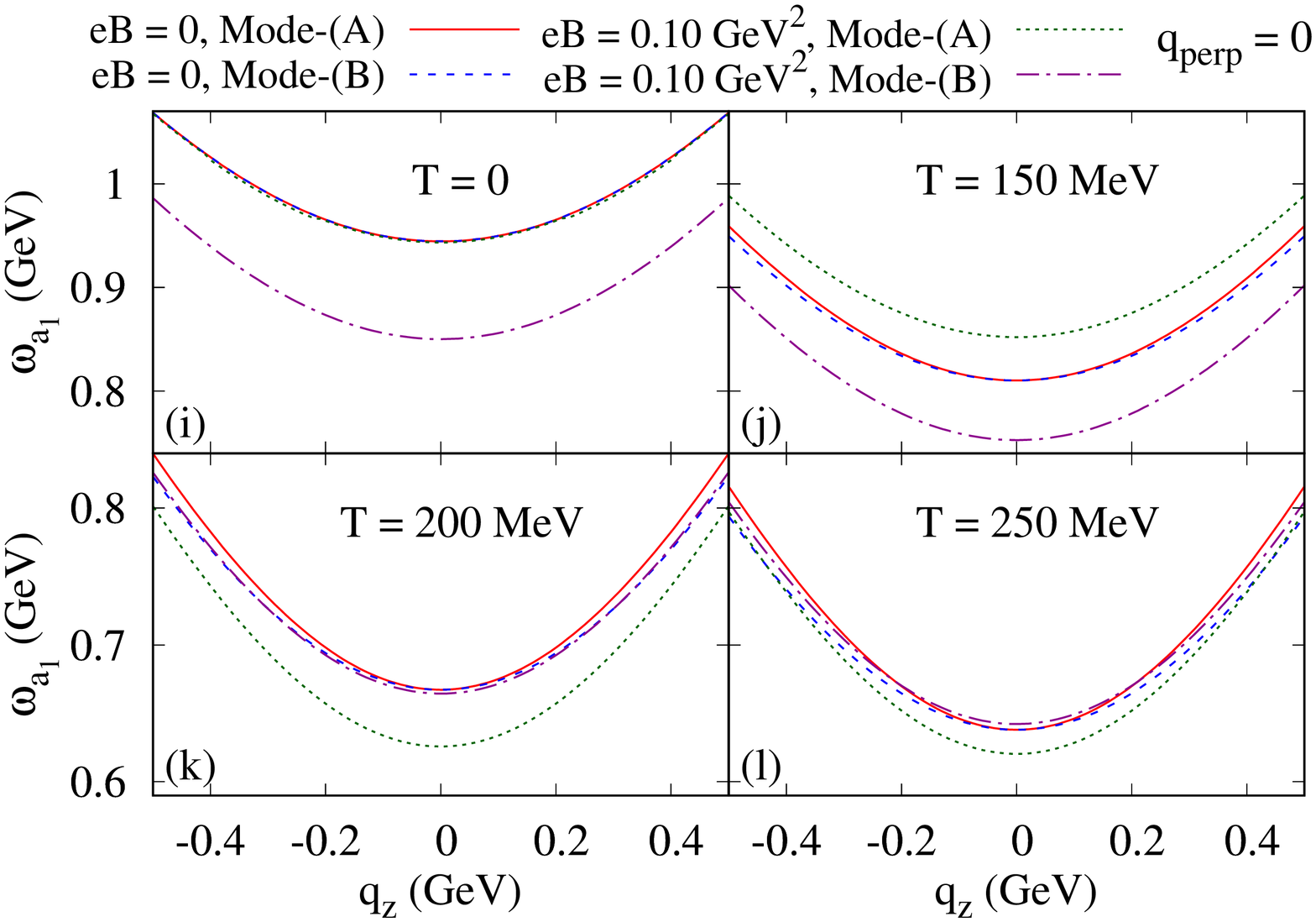} \includegraphics[angle=-90,scale=0.35]{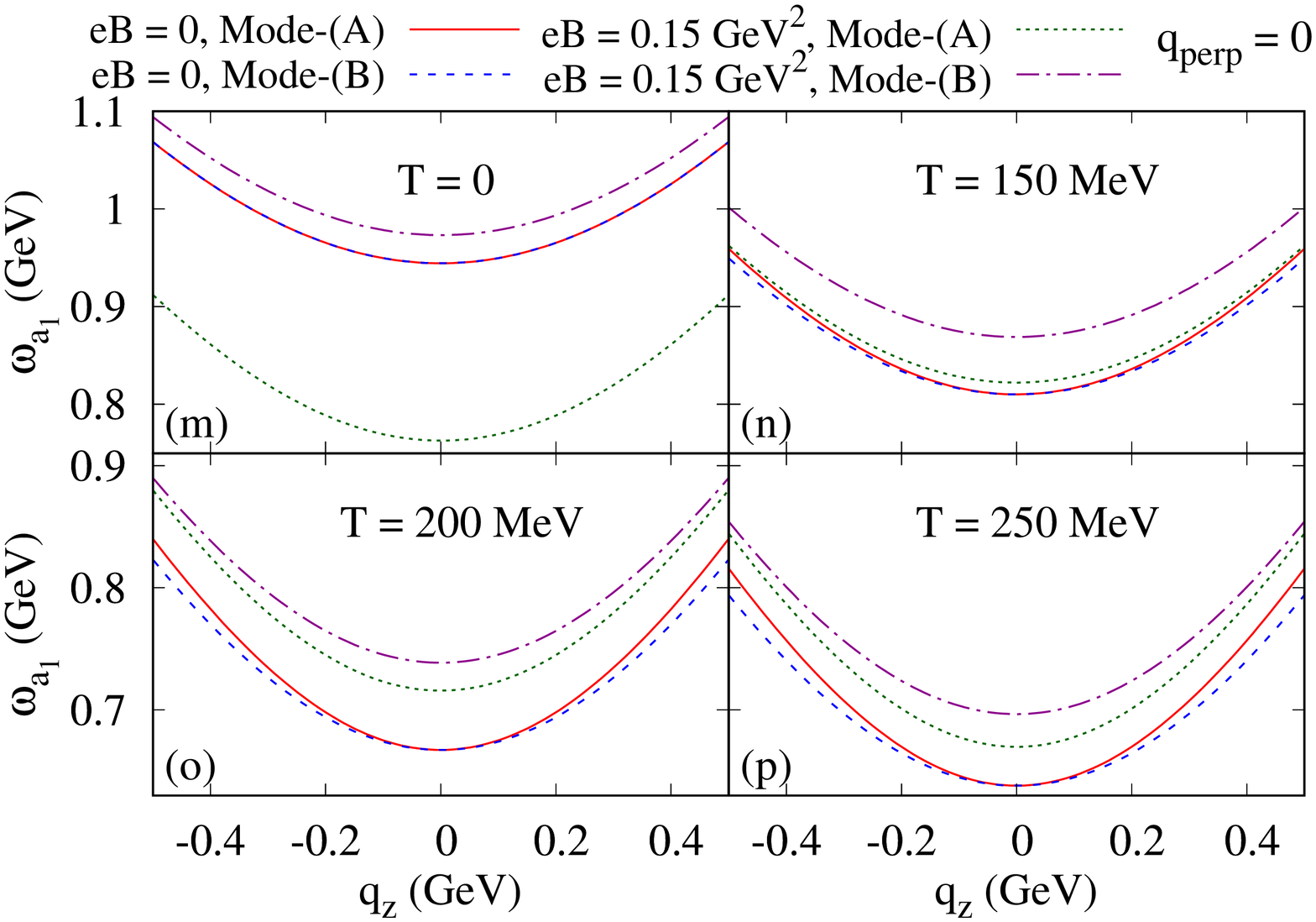}
	\end{center}
	\caption{The dispersion curves of $\rho^0$ and $a_1^0$ mesons with vanishing transverse momentum ($q_\perp=0$) for different 
		values of temperature and external magnetic field.}
	\label{fig.disp.2}
\end{figure}

Next in Figs.~\ref{fig.disp.2}(a)-(h), we have plotted the dispersion curves for the $\rho$ meson as a function of $q_z$ for different 
values of temperature and external magnetic field. The dispersion curves for Mode-(A) and (B) are degenerate at $T=0$ and $B=0$ and 
lie on top of each other. This degeneracy is lifted when we take either $T$ or $B\ne 0$. Moreover, for $B=0$ and $T\ne0$, the 
dispersion curves are identical around $q_z \simeq 0$. The nature of the dispersion curves at different values of $eB$ 
are similar to that of $\pi$ and $\sigma$ as they are mostly separated near $q_z\simeq0$ and tend to merge at high $q_z$. 
The corresponding plots for the $a_1$ meson is shown in Figs.~\ref{fig.disp.2}(i)-(p) and the nature of the curves are 
similar to that of $\rho$ meson.


\section{SUMMARY AND CONCLUSION}\label{sec.summary}
In this work,   the neutral meson properties such as mass, spectral function and dispersion relations have been studied  in the presence of a constant  background magnetic field  using two-flavor Nambu--Jona-Lasinio model. The novelty of the study lies in the detailed consideration of the general Lorentz structure for the vector and axial-vector meson polarization functions, which, to the best of our knowledge, has been ignored in  similar studies of vector mesons. Apart from the consideration of the modified Lorentz structure in presence of magnetic field,  the  Schwinger propagator expressed as sum over Landau levels has been used in the calculation of  the quark self energy and meson polarization functions. For simplicity in  the analytic calculation, only longitudinal mesons ($q_\perp=0$ ) are considered. To obtain the Lorentz structure of the vector and axial-vector meson systematically, we have  adopted a hybrid regularization scheme where as a  first step, the dimensional regularization is used to isolate  the ultra-violet divergences as the poles of   gamma functions. Subsequently, those  gamma functions    are replaced by incomplete gamma functions as usually done in  the proper time regularization scheme. We call this hybrid regularization  procedure as the incomplete gamma regularization (IGR). As a reward, the number of parameters   remain identical to that of usual cut-off regularization procedures. We have obtained two distinct modes for $\rho^0$ and $a_1$ meson. At $eB=0$ the effective mass of the modes remain degenerate, however, external  magnetic field  lifts the degeneracy. At temperatures  above the critical temperature for chiral symmetry restoration,  the   spectral functions   for each of the modes of  $\rho^0$ are observed to  overlap  with the corresponding modes of its chiral partner  $a_1^0$ meson for both zero and   non-zero values of external magnetic field.

The discontinuity in the pion mass near the Mott transition temperature is observed which is consistent with recent works~\cite{PhysRevD.99.056005,PhysRevD.99.056009}.  However, in our case,  the discontinuous mass jump is also observed in the effective mass of sigma meson which seems to be absent in Ref.~\cite{PhysRevD.96.034004}(see Fig.1). Also in~\cite{PhysRevD.99.056005}, it is mentioned that  no mass jump for $\sigma$ can exist in NJL model  as $m_\sigma$ always lies above  $2M$. In our work too,  we observe that  the condition  $m_\sigma > 2M$  is  always satisfied. Thus, we conclude that this condition may not be the correct explanation of the absence of mass jump in case of $m_\sigma$ in~\cite{PhysRevD.96.034004}. In our work, discontinuous mass jumps have also been observed in different modes of the $\rho$ and $a_1$ mesons. 
The presence of the mass jump in fact depends non-trivially  on the oscillation of the meson polarization function. This implies that the existence of a real solution for the pole  of the propagator will depend on the external parameters. For example, there can be certain  values of the magnetic fields  for which no mass jump may will occur (see for example Fig.~\ref{fig.mass.2}(a) for $eB=0.05$ GeV$^2$) within a certain range of temperature.  Moreover, one should keep in mind that the polarization function also requires a regularization prescription. In our two step regularization scheme, the dimensional regularization  is the essential first step to obtain the Lorentz structure for the vector and axial-vector  mesons. As mentioned earlier, the Lorentz structure can not be achieved systematically in thermo-magnetic case with the cut-off procedure commonly used. Thus, it is very interesting to study the similar analysis in other covariant regularization prescription such as Pauli-Villars method to conclude about the regularization scheme independent qualitative properties of the mesons .


\appendix

\section{CALCULATION OF $\RE\Sigmabarbar_\text{MFA}(M,B,T)$} \label{app.condensate}
In this appendix, we will briefly sketch the calculation of the quantity $\RE\Sigmabarbar_\text{MFA}(M,B,T)$. 
Substituting Eq.~\eqref{eq.s11.tb} into Eq.~\eqref{eq.quark.self.energy.TB} and performing the traces over colour and flavor 
spaces, we arrive at
\begin{eqnarray}
\RE\Sigmabarbar_\text{MFA}(M,B,T) = -2g_s Nc \sum_{f\in\{\text{u,d}\}}^{} \RE\TB{i\int\frac{d^4k}{(2\pi)^4}
\Tr_\text{d}\SB{S_f^{11}(k,M)}}.
\end{eqnarray}
Again substituting $S_f^{11}$ from Eq.~\eqref{eq.sf11} in the above equation and evaluating the trace over Dirac matrices, we get,
\begin{eqnarray}
\RE\Sigmabarbar_\text{MFA}(M,B,T) = -2 g_s N_c 4M \sum_{f\in\{\text{u,d}\}}^{} \RE \TB{i\int\frac{d^4k}{(2\pi)^4}
	\sum_{l=0}^{\infty}(-1)^le^{-\alpha_k^f}\SB{L_l(2\alpha_k^f)-L_{l-1}(2\alpha_k^f)} \right. \nn \\
	\left.\times \SB{\frac{-1}{\kpll^2-M_{lf}^2+i\epsilon}-2\pi i\eta(k\cdot u)\delta\FB{\kpll^2-M_{lf}^2}}}.
\end{eqnarray}
The $d^2\kper$ integral of the above equation is now performed using the orthogonality of the Laguerre polynomials and we are 
left with
\begin{eqnarray}
\RE\Sigmabarbar_\text{MFA}(M,B,T) &=& -2 g_s \frac{N_c M}{\pi} \sum_{f\in\{\text{u,d}\}}^{} |e_fB| \sum_{l=0}^{\infty} (2-\delta_l^0)
\RE \TB{i\int\frac{d^2\kpll}{(2\pi)^2}
	 \SB{\frac{-1}{\kpll^2-M_{lf}^2+i\epsilon}-2\pi i\eta(k\cdot u)\delta\FB{\kpll^2-M_{lf}^2}}} \\
 &=& \RE\Sigma^\text{Vac}_\text{MFA}(M,B) + \RE\Sigma^\text{B-Med}_\text{MFA}(M,B,T) 
\end{eqnarray}
where, 
\begin{eqnarray}
\RE\Sigma^\text{Vac}_\text{MFA}(M,B) &=& 2 g_s \frac{N_c M}{\pi} \sum_{f\in\{\text{u,d}\}}^{} |e_fB| \sum_{l=0}^{\infty} (2-\delta_l^0)
\RE \TB{i\int\frac{d^2\kpll}{(2\pi)^2}\frac{1}{\kpll^2-M_{lf}^2+i\epsilon}}, 
\label{eq.resigma.vac} \\
\RE\Sigma^\text{B-Med}_\text{MFA}(M,B,T) &=& -2 g_s \frac{N_c M}{\pi} \sum_{f\in\{\text{u,d}\}}^{} |e_fB| \sum_{l=0}^{\infty} (2-\delta_l^0)
\int\frac{d^2\kpll}{(2\pi)^2} 2\pi \eta(k\cdot u)\delta\FB{\kpll^2-M_{lf}^2}
\label{eq.resigma.bmed.1}
\end{eqnarray}
are respectively the magnetic field dependenet and both temperature as well as magnetic field dependent contributions to the self energy fucntion. Eq.~\eqref{eq.resigma.bmed.1} can be further simplified by performing the $dk^0$ integral 
using the Dirac delta function to obtain,
\begin{eqnarray}
\RE\Sigma^\text{B-Med}_\text{MFA}(M,B,T) = -2g_s \frac{N_cM}{\pi^2}\sum_{f\in\{\text{u,d}\}}^{}|e_fB|\sum_{l=0}^{\infty}(2-\delta_l^0)
\int_{0}^{\infty}dk_z \frac{1}{\omega_k^{lf}}f(\omega_k^{lf}).
\label{eq.resigma.bmed.2}
\end{eqnarray}

Note that, the quantity $\RE\Sigma^\text{Vac}_\text{MFA}(M,B)$ contains the divergent 
pure vacuum self energy $\RE\Sigma^\text{Pure-Vac}_\text{MFA}(M)$ which has to be separated out. To do this, we use 
the formalism developed in Ref.~\cite{Snigdha_2019} and simplify Eq.~\eqref{eq.resigma.vac} using the dimensional regularization. 
Going to $d$-dimension, we get
\begin{eqnarray}
\RE\Sigma^\text{Vac}_\text{MFA}(M,B) &=& 2 g_s \frac{N_c M}{\pi} \sum_{f\in\{\text{u,d}\}}^{} |e_fB| \sum_{l=0}^{\infty} (2-\delta_l^0)
\RE \TB{i\int\frac{d^d\kpll}{(2\pi)^d}\frac{\lambda^{1-d/2}}{\kpll^2-M_{lf}^2+i\epsilon}}\Bigg|_{d\rightarrow2}
\end{eqnarray}
where the scale $\lambda$ of dimension GeV$^2$ has been introduced to keep overall dimension of the equation consistent. It is now 
straightforward to perform the remaining momentum integral of the above equation to reach at
\begin{eqnarray}
\RE\Sigma^\text{Vac}_\text{MFA}(M,B) &=& 2 g_s \frac{N_c M}{4\pi^2}\Gamma(\varepsilon) 
\sum_{f\in\{\text{u,d}\}}^{} |e_fB| \sum_{l=0}^{\infty} (2-\delta_l^0)
\FB{\frac{4\pi\lambda}{M^2+2l|e_fB|}}^\varepsilon\Bigg|_{\varepsilon\rightarrow0}
\end{eqnarray}
where, $\varepsilon=(1-d/2)$ and we have used Eq.~\eqref{eq.Mlf}. The infinite sum over the index $l$ in the above equation can 
now be expressed in terms of Hurwitz-Riemann zeta function as
\begin{eqnarray}
\RE\Sigma^\text{Vac}_\text{MFA}(M,B) &=& 2 g_s \frac{N_c M}{4\pi^2}\Gamma(\varepsilon) 
\sum_{f\in\{\text{u,d}\}}^{} |e_fB| \TB{\FB{\frac{4\pi\lambda}{M^2}}^\varepsilon+2\FB{\frac{4\pi\lambda}{2|e_fB|}}^\varepsilon 
\zeta\FB{\varepsilon,1+\frac{M^2}{2|e_fB|}}}
\Bigg|_{\varepsilon\rightarrow0}.
\end{eqnarray}
An expansion of the RHS of the above equation about $\varepsilon=0$ yields,
\begin{eqnarray}
\RE\Sigma^\text{Vac}_\text{MFA}(M,B) &=& 2g_s \frac{N_cN_fM^3}{4\pi^2}\TB{-\frac{1}{\varepsilon}+\gamma_\text{E}-1-\ln\FB{\frac{4\pi\lambda}{M^2}}}
\nn \\
 && -2g_s\frac{MN_c}{4\pi^2}\sum_{f\in\{\text{u,d}\}}^{}\TB{-M^2 + \FB{M^2-|e_fB|}\ln\FB{\frac{M^2}{2|e_fB|}}
	-2|e_fB|\SB{\ln\Gamma\FB{\frac{M^2}{2|e_fB|}} - \ln\sqrt{2\pi}} }.
\end{eqnarray}
The first term on the RHS can now be identified (see Eq.~\eqref{eq.sigma.purevac.expansion}) as the magnetic field 
independent divergent pure vacuum contribution to the self energy $\RE\Sigma^\text{Pure-Vac}_\text{MFA}(M)$ which has 
been separated from the $\RE\Sigma^\text{Vac}_\text{MFA}(M,B)$ so that, we rewrite the above equation as,
\begin{eqnarray}
\RE\Sigma^\text{Vac}_\text{MFA}(M,B) &=& \RE\Sigma^\text{Pure-Vac}_\text{MFA}(M) + \RE\Sigma^\text{B-Vac}_\text{MFA}(M,B)
\end{eqnarray}
where, 
\begin{eqnarray}
\Sigma^\text{B-Vac}_\text{MFA}(M,B) = -2g_s\frac{MN_c}{4\pi^2}\sum_{f\in\{\text{u,d}\}}^{}\TB{-M^2 + \FB{M^2-|e_fB|}\ln\FB{\frac{M^2}{2|e_fB|}}
	-2|e_fB|\SB{\ln\Gamma\FB{\frac{M^2}{2|e_fB|}} - \ln\sqrt{2\pi}} }.
\label{eq.quark.self.energy.TB.final}
\end{eqnarray}


\section{EXPRESSIONS OF THE SUSCEPTIBILITIES}\label{app.sus}
In this appendix, we will specify the explicit expressions for the susceptibilities. 
We will do this for the two cases separately: (i) $B=0$ and (ii) $B\ne0$ in the following subsections.
\subsection{CASE-I: $B=0$}
A straight forward differentiation of the gap equation at $B=0$ with respect to $T$ and $m$ yields,
\begin{eqnarray}
-\frac{\del M}{\del T} &=&\frac{1}{ \FB{1+\mathcal{D_\text{Vac}}+\mathcal{D_\text{Med}}}} 2g_s \frac{2N_cN_fM}{\pi^2T^2}
\int_{0}^{\infty}d|\vec{k}|\vec{k}^2 f(\omega_k)\SB{1-f(\omega_k)}, \\
\chi &=& \frac{1}{2g_s}\FB{\frac{\del M}{\del m}-1} = -\frac{1}{2g_s}\FB{\frac{\mathcal{D_\text{Vac}}+\mathcal{D_\text{Med}} }
	{1+\mathcal{D_\text{Vac}}+\mathcal{D_\text{Med}}}}
\end{eqnarray}
where,
\begin{eqnarray}
\mathcal{D_\text{Vac}} &=& -2g_s \frac{N_cN_f}{4\pi^2}\TB{\Lambda^2 e^{-M^2/\Lambda^2}-3M^2\Gamma\FB{0,\frac{M^2}{\Lambda^2}}}, \\
\mathcal{D_\text{Med}} &=& 2g_s \frac{2N_cN_f}{\pi^2T}
\int_{0}^{\infty}d|\vec{k}|\vec{k}^2 \frac{1}{\omega_k^3} \TB{ T\vec{k}^2
 - M^2\omega_k \SB{1-f(\omega_k)} }f(\omega_k).
\end{eqnarray}

\subsection{CASE-II: $B\ne0$}
A straight forward differentiation of the gap equation at $B\ne0$ with respect to $T$ and $m$ yields,
\begin{eqnarray}
-\frac{\del M}{\del T} &=&\frac{1}{ \FB{1+\mathcal{D_\text{Vac}}+\mathcal{D_\text{BVac}}+\mathcal{D_\text{BMed}}}} 
2g_s \frac{N_cM}{\pi^2T^2} \sum_{f\in\{\text{u,d}\}} |e_fB| \sum_{l=0}^{\infty}(2-\delta_{l}^0)
\int_{0}^{\infty}dk_z f(\omega_k^{lf})\SB{1-f(\omega_k^{lf})}, \\
\chi &=& \frac{1}{2g_s}\FB{\frac{\del M}{\del m}-1} = -\frac{1}{2g_s}\FB{\frac{\mathcal{D_\text{Vac}}+\mathcal{D_\text{BVac}}+\mathcal{D_\text{BMed}} }
	{1+\mathcal{D_\text{Vac}}+\mathcal{D_\text{BVac}}+\mathcal{D_\text{BMed}}}}
\end{eqnarray}
where,
\begin{eqnarray}
\mathcal{D_\text{BVac}} &=& -2g_s \frac{N_c}{4\pi^2} \sum_{f\in\{\text{u,d}\}}\TB{
|e_fB|\SB{\ln y+2\ln\Gamma(y)-\ln(2\pi)+2\frac{}{}} + M^2\SB{1-3\ln y + 2\psi(y)\frac{}{}}
}, \\
\mathcal{D_\text{BMed}} &=& 
2g_s \frac{N_c}{\pi^2T} \sum_{f\in\{\text{u,d}\}} |e_fB| \sum_{l=0}^{\infty}(2-\delta_{l}^0)
\int_{0}^{\infty}dk_z \frac{1}{(\omega_k^{lf})^3} \TB{ T\FB{k_z^2+2l|e_fB|}
	- M^2\omega_k^{lf} \SB{1-f(\omega_k^{lf})} }f(\omega_k^{lf})
\end{eqnarray}
with $y=M^2/|2e_fB|$.


\section{FORM FACTORS OF THE POLARIZATION FUNCTION IN TERMS OF LOCAL INVARIANTS} \label{app.form.factors}
In this appendix we will enlist the different form factors in terms of the Lorentz invariant quantities. 
Let us start with the case $T=0$ and $B=0$. Substituting Eqs.~\eqref{eq.proj.vac.1} and \eqref{eq.proj.vac.2} 
into Eq.~\eqref{eq.ff.1}, we get, after some simplifications,
\begin{eqnarray}
\Pi_{H2} = \frac{1}{q^2}q_\mu q_\nu \Pi^\munu_{H} ~~~ \text{and} ~~~ 
\Pi_{H1} = \frac{1}{3}\FB{g_\munu \Pi^\munu_{H} - \Pi_{H2}}.
\end{eqnarray}

Now, at $T\ne0$ and $B=0$, we substitute Eqs.~\eqref{eq.proj.t.11} and \eqref{eq.proj.t.14} into Eq.~\eqref{eq.ff.t.1} to obtain
\begin{eqnarray}
\Pibar_{H2} &=& \frac{1}{q^2}q_\mu q_\nu \Pibar^\munu_{H}, \\
\Pibar_{H3} &=& \frac{1}{\utilde^2}\TB{ 
u_\mu u_\nu \Pibar^\munu_{H} + \frac{(q\cdot u)^2}{q^2}\Pibar_{H2} - 2\frac{(q\cdot u)}{q^2}q_\mu u_\nu \Pibar^\munu_{H}}, \\
\Pibar_{H1} &=& \frac{1}{2}\FB{g_\munu \Pibar^\munu_{H} - \Pibar_{H2} - \Pibar_{H3}}, \\
\Pibar_{H4} &=& \frac{1}{\sqrt{q^2\utilde^2}}\TB{ q_\mu u_\nu \Pibar^\munu_{H} - (q\cdot u)\Pibar_{H2} }.
\end{eqnarray}

Similarly for the case $T\ne0$ and $B\ne0$, substituting Eqs.~\eqref{eq.proj.tb.11} and \eqref{eq.proj.tb.17} 
into Eqs.~\eqref{eq.ff.tb.1} and \eqref{eq.ff.tb.2}, we get
\begin{eqnarray}
\Pibarbar_{H2} &=& \frac{1}{q^2}q_\mu q_\nu \Pibarbar^\munu_{H}, \\
\Pibarbar_{H3} &=& \frac{1}{\utilde^2}\TB{ 
	u_\mu u_\nu \Pibarbar^\munu_{H} + \frac{(q\cdot u)^2}{q^2}\Pibar_{H2} - 2\frac{(q\cdot u)}{q^2}q_\mu u_\nu \Pibarbar^\munu_{H}}, \\
\Pibarbar_{H5} &=& \frac{1}{\sqrt{q^2\utilde^2}}\TB{ q_\mu u_\nu \Pibarbar^\munu_{H} - (q\cdot u)\Pibarbar_{H2} }, \\
\Pibarbar_{H6} &=& \frac{1}{\sqrt{q^2\btilde^2}}\TB{ q_\mu b_\nu \Pibarbar^\munu_{H} - (q\cdot b)\Pibarbar_{H2} 
- \frac{(b\cdot\utilde)}{\utilde^2}\sqrt{q^2\utilde^2}\Pibarbar_{H5} }, \\
\Pibarbar_{H7} &=& \frac{1}{\sqrt{q^2\btilde^2}}\TB{ u_\mu b_\nu \Pibarbar^\munu_{H} 
-\frac{(q\cdot u)}{q^2}	q_\mu b_\nu \Pibarbar^\munu_{H} 
- \frac{(q\cdot b)}{q^2}\sqrt{q^2\utilde^2}\Pibarbar_{H5} - (b\cdot \utilde)\Pibarbar_{H3}  }, \\
\Pibarbar_{H4} &=& \frac{1}{\btilde^2}\TB{ b_\mu b_\nu \Pibarbar^\munu_{H} 
	+ \frac{1}{q2}\SB{\frac{(b\cdot\utilde)}{\utilde^2}(q\cdot u) -(\qdotb) }q_\mu b_\nu \Pibarbar^\munu_{H} 
	-\frac{(b\cdot\utilde)}{\utilde^2}	u_\mu b_\nu \Pibarbar^\munu_{H} \right. \nn \\ && \left. 
	- \frac{(q\cdot b)}{q^2}\sqrt{q^2\btilde^2}\Pibarbar_{H6} 
	- \frac{(b\cdot \utilde)}{\utilde^2}\sqrt{\utilde^2\btilde^2}\Pibarbar_{H7}  }, \\
\Pibarbar_{H1} &=& \FB{g_\munu \Pibarbar^\munu_{H} - \Pibarbar_{H2} - \Pibarbar_{H3} - \Pibarbar_{H4} }.
\end{eqnarray}

\section{CALCULATION OF THE PURE-VACUUM POLARIZATION FUNCTIONS USING DIMENSIONAL REGULARIZATION} \label{app.pola.vac}
In this appendix, we will simplify Eqs.~\eqref{eq.Pi.pi.1}-\eqref{eq.Pi.a1.1} by evaluating the momentum integral using 
dimensional regularization. Substituting $S'(q,m)$ from Eq.~\eqref{eq.dressed.propagator} and evaluating the traces 
over color, flavor and Dirac spaces we can express the polarization functions as
\begin{eqnarray}
\Pi_h(q) &=& i\int\frac{d^4k}{(2\pi)^4}\frac{N_h(q,k)}{(k^2-M^2+i\epsilon)\SB{(k+q)^2-M^2+i\epsilon}} ~~;~~ h\in\{\pi,\sigma\}, 
\label{eq.Pih.1}\\
\Pi_H^\munu(q) &=& i\int\frac{d^4k}{(2\pi)^4}\frac{N_H^\munu(q,k)}{(k^2-M^2+i\epsilon)\SB{(k+q)^2-M^2+i\epsilon}} ~~;~~ H\in\{\rho,a_1\}
\label{eq.PiH.1}
\end{eqnarray}
where 
\begin{eqnarray}
N_h &=& 4N_cN_f(k^2+k\cdot q - aM^2), \label{eq.Nh}\\
N^\munu_H &=& -4N_cN_f\TB{g^\munu(k^2 +k\cdot q - aM^2) - 2k^\mu k^\nu -(k^\mu q^\nu+k^\nu q^\mu)} \label{eq.NH}
\end{eqnarray}
with $a = \begin{cases}
1 ~~\text{for}~~ \pi,\rho \\
-1 ~~\text{for}~~ \sigma,a_1
\end{cases} 
$. 
Now using the standard Feynman parametrization, the denominators of Eqs.~\eqref{eq.Pih.1} and \eqref{eq.PiH.1} are combined to get
\begin{eqnarray}
\Pi_\pi(q) &=& i4N_cN_f \lambda^{2-d/2}\int\frac{d^dk}{(2\pi)^d}\TB{
\frac{1}{k^2-M^2+i\epsilon} -\frac{1}{2}q^2\int_{0}^{1}dx\frac{1}{\TB{(k+xq)^2-\Delta}^2}}\Bigg|_{d\rightarrow4}, \\
\Pi_\sigma(q) &=& i4N_cN_f \lambda^{2-d/2} \int\frac{d^dk}{(2\pi)^d}\TB{
	\frac{1}{k^2-M^2+i\epsilon} -\frac{1}{2}(q^2-4M^2)\int_{0}^{1}dx\frac{1}{\TB{(k+xq)^2-\Delta}^2}}\Bigg|_{d\rightarrow4}, \\
\Pi_H^\munu(q) &=& i \lambda^{2-d/2} \int_{0}^{1}dx\int\frac{d^dk}{(2\pi)^d}\frac{N_H^\munu(q,k)}{\TB{(k+xq)^2-\Delta}^2}\Bigg|_{d\rightarrow4}
 ~~;~~ H\in\{\rho,a_1\}
\label{eq.PiH.2}
\end{eqnarray}
where, $\Delta = M^2-x(1-x)q^2-i\epsilon$ and the space-time dimension has been changed from $4$ to $d$ in order to implement the 
dimensional regularization. Shifting momentum $k\rightarrow (k-xq)$, we perform the momentum integrals of the above equations to get
\begin{eqnarray}
\Pi_\pi(q) &=& \frac{N_c N_f}{4\pi^2}\TB{\frac{1}{2}q^2\Gamma(\varepsilon)\int_{0}^{1}dx\FB{\frac{4\pi\lambda}{\Delta}}^\varepsilon 
+ M^2\Gamma(\varepsilon-1) \FB{\frac{4\pi\lambda}{M^2}}^\varepsilon}, \label{eq.Pi.pi.final} \\
\Pi_\sigma(q) &=& \frac{N_c N_f}{4\pi^2}\TB{\frac{1}{2}(q^2-4M^2)\Gamma(\varepsilon)\int_{0}^{1}dx\FB{\frac{4\pi\lambda}{\Delta}}^\varepsilon 
	+ M^2\Gamma(\varepsilon-1) \FB{\frac{4\pi\lambda}{M^2}}^\varepsilon}, \label{eq.Pi.sigma.final} \\
\Pi^\munu_\rho(q) &=& -\frac{N_c N_f}{2\pi^2}\Gamma(\varepsilon) \FB{g^\munu - \frac{q^\mu q^\nu}{q^2}} q^2
\int_{0}^{1}dxx(1-x)\FB{\frac{4\pi\lambda}{\Delta}}^\varepsilon,  \label{eq.Pi.rho.final} \\
\Pi^\munu_{a_1}(q) &=& \frac{N_c N_f}{2\pi^2} \Gamma(\varepsilon)\int_{0}^{1}dx
\TB{\FB{g^\munu - \frac{q^\mu q^\nu}{q^2}}\Delta + \frac{q^\mu q^\nu}{q^2} M^2 }\FB{\frac{4\pi\lambda}{\Delta}}^\varepsilon \label{eq.Pi.a1.final}
\end{eqnarray}
where, $\varepsilon=(2-d/2)$ and note that, the UV-divergences have appeared as the pole of the Gamma functions. 
The above quantities have the following expansion about $\varepsilon=0$:
\begin{eqnarray}
\Pi_\pi(q) &=& \frac{N_cN_f}{4\pi^2}\Bigg[ 
-M^2\SB{\frac{1}{\varepsilon}-\gamma_\text{E} +1 + \ln\FB{\frac{4\pi\lambda}{M^2}}} + \frac{1}{2}q^2\int_{0}^{1}dx
\SB{\frac{1}{\varepsilon}-\gamma_\text{E} + \ln\FB{\frac{4\pi\lambda}{\Delta}}} \Bigg], \label{eq.Pi.pi.final2}\\
\Pi_\sigma(q) &=& \frac{N_cN_f}{4\pi^2}\Bigg[ 
-M^2\SB{\frac{1}{\varepsilon}-\gamma_\text{E} +1 + \ln\FB{\frac{4\pi\lambda}{M^2}}} + \frac{1}{2}(q^2-4M^2)\int_{0}^{1}dx
\SB{\frac{1}{\varepsilon}-\gamma_\text{E} + \ln\FB{\frac{4\pi\lambda}{\Delta}}} \Bigg], \label{eq.Pi.sigma.final2}\\
\Pi^\munu_\rho(q) &=& -\frac{N_c N_f}{2\pi^2} \FB{g^\munu - \frac{q^\mu q^\nu}{q^2}} q^2
\int_{0}^{1}dxx(1-x)\TB{\frac{1}{\varepsilon}-\gamma_\text{E}+\ln\FB{\frac{4\pi\lambda}{\Delta}} }, \label{eq.Pi.rho.final2}\\
\Pi^\munu_{a_1}(q) &=& \frac{N_c N_f}{2\pi^2} \int_{0}^{1}dx
\TB{\FB{g^\munu - \frac{q^\mu q^\nu}{q^2}}\Delta + \frac{q^\mu q^\nu}{q^2} M^2 }
\TB{\frac{1}{\varepsilon}-\gamma_\text{E}+\ln\FB{\frac{4\pi\lambda}{\Delta}} }. \label{eq.Pi.a1.final2}
\end{eqnarray}


\section{CALCULATION OF THERMO-MAGNETIC POLARIZATION FUNCTIONS} \label{app.pola.vbt}
In this appendix, we will briefly sketch how to obtain Eqs.~\eqref{eq.repi.tb.1}-\eqref{eq.impi.tb.2}. 
Substituting $\bm{S''}_{11}$ from Eq.~\eqref{eq.s11.tb} into Eqs.~\eqref{eq.Pi.pi.3}-\eqref{eq.Pi.a1.3}, we get 
after evaluating the traces over flavor and colour spaces for $\qper=0$
\begin{eqnarray}
\bm{\Pi}_h^{\text{B} 11}(\qpll) = i \sum_{l=0}^{\infty}\sum_{n=0}^{\infty}\sum_{f\in\{\text{u,d}\}} \int\frac{d^4k}{(2\pi)^4}\tilde{N}_h^{lnf}(\qpll,k)
\TB{\frac{-1}{\kpll^2-M_{lf}^2+i\epsilon} - 2\pi i \eta(k\cdot u)\delta(\kpll^2-M_{lf}^2)} \nn \\
\times \TB{\frac{-1}{\ppll^2-M_{nf}^2+i\epsilon} - 2\pi i \eta(p\cdot u)\delta(\ppll^2-M_{lf}^2)}, \label{eq.pih11.1}\\
\bm{\Pi}_H^{\text{B} \munu11}(\qpll) = i \sum_{l=0}^{\infty}\sum_{n=0}^{\infty}\sum_{f\in\{\text{u,d}\}} \int\frac{d^4k}{(2\pi)^4}\tilde{N}_H^{lnf\munu}(\qpll,k)
\TB{\frac{-1}{\kpll^2-M_{lf}^2+i\epsilon} - 2\pi i \eta(k\cdot u)\delta(\kpll^2-M_{lf}^2)} \nn \\
\times \TB{\frac{-1}{\ppll^2-M_{nf}^2+i\epsilon} - 2\pi i \eta(p\cdot u)\delta(\ppll^2-M_{lf}^2)} \label{eq.piH11.1}
\end{eqnarray}
where $p=(q+k)$ and 
\begin{eqnarray}
\tilde{N}_\pi^{lnf}(\qpll,k) &=& -N_c(-1)^{l+n}e^{-2\alpha_k^f}\Tr_\text{d}\TB{\scrD_{lf}(k)\gamma^5\scrD_{nf}(p)\gamma^5}, \\
\tilde{N}_\sigma^{lnf}(\qpll,k) &=& N_c(-1)^{l+n}e^{-2\alpha_k^f}\Tr_\text{d}\TB{\scrD_{lf}(k)\scrD_{nf}(p)}, \\
\tilde{N}_\rho^{lnf\munu}(\qpll,k) &=& N_c(-1)^{l+n}e^{-2\alpha_k^f}\Tr_\text{d}\TB{\scrD_{lf}(k)\gamma^\mu\scrD_{nf}(p)\gamma^\nu}, \\
\tilde{N}_{a_1}^{lnf\munu}(\qpll,k) &=& N_c(-1)^{l+n}e^{-2\alpha_k^f}\Tr_\text{d}\TB{\scrD_{lf}(k)\gamma_\mu\gamma^5\scrD_{nf}(p)\gamma^\nu\gamma^5}.
\end{eqnarray}
Evaluating the trace over Dirac matrices, the above equations become,
\begin{eqnarray}
\tilde{N}_h^{lnf}(\qpll,k) &=& -N_c(-1)^{l+n}e^{-2\alpha_k^f}8\TB{
-8\kper^2L_{l-1}^1(2\alpha_k^f)L_{n-1}^1(2\alpha_k^f) \right. \nn \\ 
&& \left.- (\kpll^2+\kpll\cdot\qpll-aM^2)\SB{L_{l}(2\alpha_k^f)L_{n}(2\alpha_k^f)+L_{l-1}(2\alpha_k^f)L_{n-1}(2\alpha_k^f)}
}, \label{eq.ntilde.1}\\
\tilde{N}_H^{lnf\munu}(\qpll,k) &=& N_c(-1)^{l+n}e^{-2\alpha_k^f}8\TB{
	8(2\kper^\mu\kper^\nu-\kper^2g^\munu)L_{l-1}^1(2\alpha_k^f)L_{n-1}^1(2\alpha_k^f) \right. \nn \\ 
	&& \left. -\SB{ (\kpll^2+\kpll\cdot\qpll-aM^2)\gpll^\munu - 2\kpll^\mu\kpll^\nu -(\kpll^\mu\qpll^\nu+\kpll^\nu\qpll^\mu) }
	\SB{L_{l}(2\alpha_k^f)L_{n}(2\alpha_k^f)+L_{l-1}(2\alpha_k^f)L_{n-1}(2\alpha_k^f)} \right. \nn \\ 
&& \left. + (\kpll^2+\kpll\cdot\qpll-aM^2)\gper^\munu \SB{L_{l}(2\alpha_k^f)L_{n-1}(2\alpha_k^f)+L_{l-1}(2\alpha_k^f)L_{n}(2\alpha_k^f)} 
} \label{eq.ntilde.2}
\end{eqnarray}
with $a = \begin{cases}
1 ~~\text{for}~~ \pi,\rho \\
-1 ~~\text{for}~~ \sigma,a_1
\end{cases} 
$. Substituting Eqs.~\eqref{eq.ntilde.1} and \eqref{eq.ntilde.2} into Eqs.~\eqref{eq.pih11.1} and \eqref{eq.piH11.1}, we 
can perform the $d^2\kper$ integral using the orthogonality of the Laguerre polynomials to obtain 
\begin{eqnarray}
\bm{\Pi}_h^{\text{B} 11}(\qpll) = i \sum_{l=0}^{\infty}\sum_{n=(l-1)}^{(l+1)}\sum_{f\in\{\text{u,d}\}} \int\frac{d^2\kpll}{(2\pi)^2}N_h^{lnf}(\qpll,k)
\TB{\frac{-1}{\kpll^2-M_{lf}^2+i\epsilon} - 2\pi i \eta(k\cdot u)\delta(\kpll^2-M_{lf}^2)} \nn \\
\times \TB{\frac{-1}{\ppll^2-M_{nf}^2+i\epsilon} - 2\pi i \eta(p\cdot u)\delta(\ppll^2-M_{lf}^2)}, \label{eq.pih11.2}\\
\bm{\Pi}_H^{\text{B} \munu11}(\qpll) = i \sum_{l=0}^{\infty}\sum_{n=(l-1)}^{(l+1)}\sum_{f\in\{\text{u,d}\}} \int\frac{d^2\kpll}{(2\pi)^2}N_H^{lnf\munu}(\qpll,k)
\TB{\frac{-1}{\kpll^2-M_{lf}^2+i\epsilon} - 2\pi i \eta(k\cdot u)\delta(\kpll^2-M_{lf}^2)} \nn \\
\times \TB{\frac{-1}{\ppll^2-M_{nf}^2+i\epsilon} - 2\pi i \eta(p\cdot u)\delta(\ppll^2-M_{lf}^2)} \label{eq.piH11.2}
\end{eqnarray}
where, 
\begin{eqnarray}
N_h^{lnf}(\qpll,k) &=& -\frac{N_c}{\pi} |e_fB| \TB{
	4|e_fB|n\delta_{l-1}^{n-1} - (\kpll^2+\kpll\cdot\qpll-aM^2)(\delta_{l}^{n}+\delta_{l-1}^{n-1})
}, \label{eq.Nh.tb}\\
N_H^{lnf\munu}(\qpll,k) &=& \frac{N_c}{\pi} |e_fB| \TB{
	4|e_fB|n\delta_{l-1}^{n-1}\gpll^\munu 
	- \SB{(\kpll^2+\kpll\cdot\qpll-aM^2)\gpll^\munu-2\kpll^\mu\kpll^\nu - (\kpll^\mu\qpll^\nu+\kpll^\nu\qpll^\mu)  }
	(\delta_{l}^{n}+\delta_{l-1}^{n-1}) \right. \nn \\ 
	&& \left.- (\kpll^2+\kpll\cdot\qpll-aM^2)\gper^\munu(\delta_{l}^{n-1}+\delta_{l-1}^{n})
} \label{eq.NH.tb}.
\end{eqnarray}
Note that, the presence of the Kronecker delta in the above equations has eliminated one of the double sums 
in Eqs.~\eqref{eq.pih11.2} and \eqref{eq.piH11.2} so that the sum over index $n$ runs from $(l-1)$ to $(l+1)$.

The calculation of the imaginary parts of Eqs.~\eqref{eq.pih11.2} and \eqref{eq.piH11.2} is trivial since the imaginary 
parts are free from any UV-divergences. Evaluating the $dk^0$ integral of Eqs.~\eqref{eq.pih11.2} and \eqref{eq.piH11.2} 
and making use of the relations 
\begin{eqnarray}
\IM\Pibarbar_h(\qpll) &=& \text{sign}(q^0)\tanh\FB{\frac{q^0}{2T}}\IM\bm{\Pi}_h^{\text{B}11}(\qpll), \\
\IM\Pibarbar^\munu_H(\qpll) &=& \text{sign}(q^0)\tanh\FB{\frac{q^0}{2T}}\IM\bm{\Pi}_h^{\text{B}\munu11}(\qpll)
\end{eqnarray}
we get
\begin{eqnarray}
\IM \Pibarbar_h(\qpll) &=& -\text{sign}(q^0)\tanh\FB{\frac{q^0}{2T}}\pi\sum_{l=0}^{\infty}\sum_{n=(l-1)}^{(l+1)}\sum_{f\in\{\text{u,d}\}} 
\int_{-\infty}^{\infty}\frac{dk_z}{2\pi} \frac{1}{4\omega_k^{lf}\omega_p^{nf}} \nn \\
&& \hspace{-1.5cm}\TB{ 
	\SB{1-f(\omega_k^{lf})-f(\omega_p^{nf})+2f(\omega_k^{lf})f(\omega_p^{nf})}
	\SB{N^{lnf}_h(k^0=-\omega_k^{lf})\delta(q^0-\omega_k^{lf}-\omega_p^{nf})+ N_h^{lnf}(k^0=\omega_k^{lf})\delta(q^0+\omega_k^{lf}+\omega_p^{nf})} \right. \nn \\ 
	&& \hspace{-1.9cm} \left.
	+ \SB{-f(\omega_k^{lf})-f(\omega_p^{nf})+2f(\omega_k^{lf})f(\omega_p^{nf})}
	\SB{N^{lnf}_h(k^0=-\omega_k^{lf})\delta(q^0-\omega_k^{lf}-\omega_p^{nf})+ N^{lnf}_h(k^0=\omega_k^{lf})\delta(q^0+\omega_k^{lf}+\omega_p^{nf})}
}, \label{eq.impi.tb.1.final} \\
\IM \Pibarbar_H^\munu(\qpll) &=& -\text{sign}(q^0)\tanh\FB{\frac{q^0}{2T}}\pi\sum_{l=0}^{\infty}\sum_{n=(l-1)}^{(l+1)}\sum_{f\in\{\text{u,d}\}} 
\int_{-\infty}^{\infty}\frac{dk_z}{2\pi} \frac{1}{4\omega_k^{lf}\omega_p^{nf}} \nn \\
&& \hspace{-1.5cm}\TB{ 
	\SB{1-f(\omega_k^{lf})-f(\omega_p^{nf})+2f(\omega_k^{lf})f(\omega_p^{nf})}
	\SB{N^{lnf\munu}_H(k^0=-\omega_k^{lf})\delta(q^0-\omega_k^{lf}-\omega_p^{nf})+ N_H^{lnf\munu}(k^0=\omega_k^{lf})\delta(q^0+\omega_k^{lf}+\omega_p^{nf})} \right. \nn \\ 
	&& \hspace{-1.5cm} \left.
	+ \SB{-f(\omega_k^{lf})-f(\omega_p^{nf})+2f(\omega_k^{lf})f(\omega_p^{nf})}
	\SB{N^{lnf\munu}_H(k^0=-\omega_k^{lf})\delta(q^0-\omega_k^{lf}-\omega_p^{nf})+ N^{lnf\munu}_H(k^0=\omega_k^{lf})\delta(q^0+\omega_k^{lf}+\omega_p^{nf})}
}. \nn \\ \label{eq.impi.tb.2.final}
\end{eqnarray}

The temperature dependent real parts of Eqs.~\eqref{eq.pih11.2} and \eqref{eq.piH11.2} are also easy to simplify because of the presence 
of the Dirac delta functions. Thus, evaluating the $dk^0$ integral of the temperature dependent real parts, and making use of the 
relations 
\begin{eqnarray}
\RE\Pibarbar_h(\qpll) = \RE\bm{\Pi}_h^{\text{B}11}(\qpll) ~~~\text{and}~~~
\RE\Pibarbar^\munu_H(\qpll) = \RE\bm{\Pi}_h^{\text{B}\munu11}(\qpll) \label{eq.bar211.real}
\end{eqnarray}
we get,
\begin{eqnarray}
\RE \Pibarbar_h(\qpll) &=& \RE\Pi_{h\text{Vac}}(\qpll,B) - \sum_{l=0}^{\infty}\sum_{n=(l-1)}^{(l+1)}\sum_{f\in\{\text{u,d}\}} 
\int_{-\infty}^{\infty}\frac{dk_z}{2\pi}\mathcal{P}\TB{ 
	\frac{N_h^{lnf}(k^0=-\omega_k^{lf})f(\omega_k^{lf})}{2\omega_k^{lf}\SB{(q^0-\omega_k^{lf})^2-(\omega_p^{nf})^2}} \right. \nn \\ 
	&& \left. \hspace{-0.5cm} + \frac{N_h^{lnf}(k^0=\omega_k^{lf})f(\omega_k^{lf})}{2\omega_k^{lf}\SB{(q^0+\omega_k^{lf})^2-(\omega_p^{nf})^2}} 
	+ \frac{N_h^{lnf}(k^0=-q^0-\omega_p^{nf})f(\omega_p^{nf})}{2\omega_p^{nf}\SB{(q^0+\omega_p^{nf})^2-(\omega_k^{lf})^2}} 
	+ \frac{N_h^{lnf}(k^0=-q^0+\omega_p^{nf})f(\omega_p^{nf})}{2\omega_p^{nf}\SB{(q^0-\omega_p^{nf})^2-(\omega_k^{lf})^2}} 
}, \label{eq.repi.tb.31}\\ 
\RE \Pibarbar^\munu_H(\qpll) &=&  \RE\Pi^\munu_{H\text{Vac}}(\qpll,B) - \sum_{l=0}^{\infty}\sum_{n=(l-1)}^{(l+1)}\sum_{f\in\{\text{u,d}\}} 
\int_{-\infty}^{\infty}\frac{dk_z}{2\pi}\mathcal{P}\TB{ 
	\frac{N_H^{lnf\munu}(k^0=-\omega_k^{lf})f(\omega_k^{lf})}{2\omega_k^{lf}\SB{(q^0-\omega_k^{lf})^2-(\omega_p^{nf})^2}} 
	\right. \nn \\ &&
	\left. 	\hspace{-0.5cm} + \frac{N_H^{lnf\munu}(k^0=\omega_k^{lf})f(\omega_k^{lf})}{2\omega_k^{lf}\SB{(q^0+\omega_k^{lf})^2-(\omega_p^{nf})^2}} 
	+ \frac{N_H^{lnf\munu}(k^0=-q^0-\omega_p^{nf})f(\omega_p^{nf})}{2\omega_p^{nf}\SB{(q^0+\omega_p^{nf})^2-(\omega_k^{lf})^2}} 
	+ \frac{N_H^{lnf\munu}(k^0=-q^0+\omega_p^{nf})f(\omega_p^{nf})}{2\omega_p^{nf}\SB{(q^0-\omega_p^{nf})^2-(\omega_k^{lf})^2}} 
} \label{eq.repi.tb.32}
\end{eqnarray}
where $\RE\Pi_{h\text{Vac}}(\qpll,B)$ and $\RE\Pi^\munu_{H\text{Vac}}(\qpll,B)$ are the temperature independent real parts of 
the analytic thermo-magnetic polarization functions. They, respectively, contain the magnetic field independent and UV-divergent pure vacuum 
polarization functions $\RE\Pi_h(\qpll)$ and $\RE\Pi_H^\munu(\qpll)$ which have to be separated. 
To this end, we will use the dimensional regularization technique as already developed in 
Ref.~\cite{Snigdha_2019}. 
We have, 
\begin{eqnarray}
\Pi_{h\text{Vac}}(\qpll,B) &=& i \sum_{l=0}^{\infty}\sum_{n=(l-1)}^{(l+1)}\sum_{f\in\{\text{u,d}\}} \int\frac{d^2\kpll}{(2\pi)^2}N_h^{lnf}(\qpll,k)
\frac{1}{(\kpll^2-M_{lf}^2+i\epsilon)(\ppll^2-M_{nf}^2+i\epsilon)}, \label{eq.pih11.31}\\
\Pi^\munu_{H\text{Vac}}(\qpll,B) &=& i \sum_{l=0}^{\infty}\sum_{n=(l-1)}^{(l+1)}\sum_{f\in\{\text{u,d}\}} \int\frac{d^2\kpll}{(2\pi)^2}N_H^{lnf\munu}(\qpll,k)
\frac{1}{(\kpll^2-M_{lf}^2+i\epsilon)(\ppll^2-M_{nf}^2+i\epsilon)}. \label{eq.piH11.32}
\end{eqnarray}
Using standard Feynman parametrization, the denominators of the above equations are combined and we get after some simplifications, 
\begin{eqnarray}
\Pi_{h\text{Vac}}(\qpll,B) &=& N_c \sum_{l=0}^{\infty}\sum_{f\in\{\text{u,d}\}}\frac{|e_fB|}{\pi}(2-\delta_l^0)
\lambda^{1-d/2} i\int\frac{d^d\kpll}{(2\pi)^d}\TB{\frac{1}{\kpll^2-M_{lf}^2+i\epsilon}-\frac{1}{2}\tilde{N}_h(\qpll)\int_{0}^{1}dx 
\SB{\frac{1}{(\kpll+x\qpll)^2-\Delta_{ll}^f}}^2}\Bigg|_{d\rightarrow2}, \nn \\ \label{eq.pih11.41} \\
\Pi^\munu_{H\text{Vac}}(\qpll,B) &=& i \sum_{l=0}^{\infty}\sum_{n=(l-1)}^{(l+1)}\sum_{f\in\{\text{u,d}\}}\int_{0}^{1}dx\lambda^{1-d/2} \int\frac{d^d\kpll}{(2\pi)^d}N_H^{lnf\munu}(\qpll,k)
\TB{\frac{1}{(\kpll+x\qpll)^2-\Delta_{ln}^f}}^2\Bigg|_{d\rightarrow2}. \label{eq.piH11.42}
\end{eqnarray}
where $\Delta_{ln}^f = M^2-x(1-x)\qpll^2+2|e_fB|(l-xl+xn)-i\epsilon$ and 
we have changed the longitudinal space-time dimension from $2$ to $d$ so that as before a scale $\lambda$ of dimention GeV$^2$ has 
been introduced. In Eq.~\eqref{eq.pih11.41}, 
$\tilde{N}_h(\qpll)=\qpll^2$ if $h\equiv\pi$ and  $\tilde{N}_h(\qpll)=(\qpll^2-4M^2)$ if $h\equiv\sigma$.
We now perform the $d^d\kpll$ integral after a momentum shift $\kpll\rightarrow(\kpll-x\qpll)$. 
After some simplifications, we arrive at,
\begin{eqnarray}
\Pi_{h\text{Vac}}(\qpll,B) &=& \frac{N_c}{4\pi^2} \sum_{l=0}^{\infty} 
\sum_{f\in\{\text{u,d}\}}|e_fB|(2-\delta_l^0) \Bigg[ \FB{\frac{4\pi\lambda}{M_{lf}^2}}^\varepsilon\Gamma(\varepsilon) 
+ \frac{1}{2}\tilde{N}_h(\qpll)\int_{0}^{1}dx \FB{\frac{4\pi\lambda}{\Delta_{ll}^f}}^{1+\varepsilon}\Gamma(1+\varepsilon)\Bigg]
\Bigg|_{\varepsilon\rightarrow0}, \label{eq.pih11.51}\\
	\Pi^\munu_{H\text{Vac}}(\qpll,B) &=& -\frac{N_c}{4\pi^2} \sum_{l=0}^{\infty}\sum_{n=(l-1)}^{(l+1)} 
	\sum_{f\in\{\text{u,d}\}}|e_fB|\int_{0}^{1}dx 	\Bigg[
	\Big[4|e_fB|n\delta_{l-1}^{n-1}\gpll^\munu + \SB{\FB{aM^2+x(1-x)\qpll^2}\gpll^\munu \right. \nn \\ && \left. -2x(1-x)\qpll^\mu\qpll^\nu}(\delta_{l}^n+\delta_{l-1}^{n-1}) 
	 + \FB{aM^2+x(1-x)\qpll^2}(\delta_{l-1}^n+\delta_{l}^{n-1})\gper^\munu		
		\Big]\Gamma(\varepsilon+1)\frac{1}{\Delta_{ln}^f} 
	\nn \\
	&&  - \SB{(\delta_{l}^n+\delta_{l-1}^{n-1})\varepsilon\gpll^\munu -(1-\varepsilon)(\delta_{l-1}^n+\delta_{l}^{n-1})\gper^\munu }
	\Gamma(\varepsilon)\Bigg]\FB{\frac{4\pi\lambda}{\Delta_{ln}^f}}^\varepsilon
	\Bigg|_{\varepsilon\rightarrow0}. \label{eq.piH11.52}
\end{eqnarray}
The sum over the indices $l$ and $n$ in the above equations can now be performed and be expressed in terms of the Hurwitz zeta function as
\begin{eqnarray}
\Pi_{h\text{Vac}}(\qpll,B) &=& \frac{N_c}{4\pi^2} \sum_{f\in\{\text{u,d}\}} \Bigg[ 
|e_fB|\SB{-\FB{\frac{2|e_fB|}{M^2}}^\varepsilon+2\zeta\FB{\varepsilon,\frac{M^2}{2|e_fB|}}}\Gamma(\varepsilon) \nn \\
&& + \frac{1}{2}\tilde{N}_h(\qpll) \int_{0}^{1}dx\SB{-\frac{1}{2}z^{-1-\varepsilon} 
	+ \zeta(1+\varepsilon,z)}\Gamma(\varepsilon+1) \Bigg]\FB{\frac{4\pi\lambda}{2|e_fB|}}^\varepsilon
\Bigg|_{\varepsilon\rightarrow0}, \label{eq.pih11.61}\\
\Pi^\munu_{H\text{Vac}}(\qpll,B) &=& -\frac{N_c}{8\pi^2}  
\sum_{f\in\{\text{u,d}\}}\int_{0}^{1}dx 	\Bigg[
\Big[4|e_fB|\FB{\zeta(\varepsilon,z)-z\zeta(1+\varepsilon,z)}\gpll^\munu
	+\SB{\FB{aM^2+x(1-x)\qpll^2}\gpll^\munu \right. \nn \\ && \left.
		-2x(1-x)\qpll^\mu\qpll^\nu} \FB{2\zeta(1+\varepsilon,z)-z^{-1-\varepsilon}}
+ 2\FB{aM^2+x(1-x)\qpll^2}\zeta(1+\varepsilon,z+x)\gper^\munu
\Big] \Gamma(\varepsilon+1) \nn \\
&& - 2|e_fB|\SB{\varepsilon\FB{2\zeta(\varepsilon,z)-z^{-\varepsilon}}\gpll^\munu 
-2(1-\varepsilon)\zeta(\varepsilon,z+x)\gper^\munu}\Gamma(\varepsilon)\Bigg]\FB{\frac{4\pi\lambda}{2|e_fB|}}^\varepsilon
\Bigg|_{\varepsilon\rightarrow0} \label{eq.piH11.62}
\end{eqnarray}
where $z=\frac{\Delta}{2|e_fB|}$. Expanding the above equations about $\varepsilon=0$, we get after some simplifications,
\begin{eqnarray}
\Pi_{h\text{Vac}}(\qpll,B) &=& \frac{N_cN_f}{4\pi^2}\Bigg[ 
-M^2\SB{\frac{1}{\varepsilon}-\gamma_\text{E} +1 + \ln\FB{\frac{4\pi\lambda}{M^2}}} + \frac{1}{2}\tilde{N}_h(\qpll)\int_{0}^{1}dx
\SB{\frac{1}{\varepsilon}-\gamma_\text{E} + \ln\FB{\frac{4\pi\lambda}{\Delta}}} \Bigg] \nn \\ 
&& + \frac{N_c}{4\pi^2}\sum_{f\in\{\text{u,d}\}} \Bigg[ M^2 + (|e_fB|-M^2)\ln\FB{\frac{M^2}{2|e_fB|}}
+ 2|e_fB|\SB{\ln\Gamma\FB{\frac{M^2}{2|e_fB|}}-\ln\sqrt{2\pi}}  \nn \\
&& + \frac{1}{2}\tilde{N}_h(\qpll)\int_{0}^{1}dx\SB{\ln z-\psi(z)-\frac{1}{2z}}
\Bigg], \label{eq.pipi.vac.1} 
\end{eqnarray}
\begin{eqnarray}
\Pi^\munu_{\rho\text{Vac}}(\qpll,B) &=& -\frac{N_cN_f}{2\pi^2}\FB{g^\munu-\frac{\qpll^\mu\qpll^\nu}{\qpll^2}}\qpll^2\int_{0}^{1}dxx(1-x)
\TB{\frac{1}{\varepsilon}-\gamma_\text{E}+ \ln\FB{\frac{4\pi\lambda}{M^2}}}  \nn \\ 
&& - \frac{N_c}{4\pi^2}\sum_{f\in\{\text{u,d}\}}\int_{0}^{1}dx \Bigg[
\FB{\qpll^2g^\munu-\qpll^\mu\qpll^\nu}2x(1-x)\ln z 
- \FB{\qpll^2\gpll^\munu-\qpll^\mu\qpll^\nu}x(1-x)\FB{2\psi(z)+1/z} \nn \\
&& + \TB{\FB{\Delta-2M^2}\psi\FB{z+x} + \Delta 
+ 2|e_fB|\SB{\ln\Gamma\FB{z+x}-\ln\sqrt{2\pi}}}\gper^\munu
\Bigg], \label{eq.pirho.vac.1} 
\end{eqnarray}
\begin{eqnarray}
\Pi^\munu_{a_1\text{Vac}}(\qpll,B) &=& \frac{N_cN_f}{2\pi^2}\int_{0}^{1}dx
\TB{\FB{g^\munu-\frac{\qpll^\mu\qpll^\nu}{\qpll^2}}\Delta + \frac{\qpll^\mu\qpll^\nu}{\qpll^2}M^2}
\TB{\frac{1}{\varepsilon}-\gamma_\text{E}+ \ln\FB{\frac{4\pi\lambda}{M^2}}}  \nn \\ 
&& - \frac{N_c}{4\pi^2}\sum_{f\in\{\text{u,d}\}}\int_{0}^{1}dx \Bigg[
\FB{g^\munu\Delta+x(1-x)\qpll^\mu\qpll^\nu}(-2\ln z)
+ \FB{\gpll^\munu\Delta+x(1-x)\qpll^\mu\qpll^\nu}\FB{2\psi(z)+1/z} \nn \\
&& + \TB{\Delta\psi\FB{z+x} + \Delta 
	+ 2|e_fB|\SB{\ln\Gamma\FB{z+x}-\ln\sqrt{2\pi}}}\gper^\munu
\Bigg]. \label{eq.pia1.vac.1}
\end{eqnarray}
Comparing the RHS of Eqs.~\eqref{eq.pipi.vac.1}-\eqref{eq.pia1.vac.1} with that of Eqs.~\eqref{eq.Pi.pi.final2}-\eqref{eq.Pi.a1.final2}, 
we find that the divergent pure vacuum contributions have completely been untangled on the RHS of the above equations. Thus making use of 
Eqs.~\eqref{eq.Pi.pi.final2}-\eqref{eq.Pi.a1.final2}, the above equations can be rewritten as
\begin{eqnarray}
\Pi_{h\text{Vac}}(\qpll,B) &=& \Pi_{h}(\qpll) + \Pi_{h\text{B}}(\qpll,B), \label{eq.pih.B}\\
\Pi^\munu_{H\text{Vac}}(\qpll,B) &=& \Pi^\munu_{H}(\qpll) + \Pi^\munu_{HB}(\qpll,B) \label{eq.piH.B}
\end{eqnarray}
where,
\begin{eqnarray}
\Pi_{\pi\text{B}}(\qpll,B) &=&  \frac{N_c}{4\pi^2}\sum_{f\in\{\text{u,d}\}} \Bigg[ M^2 + (|e_fB|-M^2)\ln\FB{\frac{M^2}{2|e_fB|}}
+ 2|e_fB|\SB{\ln\Gamma\FB{\frac{M^2}{2|e_fB|}}-\ln\sqrt{2\pi}}  \nn \\
&& + \frac{1}{2}\qpll^2\int_{0}^{1}dx\SB{\ln z-\psi(z)-\frac{1}{2z}}
\Bigg], \label{eq.pipi.bvac.final} 
\end{eqnarray}
\begin{eqnarray}
\Pi_{\sigma\text{B}}(\qpll,B) &=& \frac{N_c}{4\pi^2}\sum_{f\in\{\text{u,d}\}} \Bigg[ M^2 + (|e_fB|-M^2)\ln\FB{\frac{M^2}{2|e_fB|}}
+ 2|e_fB|\SB{\ln\Gamma\FB{\frac{M^2}{2|e_fB|}}-\ln\sqrt{2\pi}}  \nn \\
&& + \frac{1}{2}\FB{\qpll^2-4M^2}\int_{0}^{1}dx\SB{\ln z-\psi(z)-\frac{1}{2z}}
\Bigg], \label{eq.pisigma.bvac.final} 
\end{eqnarray}
\begin{eqnarray}
\Pi^\munu_{\rho\text{B}}(\qpll,B) &=& -\frac{N_c}{4\pi^2}\sum_{f\in\{\text{u,d}\}}\int_{0}^{1}dx \Bigg[
\FB{\qpll^2g^\munu-\qpll^\mu\qpll^\nu}2x(1-x)\ln z 
- \FB{\qpll^2\gpll^\munu-\qpll^\mu\qpll^\nu}x(1-x)\FB{2\psi(z)+1/z} \nn \\
&& + \TB{\FB{\Delta-2M^2}\psi\FB{z+x} + \Delta 
	+ 2|e_fB|\SB{\ln\Gamma\FB{z+x}-\ln\sqrt{2\pi}}}\gper^\munu
\Bigg], \label{eq.pirho.bvac.final} 
\end{eqnarray}
\begin{eqnarray}
\Pi^\munu_{a_1\text{B}}(\qpll,B) &=& -\frac{N_c}{4\pi^2}\sum_{f\in\{\text{u,d}\}}\int_{0}^{1}dx \Bigg[
\FB{g^\munu\Delta+x(1-x)\qpll^\mu\qpll^\nu}(-2\ln z)
+ \FB{\gpll^\munu\Delta+x(1-x)\qpll^\mu\qpll^\nu}\FB{2\psi(z)+1/z} \nn \\
&& + \TB{\Delta\psi\FB{z+x} + \Delta 
	+ 2|e_fB|\SB{\ln\Gamma\FB{z+x}-\ln\sqrt{2\pi}}}\gper^\munu
\Bigg]. \label{eq.pia1.bvac.final}
\end{eqnarray}

\bibliographystyle{apsrev4-1}
\bibliography{snigdha}

\end{document}